\documentclass[9pt,twocolumn,twoside]{pnas-new}
\templatetype{pnasresearcharticle} % Choose template 
\title{Time-resolved fast turbulent dynamo in a laser plasma}

\author[a,b,1]{A.~F.~A.~Bott}
\author[a,c,d,e]{P.~Tzeferacos} 
\author[a]{L.~Chen}
\author[a,f]{C.A.J.~Palmer}
\author[a]{A.~Rigby}
\author[a]{A.~Bell}
\author[g,h]{R.~Bingham}
\author[i]{A.~Birkel}
\author[j]{C.~Graziani}
\author[e]{D.~H.~Froula}
\author[e]{J.~Katz}
\author[k,l]{M.~Koenig}
\author[b]{M.W.~Kunz}
\author[i]{C.~K.~Li}
\author[a]{J.~Meinecke}
\author[a]{F.~Miniati}
\author[i]{R.~Petrasso}
\author[m]{H.-S.~Park}
\author[m]{B.~A.~Remington}
\author[n]{B.~Reville}
\author[m]{J.~S.~Ross}
\author[o]{D.~Ryu}
\author[m]{D.~Ryutov}
\author[i]{F.~S\'eguin}
\author[p]{T.~G.~White}
\author[a]{A.~A.~Schekochihin}
\author[c]{D.~Q.~Lamb}
\author[a,c]{G.~Gregori}

\affil[a]{Department of Physics, University of Oxford, Parks Road, Oxford OX1 3PU, UK}
\affil[b]{Department of Astrophysical Sciences, Princeton University, 4 Ivy Ln, Princeton, NJ 08544, USA}
\affil[c]{Department of Astronomy and Astrophysics, University of Chicago, 5640 S. Ellis Ave, Chicago, IL 60637, USA}
\affil[d]{Department of Physics and Astronomy, University of Rochester, 206 Bausch \& Lomb Hall, Rochester, NY 14627}
\affil[e]{Laboratory for Laser Energetics, University of Rochester, 250 E River Rd, Rochester, NY 14623, USA}
\affil[f]{School of Mathematics and Physics, Queens University Belfast, Belfast BT7 1NN, UK}
\affil[g]{Rutherford Appleton Laboratory, Chilton, Didcot OX11 0QX, UK}
\affil[h]{Department of Physics, University of Strathclyde, Glasgow G4 0NG, UK}
\affil[i]{Massachusetts Institute of Technology, 77 Massachusetts Ave, Cambridge, MA 02139, USA}
\affil[j]{Argonne National Laboratory,Mathematics and Computer Science Division, Argonne, IL, USA}
\affil[k]{LULI, CNRS, CEA, Ecole Polytechnique, UPMC, Univ Paris 06: Sorbonne Universites, Institut Polytechnique de Paris, F-91128 Palaiseau cedex, France}
\affil[l]{Graduate School of Engineering, Osaka University, Suita, Osaka 565-0871, Japan}
\affil[m]{Lawrence Livermore National Laboratory, 7000 East Ave, Livermore, CA 94550, USA}
\affil[n]{Max-Planck-Institut f\"ur Kernphysik, Postfach 10 39 80, 69029 Heidelberg, Germany}
\affil[o]{Department of Physics, School of Natural Sciences, UNIST, Ulsan 44919, Korea}
\affil[p]{Department of Physics, University of Nevada, Reno, Nevada 89557, USA}

\leadauthor{Bott} 

\significancestatement{Our laser-plasma experiment has reproduced the physical process thought to be responsible for generating and sustaining magnetic fields in turbulent plasmas (the `fluctuation dynamo'), and, for the first time in the laboratory, has accessed the parameter regime of relevance to most of the plasma in the universe. Also for the first time, these measurements are time-resolved, which provides evolutionary information about the fluctuation dynamo (including the field's growth rate) previously only available from simulations. The efficient amplification of large-scale magnetic fields seen in our experiment could explain the origin of large-scale fields that are observed in turbulent astrophysical plasmas, but are not predicted by current analytical calculations or idealized simulations of the fluctuation dynamo.}

\authorcontributions{This project was conceived by G.G., D.Q.L., P.T., A.F.A.B., and A.A.S.. The delivery of the experiment was led by G.G. and L.C.. C.-K.L. and R.P. contributed to the proton radiography development and data extraction, while D.H.F. and J.K. contributed to the Thomson scattering diagnostics. P.T. designed, executed, and analyzed the FLASH simulations. The analysis of the experimental and simulation data was led by A.F.A.B. with support from P.T., L.C., C.P., A.R., A.R.B., R.B., C.G., J.K., M.K., C.-K.L., J.M., J.M., R.P., H.-S.P.,B.A.R., B.R., J.S.R., D.Ryu, D.Ryutov, T.G.W., A.A.S., D.Q.L., and G.G. The paper was written by A.F.A.B. with contributions from all other co-authors.}

\authordeclaration{The authors declare that they have no conflicts of interest.}
\correspondingauthor{\textsuperscript{1}To whom correspondence should be addressed. E-mail: abott@princeton.edu}


\begin{abstract}
Understanding magnetic-field generation and amplification in turbulent plasma is 
essential to account for observations of magnetic fields 
in the universe. A theoretical framework attributing the origin and sustainment of these fields to the so-called fluctuation dynamo was recently validated 
by experiments on laser facilities in low-magnetic-Prandtl-number plasmas ($\mathrm{Pm} < 1$). However, the same framework proposes that the fluctuation dynamo should operate differently when $\mathrm{Pm} \gtrsim 1$, the regime relevant to many astrophysical environments such as the intracluster medium of galaxy clusters.
This paper reports a new experiment that creates a laboratory $\mathrm{Pm} \gtrsim 1$ 
plasma dynamo for the first time. We provide a time-resolved
characterization of the plasma's evolution, measuring temperatures, 
densities, flow velocities and magnetic fields, which allows us to explore various stages of the fluctuation dynamo's operation. The magnetic energy in structures with characteristic scales close to the driving scale of the stochastic motions is found to increase by almost three orders of magnitude from its initial value and saturate dynamically.  
It is shown that the growth of these fields occurs exponentially at a rate that is much greater than the turnover rate of the driving-scale stochastic motions. Our results point to the possibility that plasma turbulence produced by strong shear can generate fields more efficiently at the driving scale than anticipated by idealized MHD simulations of the nonhelical fluctuation dynamo; this finding could help explain the large-scale fields inferred from observations of astrophysical systems. 
\end{abstract}

\dates{This manuscript was compiled on \today}
\doi{\url{www.pnas.org/cgi/doi/10.1073/pnas.XXXXXXXXXX}}

\begin{document}

\maketitle
\thispagestyle{firststyle}
\ifthenelse{\boolean{shortarticle}}{\ifthenelse{\boolean{singlecolumn}}{\abscontentformatted}{\abscontent}}{}

% If your first paragraph (i.e. with the \dropcap) contains a list environment (quote, quotation, theorem, definition, enumerate, itemize...), the line after the list may have some extra indentation. If this is the case, add \parshape=0 to the end of the list environment.
%This PNAS journal template is provided to help you write your work in the correct journal format.  Instructions for use are provided below. 

%Note: please start your introduction without including the word ``Introduction'' as a section heading (except for math articles in the Physical Sciences section); this heading is implied in the first paragraphs.

%\body

\dropcap{C}osmic magnetic fields play a dynamically important role in a myriad of astrophysical environments~\citep{B15,V18}. 
Understanding how these fields 
%-- typically $\sim 1 - 10\, \mu \mathrm{G}$ in the intracluster medium (ICM) -- 
attained such strengths is a long-standing question in astrophysics~\citep{BS51}. 
Most physical processes thought to generate seed magnetic fields in initially unmagnetized plasma, such 
 as the Biermann battery mechanism~\citep{KCOR97}, predict field-strength values in astrophysical settings that are far smaller than those observed~\citep{K99,S19}, 
 necessitating the existence of some mechanism for amplifying fields and maintaining them at their observed magnitudes~\citep{SSH06,RKCD08}.
One possible mechanism is the fluctuation dynamo, whereby stochastic motions of plasma lead 
to stretching and folding of magnetic-field lines~\citep{B50,R19}. In this dynamo, fields are amplified exponentially until their  strength 
comes into approximate equipartition with the fluid kinetic energy, saturating growth. 

The fluctuation dynamo is best understood in the context of resistive magnetohydrodynamics (MHD) thanks to both analytical calculations~\citep{K67,VZ72,ZRMS84,KA92} and  simulations~\citep{MFP81,KYM91,MMAG96,CV01,SCTM04,HBD04,S07,CR09,B12,PJR15,SBSW20}. 
In resistive MHD, the fluctuation dynamo can only operate if the magnetic Reynolds number $\mathrm{Rm} \equiv u_L L/\eta$ -- where $L$ is the length scale of driving stochastic motions, $u_\ell$ the characteristic velocity of motions at a given scale $\ell$, and $\eta$ the resistivity of the plasma -- is above some critical threshold, $\mathrm{Rm}_{\rm c}$~\citep{RS81}. 
The precise value of this threshold depends on the magnetic Prandtl number $\mathrm{Pm}$ of the plasma~\citep{S07,BC04,ISCMP07}, defined by $\mathrm{Pm} \equiv \mathrm{Rm}/\mathrm{Re} = \nu/\eta$ (for $\mathrm{Re} \equiv u_L L/\nu$ the fluid Reynolds number and $\nu$ the kinematic viscosity).
If this threshold is surpassed, then any initially dynamically insignificant magnetic field is amplified, and most rapidly so near the resistive scale $\ell_{\eta} \ll L$ (for $\mathrm{Pm} \ll 1$, $\ell_{\eta} \sim \eta/u_{\ell_{\eta}}$; for $\mathrm{Pm} \gtrsim 1$, $\ell_{\eta} \sim \eta/u_{\ell_{\nu}}$).
The nature of this amplification depends on $\rm Pm$, because $\rm Pm$ determines the relative magnitudes of $\ell_\eta$ and the viscous scale $\ell_{\nu} \sim \nu/u_{\ell_{\nu}}$, and thereby whether the stochastic fluid motions driving dynamo action are smooth or chaotic. 
The $\mathrm{Pm}\ll 1$ regime is relevant to stellar and planetary dynamos, while the $\mathrm{Pm} \gtrsim 1$ regime is pertinent to hot, diffuse plasmas such as many astrophysical disks or the intracluster medium (ICM)~\citep{R19}. 

A fundamental question about the character of the fluctuation dynamo in resistive MHD concerns the rate of magnetic-field growth at a given scale. When the growing field is dynamically insignificant, its spectrum is peaked near the resistive scale~\citep{K67,KA92}; magnetic fluctuations at this scale grow exponentially, at a rate proportional to the characteristic turnover rate $\gamma_{\ell_{\nu}} \sim u_{\ell_{\nu}}/\ell_{\nu}$ of motions at the viscous scale (for $\mathrm{Pm} \gtrsim 1$). For Kolmogorov turbulence, $\gamma_{\ell_{\nu}}$ greatly exceeds the characteristic turnover rate $\gamma_{L} \sim u_L/L$ of the driving-scale stochastic motions. Once the magnetic energy at resistive scales becomes comparable to the kinetic energy at the viscous scale, MHD simulations indicate that the magnetic-energy spectrum changes, with the total energy continuing to grow -- albeit secularly rather than exponentially -- and the peak wavenumber moving to scales larger than the resistive scale~\citep{SCTM04,B12,CVB09}. Whether the peak wavenumber ultimately moves to the driving scale of the motions depends on $\rm Pm$: previous simulations of the $\mathrm{Pm} \sim 1$ dynamo (with non-helical flow) suggest that in the saturated state of the dynamo the peak wavenumber is a factor of a few larger than the driving wavenumber~\citep{HBD04,CR09}, while for $\mathrm{Pm} \gg 1$, an excess of energy remains near the resistive scale~\citep{SCTM04}. Thus, whilst simulations of the fluctuation dynamo show that magnetic fields can be amplified very quickly at the resistive scale, dynamically significant fields on the driving scales only develop after many driving-scale eddy turnover times, or possibly not at all.% (as far as numerical evidence tells us).
% RETURN TO THIS AFTER OTHER SMALL CORRECTIONS DONE

%However, there remain serious theoretical questions concerning the viability of MHD fluctuation dynamo theory as a satisfactory explanation for astrophysical magnetic fields. One such issue is the applicability of conventional MHD to actual astrophysical plasmas, which are typically only weakly collisional, and composed of electrons and ions whose Larmor radii are many orders of magnitude smaller than their respective mean-free-paths~\citep{SC06,KSS14}. It is known that dynamo cannot be possible in plasma where the magnetic moment of particles is conserved, a corollary of which being that dynamo action is impossible in certain models sometimes invoked to describe weakly collisional plasma: kinetic magnetohydrodynamics, or the Chew-Goldberger-Low fluid equations~\citep{HSS16}. Several recent simulations of collisionless plasma turbulence have found the dynamo was in fact possible~\citep{SK18}, although the necessary $\mathrm{Rm}_{\rm c}$ threshold was significantly above the equivalent MHD thresholds~\citep{RCSV16}.  

With dynamo experiments now possible, we have a method for exploring 
both the requirements for, and the properties of, the fluctuation dynamo. Until recently, experimental investigations of plasma
dynamos were limited by the practical difficulty of realizing sufficiently large values of $\mathrm{Rm}$
 in the laboratory~\citep{G12,M14,M15,GRM15}.
 However, a recent laser-plasma experiment~\citep{T17,T18} carried out on the Omega Laser Facility~\citep{B97} demonstrated the feasibility of the fluctuation dynamo in a turbulent plasma at $\mathrm{Pm} < 0.5$. In that experiment, a region of turbulent plasma was created by colliding two laser-plasma jets that had first passed through offset grids. The state of this region was characterized, and the magnetic Reynolds number $\mathrm{Rm} \approx 600$ was above the necessary threshold for the onset of the fluctuation dynamo in MHD. 
Magnetic fields were measured using both polarimetry 
and proton imaging, and the magnetic-energy density in the turbulent plasma a few turnover times after collision was found 
to be several orders of magnitude larger than that present during the turbulent region's 
formation. Most significantly, this magnetic-energy density was a finite fraction of the turbulent kinetic-energy density, a key signature of the saturated fluctuation dynamo.

In this paper, we report new experiments on the Omega Laser Facility, which employs a re-designed version of the platform described in~\citep{T18} to create the first laboratory $\mathrm{Pm} \gtrsim 1$
fluctuation dynamo. {\color{black} As before, we used three-dimensional radiation-MHD simulations with FLASH~\citep{F00,T15} to design and interpret the experiments -- see Supplementary Information for details}. Also for the first time, by carrying out multiple identical experiments, we are able to provide a time-resolved characterization of this plasma dynamo's evolution by measuring spatially averaged electron and ion temperatures, densities, flow velocities, and magnetic fields with a time resolution smaller than the turnover time of the plasma's driving-scale stochastic motions. Such a characterization is an important advance over our previous OMEGA experiment, which did not measure the growth rate of magnetic fields. 
\color{black}{Finally, the concerted analysis of the experimental data in tandem with the simulation results enabled a thorough assessment of the dynamo mechanism realized in our experiment}. 
\color{black}

%, as they could relate to the tangled magnetic fields in the ICM
%on account of the previously outlined theoretical expectation that the fluctuation dynamo proceeds through various stages, each characterized by distinct physics. 
%Such a characterization is important on account of the non-stationary nature of the plasma's interaction. 
%Most significantly, we experimentally determine a lower bound on the fluctuation dynamo's growth rate, finding that growth occurs much more rapidly that the turnover rate of driving-scale stochastic motions in the plasma. 

\section*{Experimental Design}\label{exp_design}

\begin{figure}[t!]
  \centering
\includegraphics[width=\linewidth]{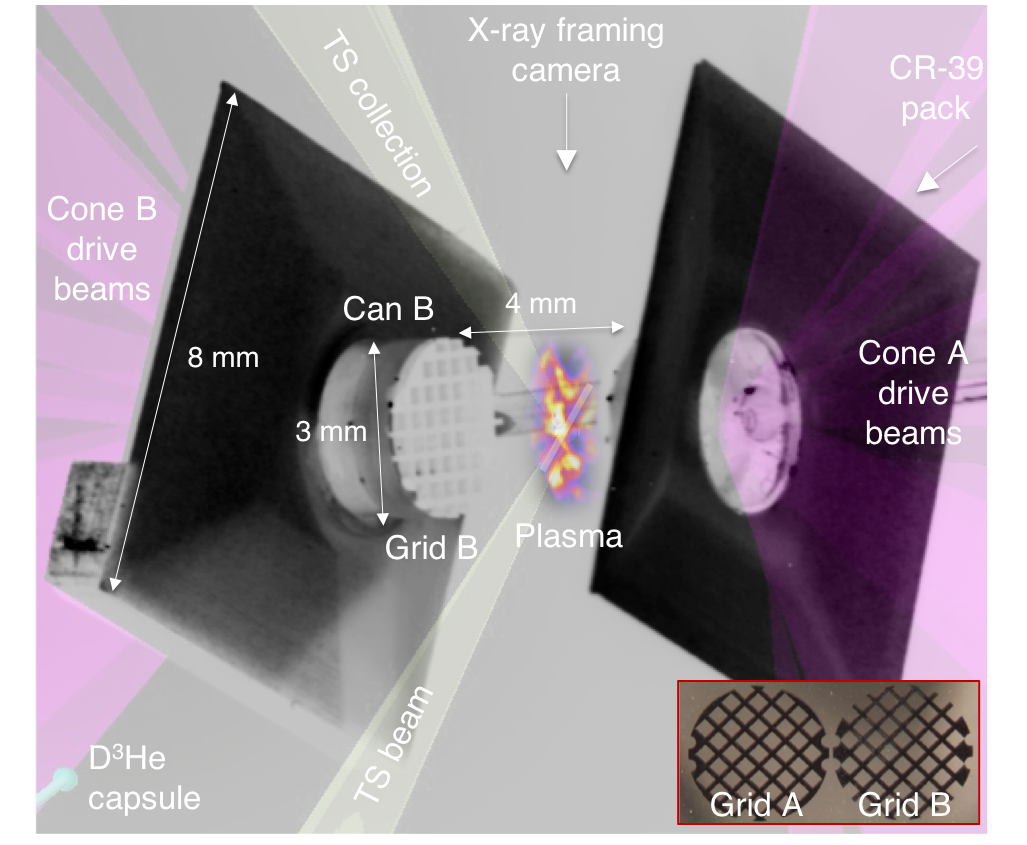}
\caption{\textbf{Experimental set-up}. An annotated photograph of a target used in our experiment. 
The laser-beam-driven foils are composed of CH plastic (i.e., 50\% carbon, 50\% hydrogen by atom number) and are 3 mm in diameter and 50 
$\mu$m in thickness; attached to the front sides of each foil are 230 $\mu$m thick, 3 mm diameter annular 
`washers', also composed of CH plastic, with a 400 $\mu$m central hole.
The separation between the two opposing foils is 8 mm. 
The shields (which prevent direct interaction between the 
front- and rear-side blow-off plasmas) are also CH plastic. CH plastic cans 
attach polyimide grids to the foils; the grids themselves are 250 $\mu$m thick, with a 3 mm diameter, 
300 $\mu$m holes and 100 $\mu$m wires. The holes in the opposing grids are chosen to be offset (see bottom right); grid A has a hole located at its center, while grid B has crossing rods.
Ten 500 J drive beams (individual pulse length 1 ns) with $351$ nm wavelength and 800 $\mu$m 
focal spot size were applied to each foil, configured to deliver a 10 ns staggered flat pulse shape with a total 
energy per foil of 5 kJ. The orientation of the Thomson scattering (TS) beam is denoted, as well as the 
cylindrical scattering volume and collection direction. A D$^3$He capsule is attached to the target 
for the proton imaging diagnostic (see Materials and Methods for details): 
fusion protons are generated by the capsule's implosion, pass between the target grids, and
are detected via a CR-39 pack positioned as shown.} 
\label{fig:expsetup}
\end{figure}

The experimental platform employed for the experiment (see Figure \ref{fig:expsetup} for a schematic of the experimental target) generates a turbulent plasma in the following manner.
Ten long-pulse laser beams illuminate two opposing CH foils, 
creating counter-propagating supersonic plasma jets. These jets then pass through offset grids
before colliding at the experimental target's center. On collision, the jets coalesce, forming an `interaction region' of 
plasma (demarcated by two shocks) whose density and temperature are significantly greater than that of either jet. 
The inhomogeneity and asymmetry of the initial plasma-jet density and flow profiles  
gives rise to significant shearing motions in the interaction region; this 
facilitates Kelvin-Helmholtz (KH) instabilities over a range of length scales, 
and thus significant stochasticity emerges in the flow profile as the interaction region develops. 
In contrast to the initial jet motion, stochastic motions in the interaction region are subsonic, 
because of their reduced characteristic speeds and the higher temperature of the plasma in the interaction-region (a result of compressive heating). At a given instant, we characterize this plasma using various 
experimental diagnostics: X-ray imaging for investigating the spatial distribution of the plasma in the interaction region plasma (see Section \ref{Xray_results}),
Thomson scattering for measuring the plasma properties (Section \ref{TS_results}), and proton imaging 
for quantifying magnetic fields (Section \ref{protonimag_results}). 
%To aid the design, execution and interpretation of the experiment, three-dimensional MHD FLASH simulations of the experiment were also performed (see Supplementary Information). 

Despite some similarities with the previous OMEGA experiment investigating dynamo processes~\citep{T18}, the design of the new experiment was different in a key regard. In order to realize a larger $\rm Pm$, chlorine dopants 
previously introduced into the CH foils to enhance X-ray emissivity of the plasma 
were removed. Their presence in even 
moderate quantities was found to reduce initial plasma-jet velocities, cool the plasma radiatively and increase the effective ion charge; all three effects in combination reduced $\mathrm{Pm}$ significantly.
We also made a number of other improvements to the target's design. The thickness of the grid wires 
was decreased to 100 $\mu$m, whilst the hole width was kept at 300 $\mu$m (see Figure \ref{fig:expsetup}, bottom 
right). This change was made in order to deliver more kinetic energy to the interaction region and reduce the inhomogeneity of the 
interaction region's global morphology arising from the asymmetry of the grids.
Finally, rod supports connecting the grids to the CH foils were 
removed and the grids instead attached via CH `cans' (see Figure~\ref{fig:expsetup}). This 
alteration provided both the X-ray framing camera and proton imaging 
diagnostics with unobstructed views of the interaction region. Further discussion of these target modifications is given in~\citep{M17}.

%In spite of some similarities with the previously mentioned OMEGA experiment investigating dynamo processes~\citep{T18}, the design of the new experiment was distinct in several key regards. The thickness of the grid wires 
%was decreased to 100 $\mu$m, whilst the hole width was maintained at 300 $\mu$m (see Figure \ref{fig:expsetup}, bottom 
%right). This change was made in order to reduce inhomogeneity of the 
%interaction region's global morphology arising from the asymmetry of the grids. Chlorine dopants 
%previously introduced into the CH foils in order to enhance X-ray emissivity of the plasma 
%were removed. Their presence in even 
%moderate quantities was found both to cool the plasma radiatively and to increase the effective ion charge: both effects reduced the magnetic Prandtl number significantly. 
%Finally, rod supports connecting the grids to the CH foils were 
%removed and the grids instead attached via CH `cans' (see caption of Figure~\ref{fig:expsetup}). This 
%modification increased the field of view of both the X-ray framing camera and proton imaging 
%diagnostics.
% recent improvements in target fabrication procedures mitigated the risk of target warping, which previously prohibited such a modification from being introduced. 

We also changed somewhat our methodology for diagnosing the plasma state. Instead of employing the Thomson-scattering diagnostic to measure polarization, we used it to measure the 
spectra of high-frequency fluctuations [the electron-plasma-wave (EPW) feature] as well as low-frequency 
fluctuations [the ion-acoustic-wave (IAW) feature] concurrently. Furthermore, instead of the previous
setup that measured the scattering spectrum in a small volume during a 1-ns 
time window, we employed a spatially resolved, 1-ns time-integrated set-up that 
measured the plasma parameters in a cylindrical region passing through the grids' midpoint, with length 1.5 mm and a 
50 $\mu \mathrm{m}^2$ cross-sectional area (see Figure~\ref{fig:expsetup}). This enabled us to measure simultaneously the values of a number of plasma parameters characterizing the interaction-region plasma: mean electron 
number density $\bar{n}_e$, fluctuating electron number density $\Delta n_e$, electron temperature 
$T_e$, ion temperature $T_i$, inflow velocity $\bar{u}_{\rm in}$ and small-scale stochastic velocity $\Delta 
u$. Removing polarimetry from this experiment did not inhibit our ability to measure magnetic fields, because we had previously validated the accuracy of such measurements obtained using proton imaging~\citep{R18}.

In order to characterize the growth of the magnetic fields in our experiment with the requisite time resolution, 
we began to collect data prior to collision and continued to do so at 1.5-ns intervals (on different experimental shots). This time interval was correctly anticipated to be smaller than the turnover time of driving-scale eddies, based on FLASH simulations that were validated by our earlier experiment~\citep{T17,T18}. 
Detailed specifications of the X-ray framing camera diagnostic, the Thomson-scattering diagnostic and the proton-imaging diagnostic are given in Materials and Methods. 

\section*{Measurements}

\subsection{Measuring turbulence: self-emission X-ray imaging} \label{Xray_results}

\begin{figure}
  \centering
\includegraphics[width=\linewidth]{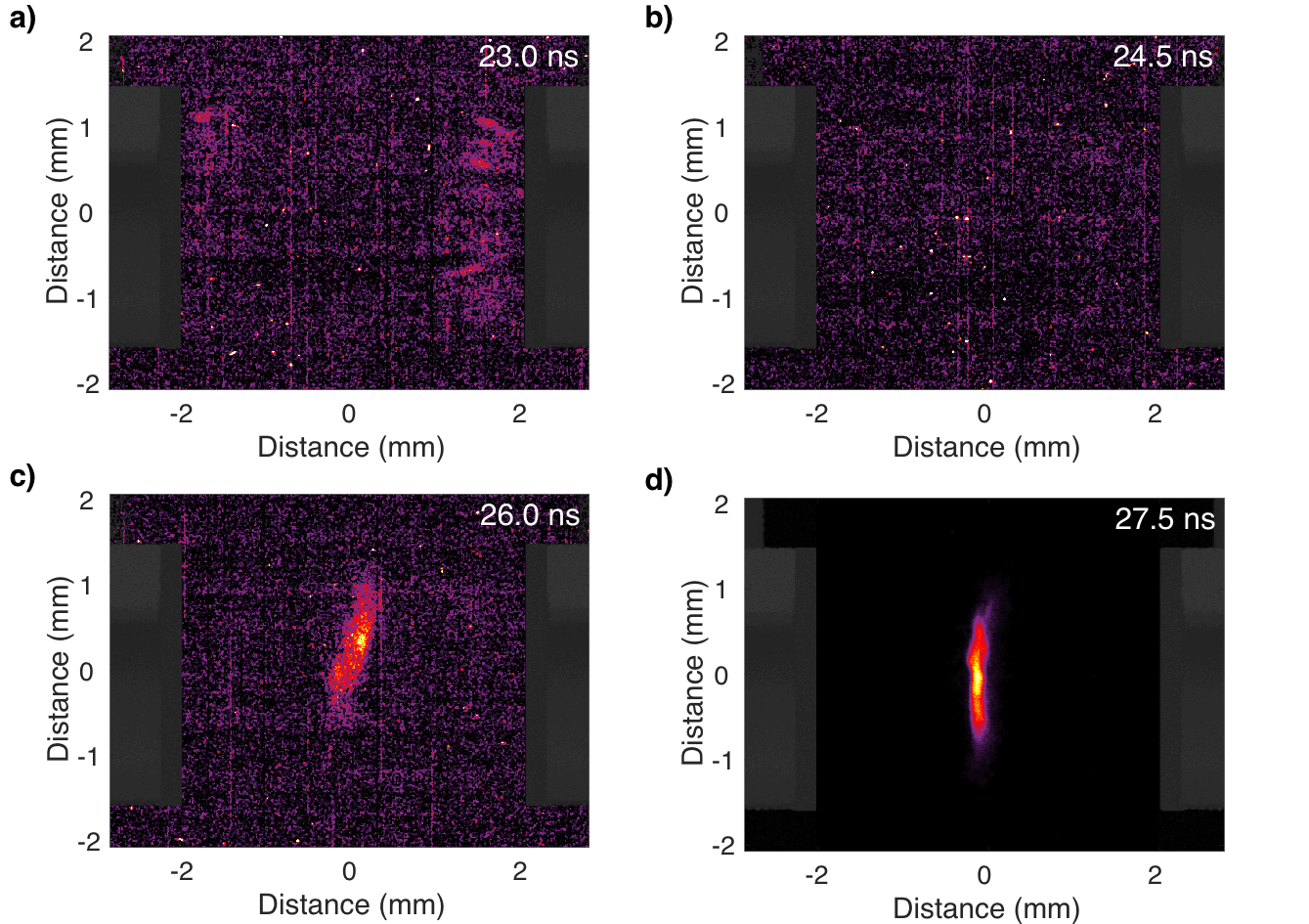}
\caption[X-ray self-emission prior to and at formation of the interaction-region]{\textbf{X-ray self-emission prior to and at formation of the interaction region.} The featured sequence of X-ray images are taken on different experimental shots. The first three images are adjusted to have the same color map, normalized to the 
maximum pixel count (56 counts) of c); the final image is normalized to its own
maximum pixel count. We note that the absence of noise in d) is due 
to the much higher signal-to-noise ratio. 
To aid interpretation of the images, a projection of the 
target is superimposed in dark gray on each image. The respective timings (in ns) of the images after drive-beam laser-pulse initiation are \textbf{a)} 23.0 ns, 
\textbf{b)} 24.5 ns, \textbf{c)} 26.0 ns, and \textbf{d)} 27.5 ns.} 
\label{fig:Xrays_earlytime}
\end{figure}

\begin{SCfigure*}[\sidecaptionrelwidth][t!]
  \centering
\includegraphics[width=11.4cm,height=11.4cm]{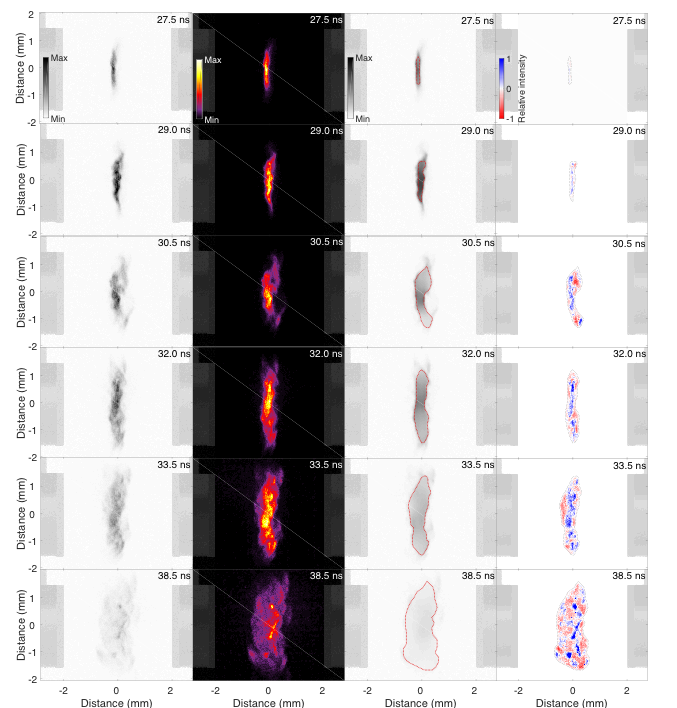}
\caption[The interaction-region plasma's evolution]{\textbf{The interaction-region plasma's evolution.} Self-emission X-ray images of the interaction-region plasma.
Each image was recorded at the indicated time in a different experimental shot. First column: absolute X-ray intensity images, 
normalized to a maximum count value of 1,050 (the maximum count value associated with the interaction-region plasma in any of the images). Second column: X-ray intensity images normalized
by the maximum pixel value in the image. Third column: mean emission profiles calculated from the far-left column; the boundary denoted in red in each image is 
that used to calculate the two-dimensional (2D) Gaussian window function discussed in the main text and the gray-scale map is the same as in the far-left images. Fourth column: 
relative X-ray intensity map calculated from the mean emission profile. Fluctuations with a positive value with respect to the mean intensity are denoted in blue, negative in red, with maximum 
and minimum values set at $\pm 100\%$ of the mean value. Self-emission images for the FLASH simulations, as well as mean emission profiles and relative X-ray intensity maps associated with those images, are shown in Figure S15 of the Supplementary Information.} 
\label{fig:Xrays2} 
\end{SCfigure*}

With the fixed X-ray framing camera's bias employed in our experiment (see Materials and Methods), we find that for times $\lesssim 25 \, 
\mathrm{ns}$, self-emitted X-rays from the individual plasma jets are barely detectable
(see Figure \ref{fig:Xrays_earlytime}a and Figure~\ref{fig:Xrays_earlytime}b). 
However, around 26 ns after the onset of the driving laser pulses, a 
region of emission situated approximately halfway between the grids emerges (Figure \ref{fig:Xrays_earlytime}c).
 1.5 ns later, the total intensity of the region is significantly higher (Figure \ref{fig:Xrays_earlytime}d). 
We conclude that the two plasma flows collide and form the interaction region at around 26 ns. 
Subsequent to the formation of the interaction region, the size of the 
region of bright emission increases both in the direction parallel to the 
`line of centers' (that is, to the line connecting the midpoints of grid A and grid B) and perpendicular to it (see Figure \ref{fig:Xrays2}). 
Emission peaks 3 ns after the interaction-region's coalescence, before decaying
away at later times (first column of Figure \ref{fig:Xrays2}). Random fluctuations in the detected X-ray intensity 
across the emitting region appear concurrently with the peak emission (second column, Figure \ref{fig:Xrays2}) and subsequently become clearly noticeable by eye. 

In order to distinguish fluctuations in emission from global inhomogeneities in the total self-emission from the interaction-region
plasma, we construct relative X-ray intensity maps based on experimentally derived mean emission profiles (a technical description of how these profiles are derived is given in the Supplementary Information). 
The mean emission profiles calculated for the X-ray images shown in the first column of Figure~\ref{fig:Xrays2} 
are given in the third column of the same figure and the corresponding 
relative-intensity images are presented in the fourth column. 

Quantitative analysis of the X-ray images can be carried out by noting that the plasma jets are 
fully ionized even prior to collision ($T_e \approx 180 \, \mathrm{eV}$), and so X-ray emission from the plasma during the interaction is 
dominated by free-free bremsstrahlung. Assuming a thermal distribution of particles, 
the bremsstrahlung spectral density $\epsilon_{\omega}^{\mathrm{ff}}$ for a CH plasma is given by~\citep{RL86}
\begin{equation}
\epsilon_{\omega}^{\mathrm{ff}} =  1.1 \times 10^{-38} Z_{\mathrm{eff}} n_e^2 
T_e^{-1/2} \exp{\left(-\frac{\hbar \omega}{k_B T_e}\right)} \bar{g}_{\mathrm{ff}} \; \mathrm{erg} \, \mathrm{cm}^{-3} , 
\label{bremsstrahlungemiss}
\end{equation}
where $Z_{\mathrm{eff}} = (Z_\mathrm{C}^2+Z_\mathrm{H}^2)/(Z_\mathrm{C}+Z_\mathrm{H})$ 
is the effective ion charge seen by electrons ($Z_\mathrm{H}$ and $Z_{\mathrm{C}}$ 
being the charges of hydrogen and carbon ions, respectively), $\omega$ the frequency of radiation, 
$k_B$ Boltzmann's constant, and $\bar{g}_{\mathrm{ff}}$ 
the velocity-averaged Gaunt factor.
Since the interaction-region plasma is optically thin to X-rays detected by the framing camera, 
the measured (optical) intensity $I$ on the CCD camera satisfies 
$I \propto \int \mathrm{d} s \int \mathrm{d} \omega \, \epsilon_{\omega}^{\mathrm{ff}} \hat{R}(\omega)$, 
where the integral is performed along the line of sight, and $\hat{R}(\omega)$ is a 
function incorporating the (relative) frequency-dependent responses of both the X-ray camera filter and the 
microchannel plate (MCP) (see Supplementary Information, Figure S1). Substituting Eq. [\ref{bremsstrahlungemiss}] into this proportionality relation, we find $I = I(n_e,T_e) \propto \int \mathrm{d} s \, n_e^2 \hat{f}(T_e)$, where 
\begin{equation}
 \hat{f}(T_e) =  \frac{\hat{\mathcal{A}}}{T_e^{-1/2}} \int \mathrm{d} \omega \, \hat{R}(\omega) \exp{\left(-\frac{\hbar \omega}{k_B T_e}\right)} \, , 
 \label{fresponse}
\end{equation}
and $\hat{\mathcal{A}}$ is a normalization constant. The function $\hat{f}(T_e)$ is plotted in the Supplementary Information (Figure S1b); its key property is that for temperatures ${\sim}300$--$500$ eV 
(the characteristic temperature of the plasma just after interaction-region formation -- see Section \ref{TS_results}), 
the measured X-ray intensity is only weakly dependent on temperature. However, the X-ray intensity is a sensitive function of the electron number
density: in short, our X-ray images essentially provide electron-density measurements.

\begin{figure}[t!]
  \centering
\includegraphics[width=\linewidth]{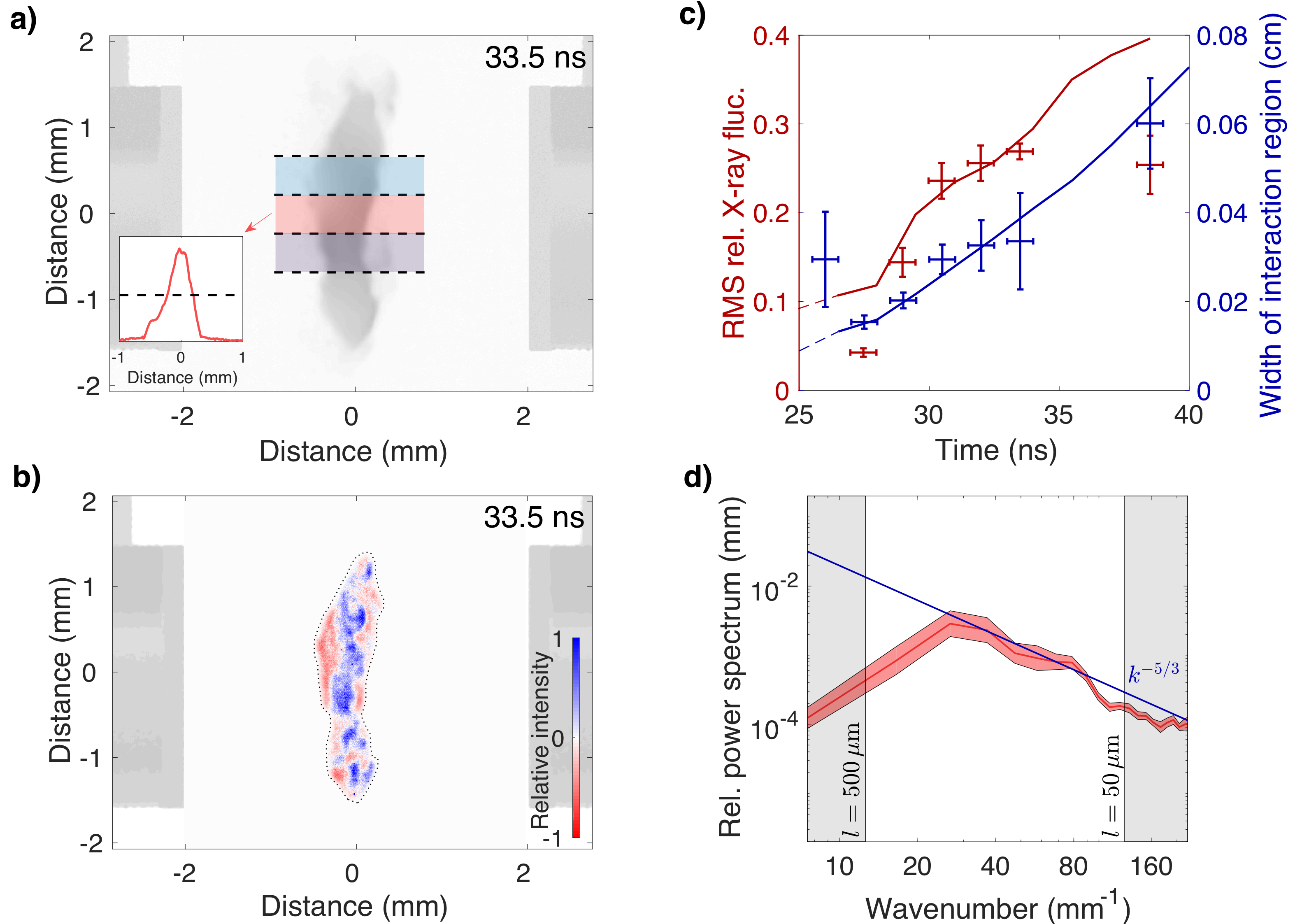}
\caption[Characterizing the interaction-region plasma using X-ray imaging]{\textbf{Characterizing the interaction-region plasma using X-ray imaging.} \textbf{a)} Mean emission profile of an X-ray image,
 recorded 33.5 ns after drive-beam pulse initiation, shown with regions used to
calculate average one-dimensional (1D) parallel profiles. One such profile, along with the half-maximum value, is also depicted. 
\textbf{b)} Relative X-ray intensity map associated with mean emission profile given in a). \textbf{c)} Root-mean-square (RMS) of relative X-ray fluctuations (in red) and the width of the interaction region $l_n$ 
over time (in blue). The behavior of both quantities in the FLASH simulations is also shown (red/blue curves). The dashed portion of the curves correspond to times when the interaction-region plasma is not yet fully collisional and so the simulations are not yet formally valid (see Supplementary Information). To determine an error of the RMS fluctuation measurement, the RMS values of fluctuations in images recorded at the 
same time are employed. 
\textbf{d)} 1D power spectrum of the relative density fluctuations (red line), calculated from the relative X-ray intensity map given in b).
 The error on the spectrum (pink patch) is determined using the power spectrum of b) and 
 the power spectrum of the relative X-ray intensity map derived 
 from the perturbed X-ray image at 33.5 ns equivalent to b). } 
\label{fig:Xrays}
\end{figure}

This conclusion is significant for several reasons. First, the full-width-half-maximum (FWHM) of the 
emitting region can be used as a reasonable measure of the width $l_n$ of the interaction region, 
on account of its increased density compared to either jet. Determining this width 
is essential for extracting magnetic-field estimates from the 
proton-imaging diagnostic (see Section \ref{protonimag_results}). Figure~\ref{fig:Xrays}a illustrates how this measurement is carried 
out in practice: we consider three vertically averaged lineouts of the 
mean emission profile, calculate the FWHMs of these lineouts, and then estimate
the error of the measurement from the standard error of the FWHMs. 
The mean emission profile is marginally more robust than the original X-ray image for calculating 
$l_n$ because fluctuations distort the measured maximum value of the 
vertically averaged profile. The resulting values of $l_n$ are shown in  
Figure \ref{fig:Xrays}c, in blue. Following an initial decrease in value immediately after the two plasma flows collide to form the interaction region, $l_n$ increases steadily over time. 
%Also shown in Figure \ref{fig:Xrays}c is the behavior of $l_n$ as a function of time given by the FLASH simulations (blue curve): reasonable agreement is obtained. 

Secondly, relative fluctuations $\delta I$ in X-ray intensity (such as those shown in Figure \ref{fig:Xrays}b) are closely correlated 
with fluctuations $\delta n_e$ of electron density; indeed, for intensity fluctuations that are small compared to the mean intensity $\bar{I}$,
 $\delta I/\bar{I} \approx 2 \int \mathrm{d}s \, \delta 
n_e/\bar{n}_e$ (assuming that $\delta T_e/\bar{T}_e \lesssim \delta 
n_e/\bar{n}_e$, an assumption justified by the small P\'eclet number of the interaction-region plasma). 
The root-mean-square (RMS) of the relative X-ray fluctuations 
therefore provides a simple measure of the onset of stochasticity in the 
interaction region. The increase in relative X-ray fluctuation magnitude $(\delta I/\bar{I})_{\rm rms}$ shown in Figure \ref{fig:Xrays}c (in red)
illustrates that significant fluctuations develop in a 5-ns interval following formation of the interaction region, 
after which their magnitude saturates at a finite fraction of the mean X-ray intensity of 
the region: $\delta I \lesssim 0.3 \bar{I}$.  Under the additional assumption that density fluctuations are statistically isotropic and homogeneous, and therefore contribute to the line-of-sight
integral as a random walk provided many fluctuations are sampled, we find $\delta n_e/\bar{n}_e \lesssim (l_{n\perp}/L)^{1/2} \delta I/2 
\bar{I}$, where $l_{n\perp}$ is the perpendicular extent of the interaction 
region and $L$ the scale of dominant density fluctuations in the
plasma. Taking $l_{n\perp} \lesssim 0.3$~cm and $L\approx 0.04$~cm (corresponding to the grid 
periodicity), we deduce that $\delta n_e/\bar{n}_e \lesssim 0.5$. Thus, it
follows that density fluctuations are not large compared to the mean 
density and thus the stochastic motions of the plasma are subsonic. 

Thirdly, under the same statistical assumptions, the power 
spectrum of the path-integrated density fluctuations derived from the X-ray intensity fluctuations 
can be directly related to the power spectrum of the density fluctuations~\citep{C12}.
Because fluctuating density in a subsonic plasma behaves as a passive scalar~\citep{Z14}, 
this in turn allows for the measurement of the velocity power spectrum~\citep{T18}.
The result of such a calculation applied to Figure \ref{fig:Xrays}b is shown in Figure \ref{fig:Xrays}d: the spectrum extends across the full range of resolved wavenumbers and, 
for characteristic wavenumbers $2 \pi/L \lesssim k < k_{\rm res} = 127 \, \mathrm{mm}^{-1}$, 
the spectral slope is consistent with the Kolmogorov power law, as expected for a turbulent, subsonic plasma~\citep{W19}.

\subsection{Measuring plasma parameters: Thomson-scattering diagnostic} \label{TS_results}

\begin{figure}[t!]
  \centering
\includegraphics[width=\linewidth]{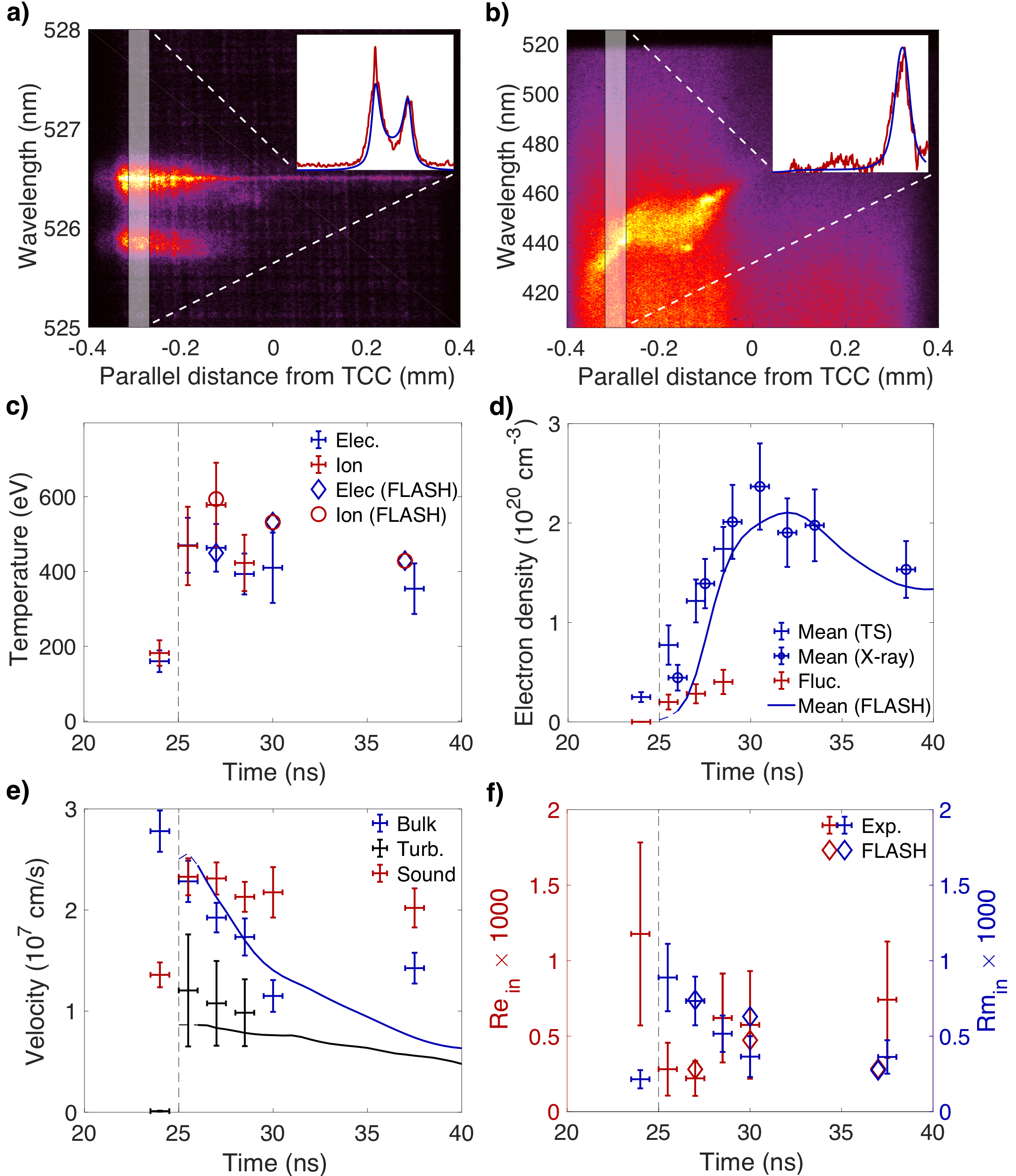}
\caption[Time evolution of interaction-region plasma parameters]{\textbf{Time-evolution of interaction-region plasma parameters.} \textbf{a)} Low-frequency, spatially resolved spectrum 
(IAW feature) obtained at 27.2 ns. A sample spectral fit (for the white highlighted region) is shown in the inset. \textbf{b)} High-frequency, spatially resolved spectrum (EPW feature) obtained on the same shot.
\textbf{c)} Evolution of electron and ion temperatures over time in the Thomson scattering volume. The experimental values for the electron (blue) and ion (red) temperatures are shown as time intervals with vertical error bars.
All values are determined as described in the main text; errors for each time are determined by regarding each spatially resolved measurement as a sample of the mean temperature value for the interaction region, with the uncertainty on each sample determined by the fit sensitivity. The results of the FLASH simulations (see the Supplementary Material) for the electron temperature are shown as blue diamonds, those for the ion temperature as red circles.
%errors are determined by combining in quadrature the fit sensitivity and the variability of parameters as measured across the Thomson-scattering volume for each time. 
%The alternative electron temperature history at later times (blue, dotted) 
%is a projected lower bound calculated from the difference in mean self-emitted X-ray intensities in the presence and absence of the Thomson scattering diagnostic (see Supplementary Information, Figure S8).
\textbf{d)} Evolution of mean electron density $\bar{n}_e$ (blue) and the fluctuating density $\Delta n_e$ (red) with time in the interaction region. 
Also shown are experimental values of $\bar{n}_e$ derived from the self-emission X-ray images (open blue circles). 
%under the assumption that bremsstrahlung emission is dominant 
%In both cases, the horizontal lines correspond to the time intervals of the measurements while the vertical lines are error bars. 
The error bars are calculated in the same manner as for the temperature. 
The blue curve shows the results of the FLASH simulations.
\textbf{e)} Evolution of bulk flow speed $\bar{u}_{\rm in}$ (blue), sound speed $c_s$ (red) and turbulent velocity $u_{\rm rms}$ (black) with time in the Thomson-scattering volume. 
%The experimental values are shown as time intervals with vertical error bars.
Errors are calculated in the same way as those for the temperature. 
Also shown are the results of the FLASH simulations for the bulk flow speed (blue curve) and turbulent velocity (black curve).
\textbf{f)} Evolution of the (bulk) fluid Reynolds number $\mathrm{Re}_{\rm in} \equiv \bar{u}_{\rm in} L/\nu$ (red) and magnetic Reynolds number $\mathrm{Rm}_{\rm in} \equiv \bar{u}_{\rm in} L/\eta$ (blue) over time. 
%The experimental values are shown as time intervals with vertical error bars.  
The kinematic viscosity $\nu$ and resistivity $\nu$ are calculated using the formulae given in Table S2 of the Supplementary Information. The input plasma state variables are the experimentally-determined values in the Thomson-scattering 
volume and $L = 400 \, \mu \mathrm{m}$; at later times (30 ns, 37.5 ns), $\mathrm{Re}_{\rm in}$ is instead calculated using an extrapolated density derived from the X-ray measurements, and assuming $T_i = T_e$.
Errors are calculated in the same way as those for the temperature.
Also shown are the results of the FLASH simulations for $\mathrm{Re}_{\rm in}$ and $\mathrm{Rm}_{\rm in}$ (red/blue diamonds).}
%An estimate of $\mathrm{Rm}$ at later times assuming lower bound on electron temperature $T_e \approx 50 \, \mathrm{eV}$ is also shown (blue, dotted) 
\label{fig:TS_paramsresults}
\end{figure}

For experimental times approximately coincidental with the collision of the two plasma flows, and just after, clear scattering spectra at both low and high frequencies were obtained. Unprocessed IAW and EPW features for a sample time close to the formation of the interaction region are shown in Figures \ref{fig:TS_paramsresults}a and \ref{fig:TS_paramsresults}b, respectively; the complete data set used for these results is given in the Supplementary Information (Figure S2). Measurements of the bulk plasma parameters listed in Experimental Design were then derived 
at a given position by fitting the spectral density function (see Materials and Methods). We averaged the parameters obtained from fits at each position over the complete spatial extent of
the observed IAW and EPW features. 
The time evolution of the physical parameters was obtained
by repeating the experiment and firing the Thomson-scattering diagnostic at different times with respect to the activation of the drive-beam.

The evolution of the average electron and ion temperatures in the Thomson-scattering volume is shown in Figure \ref{fig:TS_paramsresults}c, density in Figure \ref{fig:TS_paramsresults}d, 
and bulk and turbulent velocities in Figure \ref{fig:TS_paramsresults}e. 
At 24 ns, the characteristic electron and ion temperatures were 
$T_e \approx T_i \approx 180 \, \mathrm{eV}$, the characteristic flow
speed $\bar{u}_{\rm in} \approx 260 \, \mathrm{km} \; \mathrm{s}^{-1}$, and the mean electron number density $\bar{n}_e \approx 2.5 \times 10^{19} \, 
\mathrm{cm}^{-3}$. These values are similar to those previously obtained for
a single plasma jet~\citep{T18}, a finding consistent with the observation from the X-ray imaging diagnostic that
the two plasma flows have not yet collided to form the interaction-region plasma at this time (see Figure \ref{fig:Xrays_earlytime}). 
By contrast, 1.5 ns later the electron and ion temperatures were found to be much larger than their jet pre-collision values: 
$T_e \approx T_i \approx 450 \, \mathrm{eV}$. The measured mean electron number 
density also increased to $\bar{n}_e \approx 8 \times 10^{19} \, 
\mathrm{cm}^{-3}$. In fact, a range of densities were observed, with $\Delta n_e \approx 2 \times 10^{19} \, 
\mathrm{cm}^{-3}$, suggesting chaotic motions. For a measured characteristic sound speed of $c_s \approx 220 \, 
\mathrm{km} \; \mathrm{s}^{-1}$, this range of densities implies small-scale stochastic 
velocities $\Delta u \approx 55 \, \mathrm{km} \; \mathrm{s}^{-1}$ (see Materials and Methods). Assuming Kolmogorov scaling for 
the random small-scale motions -- as is consistent with the spectrum in Figure \ref{fig:Xrays}d -- 
the characteristic velocity $u_{\ell}$ at scale $\ell$ satisfies $u_{\ell} \sim u_{\rm rms}
(\ell/L)^{1/3}$. Because the dominant contribution to $\Delta u$ arises 
from stochastic motions with scale comparable to the Thomson scattering cross-section width 
$l_{\mathrm{TS}} \approx 50 \, \mu \mathrm{m}$, we conclude that $\Delta u \approx u_{l_{\mathrm{TS}}}$, and so $u_{\rm rms} \approx 110 \, 
\mathrm{km} \, \mathrm{s}^{-1}$. 

In the 3-ns interval subsequent to the two plasma flows colliding to form the interaction region, the ion 
temperature increased above the electron temperature ($T_i \approx 600 \, \mathrm{eV}$), before both fell 
to lower values ($T_e \approx T_i \approx 400 \, \mathrm{eV}$). The mean electron number density increased monotonically 
over the same interval, with a final measured value of $\bar{n}_e \approx 1.8 \times 10^{20} \, 
\mathrm{cm}^{-3}$. The relative magnitude of density fluctuations remained 
the same ($\Delta n_e/\bar{n}_e \approx 0.25$) over the interval. 

At later times, no EPW feature was observed and the IAW feature manifested 
itself erratically (see Figure S3 in the Supplementary Information). 
We believe that this was due to the increased density 
of the interaction region (as well as substantial density gradients) resulting 
in significant refraction of the Thomson-scattering probe beam. We were 
therefore unable to measure $\bar{n}_e$ or $\Delta n_e$ for times $ \gtrsim 30 \,$ns using 
the Thomson-scattering diagnostic. A reasonable estimate of $\bar{n}_e$ can still be obtained, however,
using the X-ray framing camera diagnostic. More specifically, assuming 
that the X-ray emission from the plasma is dominated by bremsstrahlung, 
we can estimate the mean electron number density $\bar{n}_e(t_1)$ at time $t_1$
in terms of the mean electron number density $\bar{n}_e(t_2)$ at time $t_2$ via the following 
relationship: $\bar{n}_e(t_1) \approx \bar{n}_e(t_2) [I(t_1)/I(t_2)]^{1/2}$. 
Thus, assuming a reference value for $\bar{n}_e(t_2)$ at $t_2 = 29.0\,$ns (derived via linear interpolation from the
Thomson-scattering density measurements), we obtain the evolution profile shown
in Figure \ref{fig:TS_paramsresults}d. The results imply that the density continues to rise for $\sim2$ ns after the final Thomson-scattering measurement of density is 
obtained, reaching a peak value $\bar{n}_e \approx 2.4 \times 10^{20} \, 
\mathrm{cm}^{-3}$ at $t = 30 \,$ns before falling slightly at later times.

We were still able to use the IAW feature to measure the bulk flow velocity and the electron temperature in some spatial locations at later times. The 
bulk flow velocity was found to drop to ${\sim}100$ km $\mathrm{s}^{-1}$ at 30 ns. At 37.5 ns a 
similar value was obtained but with a reversed sign; this is possibly due to the Thomson-scattering diagnostic measuring the inflow velocity at a position displaced from the line of centers, which could have an opposite velocity. The electron temperature 
measured by the Thomson scattering diagnostic remained ${\sim}400$ eV at later times. However, this is due to heating of the interaction region by the Thomson-scattering beam, which is significant at later times because of the high densities and reduced temperatures. We discuss this effect at greater length in the Supplementary Information with the aid of FLASH simulations.

\subsection{Measuring magnetic fields: proton-imaging diagnostic} \label{protonimag_results}

\begin{SCfigure*}[\sidecaptionrelwidth][t]
  \centering
\includegraphics[width=10.9cm,height=10.9cm]{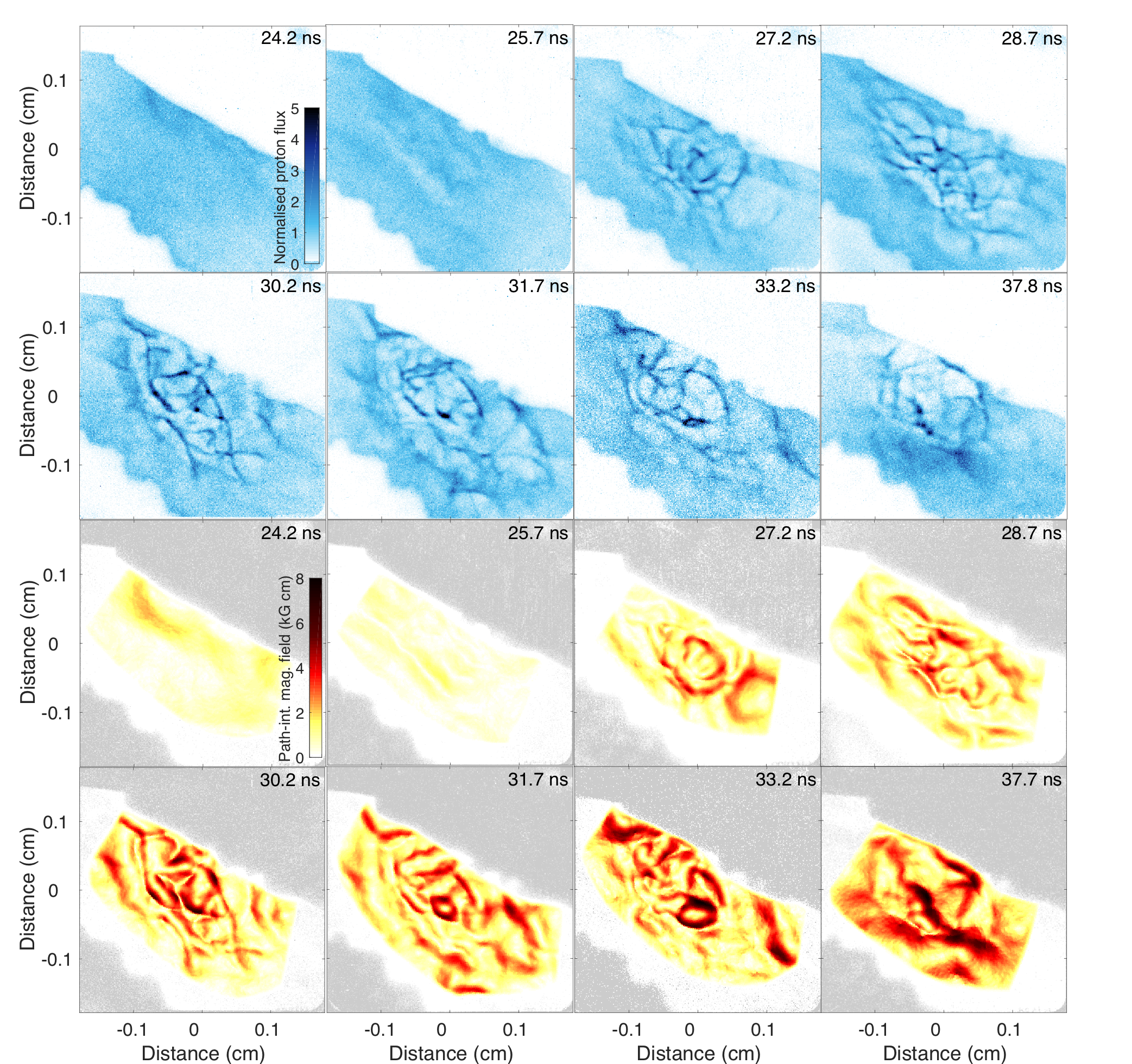}
\caption{\textbf{15.0 MeV proton images of interaction-region plasma, and extracted path-integrated magnetic fields}. The top two rows show the proton images. Each image is approximately 300 $\times$ 300 pixels, with an effective
pixel size of $12 \, \mu$m; by comparison, the proton-source size is $\sim$40 $\mu$m. To prevent confusion, all images are presented with the magnification removed.
The grid outline evident on the bottom left of each image is grid A, and the top-right grid is grid B. The mean proton flux $\Psi_0$ per pixel in these images is $\sim50$ protons per pixel
%; the signal-to-noise ratio is thus $\sim7$. 
The bottom two rows show the magnitude of the 
path-integrated perpendicular magnetic field, extracted using the field-reconstruction algorithm. The method for applying the field-reconstruction algorithm is as follows. We first select a 
region of the proton image to analyze; this region is chosen to be as large as possible,
within the requirements of staying inside the region of high detected proton flux 
between the grids, maintaining an approximately rectangular shape, and 
choosing a boundary that does not intersect regions with high proton 
flux. We then embed the cropped region of proton flux 
inside a larger rectangular region, whose size is chosen to be as small as 
possible while still containing the former region. Values of proton flux are then systematically assigned to 
pixels outside the cropped region: these values are calculated by 
linearly interpolating between the nearest actual pixel value and the mean flux of 
the cropped region of protons. The resulting image is then subjected to a 
Gaussian high-pass filter, with scale $0.1 \, \mathrm{cm}$. This image is then processed with the field-reconstruction algorithm. Subsequent to 
convergence of the algorithm, the path-integrated field is only retained for pixels inside the 
original cropped region, with other values removed via a Gaussian window function. These steps are all necessary in order to prevent systemic errors affecting the algorithm~\citep{AGGS16}.} 
\label{fig:protonimaging}
\end{SCfigure*}

The 15.0-MeV proton images for our experiment are presented as a time sequence in the top two rows of Figure \ref{fig:protonimaging}. 
The proton image before the formation of the interaction-region plasma (Figure \ref{fig:protonimaging}, 24.2 ns) shows little
structure at the center of the grids, which is consistent with the absence of significant magnetic fields. 
Around the time when the interaction region forms, a moderate 
diminution of the proton flux is observed in a central region between the grids (Figure \ref{fig:protonimaging}, 25.7 ns), 
with characteristic magnitude $\Psi$ similar to the mean proton flux $\Psi_0$: $|\Psi - \Psi_0| \lesssim 0.3 \Psi_0$. 
In contrast, in all subsequent proton images (beginning at $t \gtrsim 27.2 \, 
\mathrm{ns}$), order-unity variations in the proton flux are measured ($|\Psi - \Psi_0| \gtrsim \Psi_0$) whose 
structure and position are (at least partially) stochastic -- see Figure \ref{fig:protonimaging}, 27.2 ns, for an example. This is 
consistent with a dramatic change in the morphology and strength of the magnetic field.

Further analysis can be performed by reconstructing directly from the measured proton image the (perpendicular) path-integrated field 
experienced by the imaging proton beam 
-- quantities that are related to each other via a well-known relation~\citep{K12,AGGS16}. 
Provided the gradients in the magnetic-field strength are not so large as to cause 
the proton beam to self-intersect before arriving at the detector, this relation 
leads to an equation of Monge-Amp\`ere type, the unique inversion of which is a well-posed mathematical 
problem~\citep{GM96} and for which an efficient inversion algorithm exists~\citep{AGGS16} (we refer to this algorithm 
as the `field-reconstruction algorithm').  
The results of applying this algorithm to the proton images shown in Figure \ref{fig:protonimaging} are presented in the same figure.
The strength and morphology of the reconstructed path-integrated fields after the jet collision are quite different
from those at collision, with peak values reaching ${\sim}8$ kG cm (as opposed to ${\sim}1$ kG cm at collision) 
and randomly orientated filamentary structures evident. 

\begin{figure}[t!]
  \centering
\includegraphics[width=\linewidth]{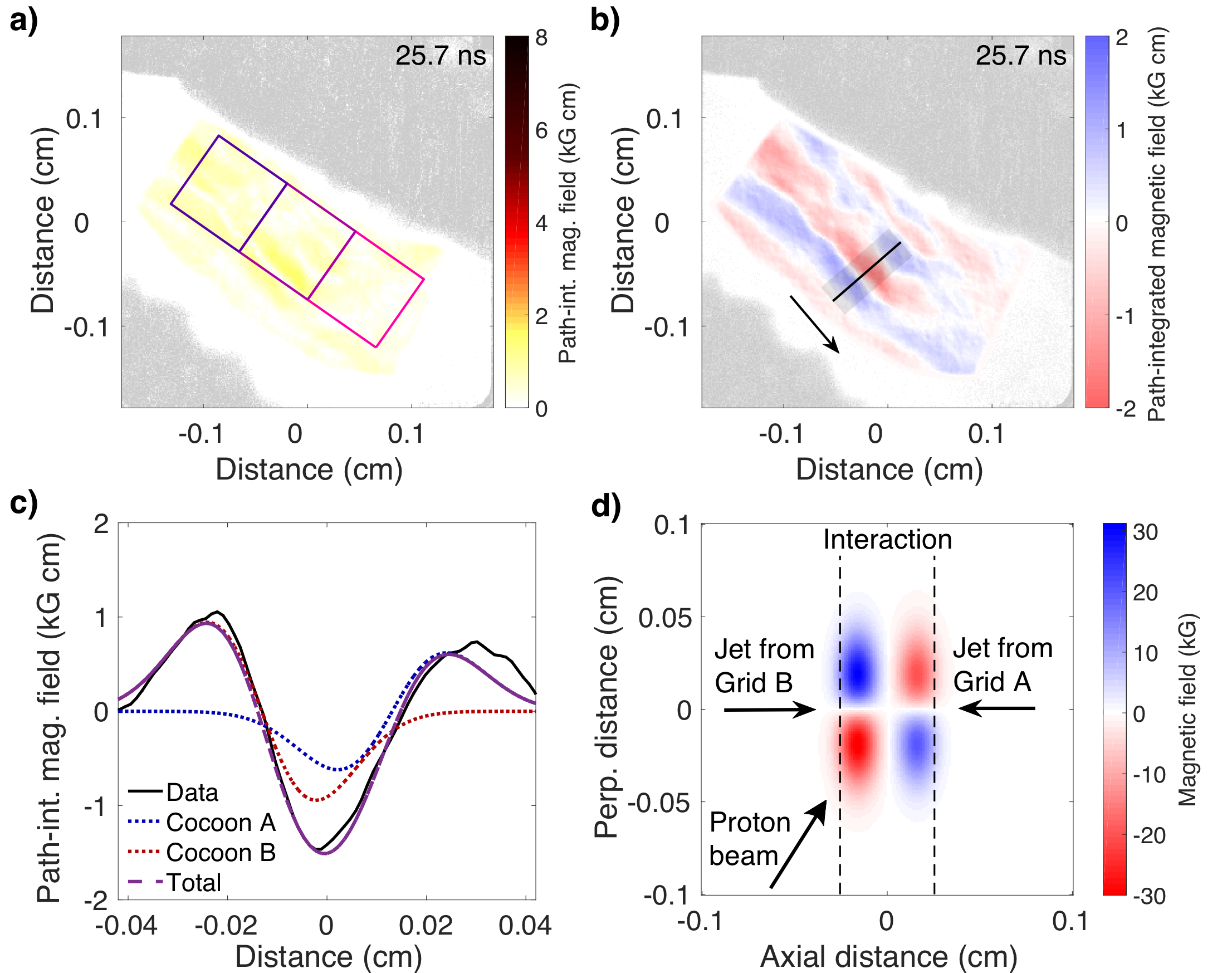}
\caption[Path-integrated magnetic fields at interaction-region-plasma's formation]{\textbf{Path-integrated magnetic fields at the moment of the interaction-region plasma's coalescence}. \textbf{a)} 
Magnitude of path-integrated perpendicular magnetic field 25.7 ns after drive-beam pulse initiation. 
The three square regions in which the average path-integrated field is evaluated 
have an edge length of 800 $\mu$m, and are orientated at $35^{\circ}$ to the 
horizontal axis of the path-integrated field map. The center of the middle square region 
corresponds to the center of the proton image.
\textbf{b)} Component of the
path-integrated magnetic field in the direction perpendicular to the projected line of centers. 
This component is calculated from the full 2D perpendicular path-integrated magnetic field. The arrow indicates the (positive) direction of the chosen path-integrated field component. 
\textbf{c)} 1D lineout of the path-integrated field component given in b) (black, solid line) calculated by averaging across its width the semi-transparent rectangular region 
denoted in a). The path-integrated field associated with model Eq. [\ref{doublecocoonfull}] is also plotted, using optimized parameters $B_{\rm max}^{+} b = -0.31 \, \mathrm{kG}$ cm, $B_{\rm max}^{-} b = 0.20 \, \mathrm{kG}$ cm, $a = 270 \, \mu \mathrm{m}$, and $l_c = 131 \, \mu
\mathrm{m}$. The total contribution is plotted (purple, dashed), as well as the individual contributions from the cocoons nearer grid A (blue, dotted), and nearer grid B (red, dotted). 
\textbf{d)} Slice plot (in the plane of basis vectors $\hat{\boldsymbol{y}}$ and $\hat{\boldsymbol{z}}$) of $B_x$ 
component associated with 3D double-cocoon magnetic-field model given by Eq. [\ref{doublecocoonfull}], with the same
model parameters as shown in c), and $b = 0.01 \, \mathrm{cm}$. The width of the plotted interaction region is obtained from the X-ray image recorded at the equivalent time (cf. Figure \ref{fig:Xrays_earlytime}c). }
\label{fig:protoninitfields}
\end{figure}

With the path-integrated magnetic field having thus been determined, the correct method of estimating 
the characteristic magnetic-field strength 
depends on the field structure. 
The path-integrated field structures evident at early times (i.e., Figure \ref{fig:protoninitfields}a) 
are non-stochastic. We therefore follow a standard method for analyzing proton 
images of non-stochastic magnetic fields~\citep{S12} and consider parameterized models of known 
three-dimensional magnetic-field structures. 
To motivate a relevant model for our experimental data, we invoke the expected physical origin of the early-time magnetic fields in the interaction-region plasma: the action of the Biermann battery during the interaction of the drive-beam lasers with the target's foils. This process generates azimuthal magnetic fields in the plane perpendicular to the target's line of centers that are opposite in sign for the two foils~\citep{S71}. 
These fields are then advected by the two counter-propagating plasma flows towards the midpoint between the two foils. We therefore consider two `cocoon' structures with magnetic fields of opposite sign, with their symmetry axis parallel to the line of centers.

A simple parameterized model for a double-cocoon configuration considered in~\citep{K13} takes the form
\begin{equation}
\boldsymbol{B} = \sqrt{2 \mathrm{e}} \Bigg[B_{\rm max}^{+} \mathrm{e}^{-\frac{(z+\ell_c)^2}{b^2}} +B_{\rm max}^{-} \mathrm{e}^{-\frac{(z-\ell_c)^2}{b^2}} \Bigg] \frac{r}{a} \mathrm{e}^{-\frac{r^2}{a^2}} \boldsymbol{e}_\phi  \, 
, \label{doublecocoonfull}
\end{equation}
where $(r,\phi,z)$ is a cylindrical coordinate system with symmetry axis $z$, 
$B_{\rm max}^{+}$ is the maximum magnetic-field strength of the cocoon centered at $z = -\ell_c < 0$, $B_{\rm max}^{-}$ is the maximum magnetic-field strength of the cocoon centered at $z = \ell_c > 0$, $a$ the characteristic 
perpendicular size of both cocoons, $b$ their characteristic parallel size, and $\boldsymbol{e}_\phi$ the azimuthal unit vector. It can be shown (see Supplementary Information) that, if $a \gtrsim b$, then the path-integrated magnetic field associated with the double-cocoon configuration, when viewed 
at angle $\theta \approx 55^{\circ}$ with respect to the $z$ axis, is  orientated predominantly perpendicularly to the direction of the line of centers projected onto the proton image, and its strength varies predominantly in the parallel direction (viz., the path-integrated field is quasi 1D). Both of these findings are consistent with the observed structure at the point of maximum path-integrated field (see Figure~\ref{fig:protoninitfields}b), validating our choice of model. 

Having obtained a quasi-1D model for the path-integrated 
magnetic field (which has four free parameters: $B_{\rm max}^{+} b$, $B_{\rm max}^{-} b$, $a$ and $\ell_c$ -- see Supplementary Information), 
we compare it with a lineout across the strongest path-integrated magnetic-field
structure (see Figure \ref{fig:protoninitfields}b). Figure~\ref{fig:protoninitfields}c shows 
the lineout, as well as the model with an optimized fit: $B_{\rm max}^{+} b = -0.31 \pm 0.02 \, \mathrm{kG}$ cm, $B_{\rm max}^{-} b = 0.20 \pm 0.02 \, \mathrm{kG}$ cm, $a = 270 \pm 19 \, \mu \mathrm{m}$, and $\ell_c = 131 \pm 9 \, \mu 
\mathrm{m}$ (here the errors in the model parameters correspond to the 95\% 
confidence intervals). 
The agreement of the model with these parameters is reasonable, 
with an adjusted R-squared value of 0.97. Further validation is provided in the Supplementary Information (Figure S9). 
The parameterized magnetic-field model itself has an additional free parameter $b$ to be determined; this is done by assuming that the entire magnetic-field configuration is contained inside the interaction-region plasma, and so $b = \ell_n/2 \approx 0.01 \, \mathrm{cm}$. The double-cocoon configuration for this choice of $b$ is shown in Figure \ref{fig:protoninitfields}d. 
The mean magnetic-field strength associated with the double-cocoon configuration can then be shown to be ${\sim}6 \, \mathrm{kG}$. This magnetic-field structure and its strength are reproduced successfully by FLASH simulations (see Supplementary Information). 

%The origin of this magnetic field structure and its strength are described in the Supplementary Information.  As the FLASH simulations of the experiment show (and is well known), the interaction of the laser drive beams with the foil targets generates strong magnetic fields via the Biermann battery effect that are toroidal in the x-y plane and opposite in sign for the two targets.  The resulting largely helical fields are advected by the two counter-propagating plasma flows toward the midpoint between the two targets.  The strengths of the helical fields are significantly reduced as this happens because the two plasma flows expand laterally (see Figures S11d, S11e, S11h, and Slli). As a result, their field strengths are $\sim 10$ kG just before the plasma flows merge to form the interaction region.

For the stochastic path-integrated magnetic fields that emerge after the jet collision (due to the interaction of the initial seed fields with stochastic fluid motions), 
a different approach is required: 
we assume statistically isotropic, homogeneous, tangled magnetic fields in the interaction-region plasma (an assumption verified in the Supplementary Information -- see Figure S10), 
which in turn allows for the unique extraction of the RMS 
magnetic field strength $B_{\rm rms}$ via the following formula:
\begin{equation}
B_{\rm rms}^2 =\frac{2}{\pi l_p} \int \mathrm{d}k \, k E_{\rm path} (k), 
\end{equation}
 where $l_p$ is the path length of 
the protons through the interaction region, $E_{\rm path}(k)$ is the 1D spectrum of a given of
path-integrated field under normalization condition $\int \mathrm{d}k \, E_{\rm path} (k) = (\int \mathrm{d}^2 \boldsymbol{x} \, \boldsymbol{B}_{\perp})_{rms}^2$~\citep{AGGS16}. 
We estimate $l_p$ at a given time using our measurements of the 
average interaction-region width $l_n$ derived from the X-ray 
imaging diagnostic, combined with the known angle $\theta_{p} = 55^{\circ}$ of the proton beam through the
interaction region (with respect to the line of centers): it follows that $l_p \approx l_n/\cos{\theta_{p}} \approx 1.7 
l_n$. We can then calculate the characteristic correlation length $\ell_B$ of the 
stochastic magnetic field via
\begin{equation}
\ell_B = \frac{1}{\ell_p B_{\rm rms}^2} \int \mathrm{d}k \, E_{\rm path} (k) \, 
\end{equation}
and determine the complete magnetic-energy spectrum $E_B(k)$ from $E_{\rm path} (k)$ 
via
\begin{equation}
E_B(k) = \frac{1}{4 \pi^2 \ell_p}  k E_{\rm path} (k) \, . \label{Especform}
\end{equation}
However, we caution that due to the likely presence of strong, small-scale magnetic fields leading to 
self-intersection of the imaging beam, the power spectrum at wavenumbers $k \gtrsim \pi \ell_B^{-1}$ 
determined via Eq. [\ref{Especform}] is not a faithful representation of the 
true magnetic-energy spectrum~\citep{AGGS16}. We therefore focus on measuring $B_{\rm rms}$ 
and $\ell_B$. We consider the three fixed regions of the 
path-integrated magnetic field images introduced in Figure \ref{fig:protoninitfields}a, and calculate $B_{\rm rms}$ 
and $\ell_B$ for those regions.

\begin{figure}[h]
  \centering
\includegraphics[width=0.9\linewidth]{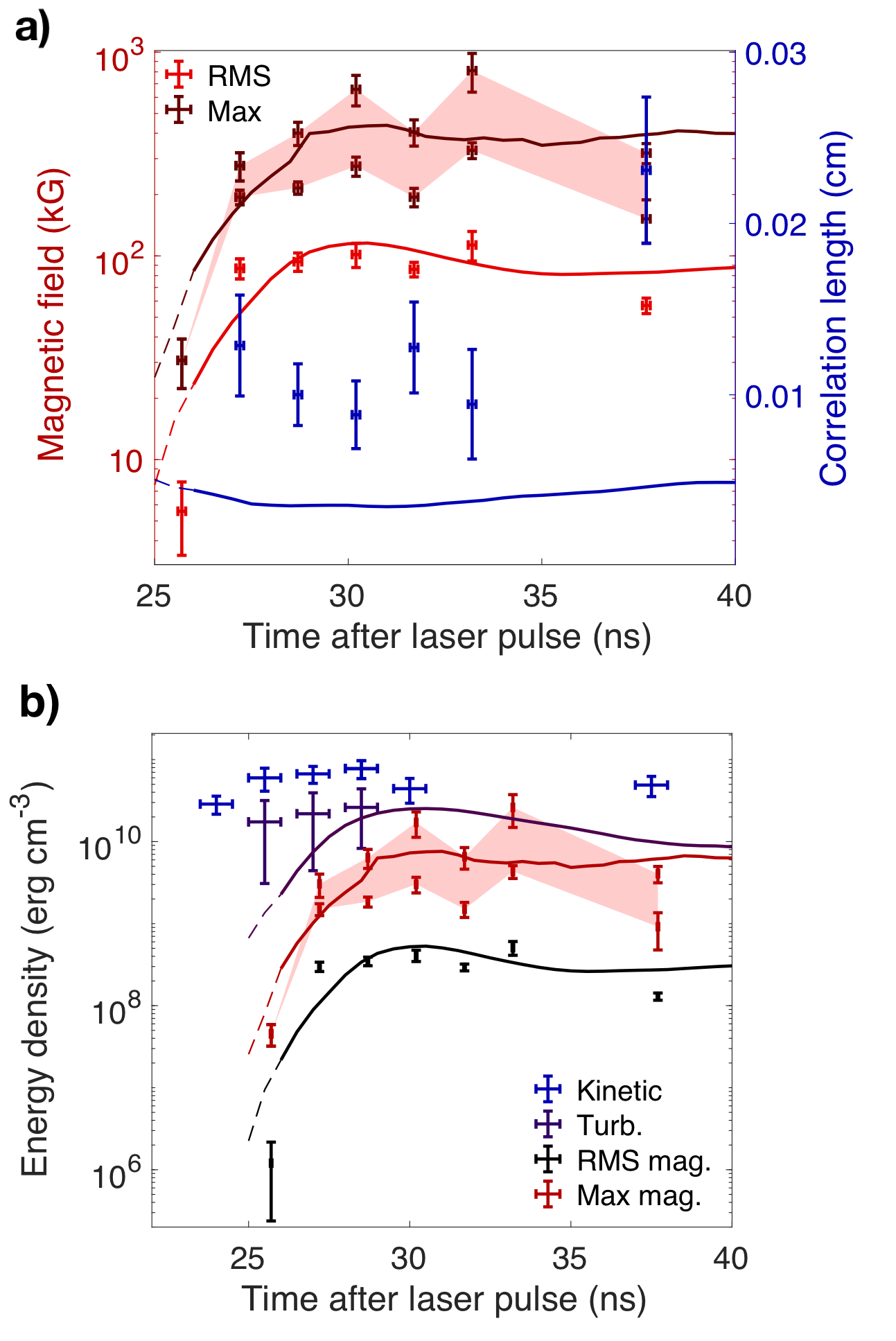}
\caption[Magnetic fields subsequent to formation of the interaction-region plasma]{\textbf{Magnetic fields subsequent to formation of the interaction-region plasma.} \textbf{a)} RMS magnetic-field strength (red data points) and the bounds on the maximum 
magnetic field (maroon band bounded by maroon data points) versus time, as well as the correlation length $\ell_B$ (blue data points). We emphasize that the mean and maximum field strengths at 25.7 ns 
are calculated differently than at the other times, on account of the non-stochastic
field structure (see Figure \ref{fig:protoninitfields}). Also shown are the evolution of the RMS magnetic field (red curve), maximum magnetic field (maroon curve) and correlation length (blue curve) versus time given by FLASH simulations of the experiment. The dashed portions of these curves correspond to times when the plasma in the interaction region is not yet fully collisional and therefore the simulations are not formally valid (see Supplementary Information). \textbf{b)} Evolution of energy densities in the plasma-interaction region versus time. For times ${\leq}30 \, \mathrm{ns}$, the bulk 
kinetic energy and turbulent kinetic energy densities are calculated using the values of the plasma state variables derived from the Thomson-scattering diagnostic; at later times, the plasma density required to calculate these energies is determined using the X-ray imaging diagnostic. Also shown are the evolution of the RMS magnetic energy (black curve), maximum magnetic energy (red curve) and turbulent kinetic energy (purple curve) versus time for the FLASH simulations. The dashed portions of these curves have the same meaning as in b). In both a) and b), the experimental values are shown as time intervals with vertical error bars.}
%The bounds on the turbulent kinetic energy derive from the equivalent bounds on the turbulent velocity. The RMS and maximum magnetic energy derive from proton imaging diagnostic.}
\label{fig:protonimaging_stocmagfields}
\end{figure}

The mean values of $B_{\rm rms}$ and $\ell_B$ arising from each path-integrated
field image (and the errors on those measurements) for the full time-sequence of path-integrated field images (see Figure \ref{fig:protonimaging}) are shown in Figure \ref{fig:protonimaging_stocmagfields}a. $B_{\rm rms}$ jumps significantly in a 1.5-ns interval subsequent to collision, 
reaching a peak value ${\sim}120$ kG, before decaying somewhat, to around ${\sim}70$ kG.
%The FLASH simulations give similar results.
The correlation length has characteristic value $\ell_B \approx 0.01 \, 
\mathrm{cm}$ for all measured times, except at 38 ns.  The FLASH simulations, which give similar values for the magnetic-field strength, give a significantly smaller value for the correlation length ($\ell_B \approx 0.004 \, \mathrm{cm}$), a discrepancy discussed in Interpretation of Results.   

%The larger experimental value is likely due to smearing of the proton radiography images by stochastic small-angle scattering of the protons as discussed in the Supplementary Information.

We can also calculate reasonable upper and lower bounds of the maximum magnetic-field 
strength realized in the stochastic field, via two different methods. For the 
lower bound, we note that the kurtosis of the 
path-integrated magnetic field will always be smaller than the kurtosis of the actual magnetic 
field. 
Therefore, the ratio 
between the maximum path-integrated field and the RMS path-integrated field will always
be smaller than the equivalent ratio for the magnetic field: in other words, a 
reasonable lower bound is $B_{\rm max,l} = B_{\rm rms} (\int \mathrm{d}^2 \boldsymbol{x} \, \boldsymbol{B}_{\perp})_{\rm max}/(\int \mathrm{d}^2 \boldsymbol{x} \, \boldsymbol{B}_{\perp})_{\rm 
rms}$. The upper bound is derived by assuming that the maximum measured path-integrated 
magnetic field is obtained when the imaging protons cross just a single magnetic 
structure: $B_{\rm max,u} = (\int \mathrm{d}^2 \boldsymbol{x} \, \boldsymbol{B}_{\perp})_{\rm 
max}/\ell_B$. These bounds are shown in Figure \ref{fig:protonimaging_stocmagfields}a. At the time corresponding to maximal $B_{\rm rms}$, we find
$310 \, \mathrm{kG} < B_{\rm max} < 810 \, \mathrm{kG}$. 
%The FLASH simulations give similar results.

\section*{Interpretation of Results} \label{Discussion}

We conclude that our experimental platform does produce a 
plasma that manifests stochastic motion across a range of scales.
In spite of some uncertainty about the late-time physical properties of the 
turbulent plasma, there exists a 4-ns time interval that starts from the formation 
of the interaction region and during which the plasma state can be thoroughly 
characterized by our experimental diagnostics. In this interval, we find that the plasma is fairly well described 
as classical and collisional ($\lambda_{e} \approx 10 \, \mu \mathrm{m}$, $\lambda_{\mathrm{CC}} \approx 0.6 \, \mu 
\mathrm{m}$, $\lambda_{\mathrm{HC}} \approx 16 \, \mu \mathrm{m}$, where $\lambda_e$, $\lambda_{\mathrm{CC}}$, and $\lambda_{\mathrm{HC}}$ are the electron, carbon-carbon 
and hydrogen-carbon mean free paths respectively), so its transport coefficients can be 
estimated (see Supplementary Information) using collisional 
transport theory~\citep{B65,H94,R99}. Momentum transport in the plasma is dominated by hydrogen ions, on account of 
their long mean free path compared to carbon ions~\citep{SM14,SM16}, 
while heat transport is dominated by electrons. 

The time history of the fluid Reynolds number $\mathrm{Re}_{\rm in} = \bar{u}_{\rm in} L/\nu$ and the magnetic 
Reynolds number $\mathrm{Rm}_{\rm in} = \bar{u}_{\rm in} L/\eta$ in our experiment (which are defined here using the inflow velocity $\bar{u}_{\rm in}$
in order to enable comparisons between the state of the plasma both before and after the two plasma flows collide to form the interaction-region plasma) 
is shown in Figure \ref{fig:TS_paramsresults}f. Prior to the collision of the plasma flows, 
$\mathrm{Re}_{\rm in} \approx 1.2 \times 10^{3}$, which exceeds $\mathrm{Rm}_{\rm in} \approx 
200$. However, after the formation of the interaction-region plasma, the rapid collisional shock heating of both ions 
and electrons simultaneously decreases the resistivity and enhances the 
viscosity, leading to the opposite ordering of dimensionless numbers: $\mathrm{Re}_{\rm in} \approx 300$ and 
$\mathrm{Rm}_{\rm in} \approx 900$, so $\mathrm{Pm} = \mathrm{Rm}_{\rm in}/\mathrm{Re}_{\rm in} > 1$. The characteristic velocity $u_{\rm rms}$ of stochastic 
motions is smaller than the in-flow velocity, and thus the fluid Reynolds number $\mathrm{Re} = u_{\rm rms} L/\nu$ and 
magnetic Reynolds number $\mathrm{Rm} = u_{\rm rms} L/\eta$ of the driving-scale stochastic motions 
are somewhat smaller than $\mathrm{Re}_{\rm in}$ and $\mathrm{Rm}_{\rm in}$: $\mathrm{Re} \approx 150$ and 
$\mathrm{Rm} \approx 450$. We observe that at such $\mathrm{Re}$, turbulence is not `fully developed' in the asymptotic sense. However, this is not necessary for the fluctuation dynamo to operate: the fluid motions need only be stochastic~\citep{SCTM04}. 
$\rm Pm$ remains order unity 
for $t \lesssim 30 \, \mathrm{ns}$; since the turnover time $\tau_L$ of the largest stochastic 
motions is $\tau_L = L/u_{rms} \approx 4 \,  \mathrm{ns}$, we conclude that 
the experimental platform does indeed produce a region of plasma with $\mathrm{Pm} \gtrsim 1$, 
which survives longer than the timescale 
on which the largest-scale stochastic motions decorrelate. 

We have measured the
magnetic field's evolution with time in the interaction-region plasma, 
and found that field strengths are amplified tenfold from their initial values during the 4-ns 
time window after collision. Having measured both the magnetic field and dynamical properties of the 
interaction-region plasma, we can compare the time history of the turbulent and magnetic energy densities (see Figure \ref{fig:protonimaging_stocmagfields}b). 
When the interaction-region plasma initially coalesces, the turbulent kinetic energy density $\varepsilon_{\rm turb} \equiv \rho u_{\rm rms}^{2}/2 \approx 2 \times 10^{10} \, \mathrm{erg}/\mathrm{cm}^3$ is over four orders of magnitude larger than the 
average magnetic-energy density associated with seed Biermann fields ($\varepsilon_{B} = B^2/8 \pi \approx 1 \times 10^{6} \, \mathrm{erg}/\mathrm{cm}^3$). However, 1.5 ns later, 
the relative magnitude of the magnetic energy is significantly larger: $\varepsilon_{B}/\varepsilon_{\rm turb} \approx 0.02$.
Furthermore, the FLASH simulations of our experiment \color{black}{-- which successfully reproduce the evolution of hydrodynamic variables and exhibit dynamo action that results in similar energy ratios -- }\color{black} indicate that the magnetic field
at the end of the 4-ns time window is dynamically significant in at 
least some locations in the plasma (see Supplementary Information). 
We therefore claim to have demonstrated the operation of a fluctuation dynamo in a $\rm Pm \gtrsim 1$ 
plasma. 

We can use the experimental data to estimate the growth rate $\gamma$
of the observed magnetic-field strength. Noting its value both at collision ($B_{t = 25.7 \, \mathrm{ns}} \approx 6 \, \mathrm{kG}$) and 1.5 ns later 
($B_{t = 27.2 \, \mathrm{ns}} \approx 86 \, \mathrm{kG}$), we find $\gamma \gtrsim 6.7 \log{(B_{t = 27.2 \, \mathrm{ns}}/B_{t = 25.7 \, \mathrm{ns}})}  \times 10^{8} \, \mathrm{s}^{-1} \approx 1.8 \times 10^{9} \, \mathrm{s}^{-1} \approx 6 u_{\rm rms}/L$.
\color{black}
This growth is more efficient than that predicted by periodic-box MHD simulations of the $\mathrm{Pm} \approx 1$ fluctuation dynamo with similar parameters (e.g.~\citep{PJR15}, where $\gamma \approx 2 u_{\rm rms}/L$ for $\mathrm{Rm} = 556$). 
We attribute this to strong shear flows in the interaction-region plasma, directed parallel to the line of centers, in addition to stochastic motions.
While a 2D uni-directional shear flow cannot account for sustained amplification of magnetic fields, its coupling to other stochastic plasma motions (including KH-unstable modes associated with the shear flow) can enable dynamo action. 
The FLASH simulations -- which reproduce similar field growth rates to those found experimentally -- support this interpretation (see Supplementary Information): the
RMS rate of strain of the simulated velocity field, which follows the growth rate of the magnetic energy, is comparable to the rate of strain of the directed shear flows. Such flows are common in astrophysical plasmas, so enhanced magnetic-field amplification on account of their interaction with turbulence may be relevant to astrophysical systems such as galaxy clusters~\citep{Si19}.
 
%suggesting that there are flow reversals of characteristic magnitude $\Delta U_{\|} \approx 200 \, \mathrm{km} \, s^{-1}$ over a (perpendicular) length scale $\Delta x_{\perp} \approx 100 \, \mu \mathrm{m}$. The resulting shear rate $\gamma_S \approx \Delta U_{\|}/\Delta x_{\perp} \approx 2 \times 10^{9} \, \mathrm{s}^{-1}$, is comparable to the growth rate of the fields. 

%This growth is surprisingly efficient when compared to periodic-box MHD simulations of the $\mathrm{Pm} = 1$ fluctuation dynamo (e.g.~\citep{SBSW20}, in which $\gamma \approx 0.4 u_{\rm rms}/L$), indicating that another field-amplification mechanism must be at play in our experiment. We suggest this mechanism is the KH instability, which arises due to the strong shear flows produced by the interleaved plasma fingers that flow into the interaction region.  While the KH instability associated with a single two-dimensional flow reversal cannot account for sustained amplification of magnetic fields, we argue that the presence of many such reversals in combination with other stochastic motions significantly enhances the initial growth of the field. Detailed examination of the validated FLASH simulations of the experiment -- which reproduce similar field growth rates to those found experimentally -- support this interpretation (see Supplementary Information). KH instabilities are common in astrophysical plasmas, so enhanced amplification of seed magnetic fields via this mechanism may in fact be relevant to astrophysical systems such as galaxies and galaxy clusters~\citep{Si19}.

\color{black}

Another noteworthy finding of our experiments is the characteristic scale 
of the amplified stochastic magnetic fields, which is a factor of ${\sim}2$--$3$ times larger 
than is measured in periodic-box MHD simulations. The
integral scale $L_{\rm int,B} \equiv 4 \ell_B$ of the magnetic fields we measure 
is the same as the driving scale $L$ of the stochastic motions: $L_{\rm int,B}  \approx L \approx 400 \, \mu 
\mathrm{m}$; the comparable value in the saturated state of periodic-box MHD simulations is robustly found to be 
$L_{\rm int,B} \approx 0.3  
L$ at similar $\mathrm{Rm}$ and $\mathrm{Pm}$~\citep{CR09,SBSW20}. 
\begin{figure}[ht]
  \centering
\includegraphics[width=0.95\linewidth]{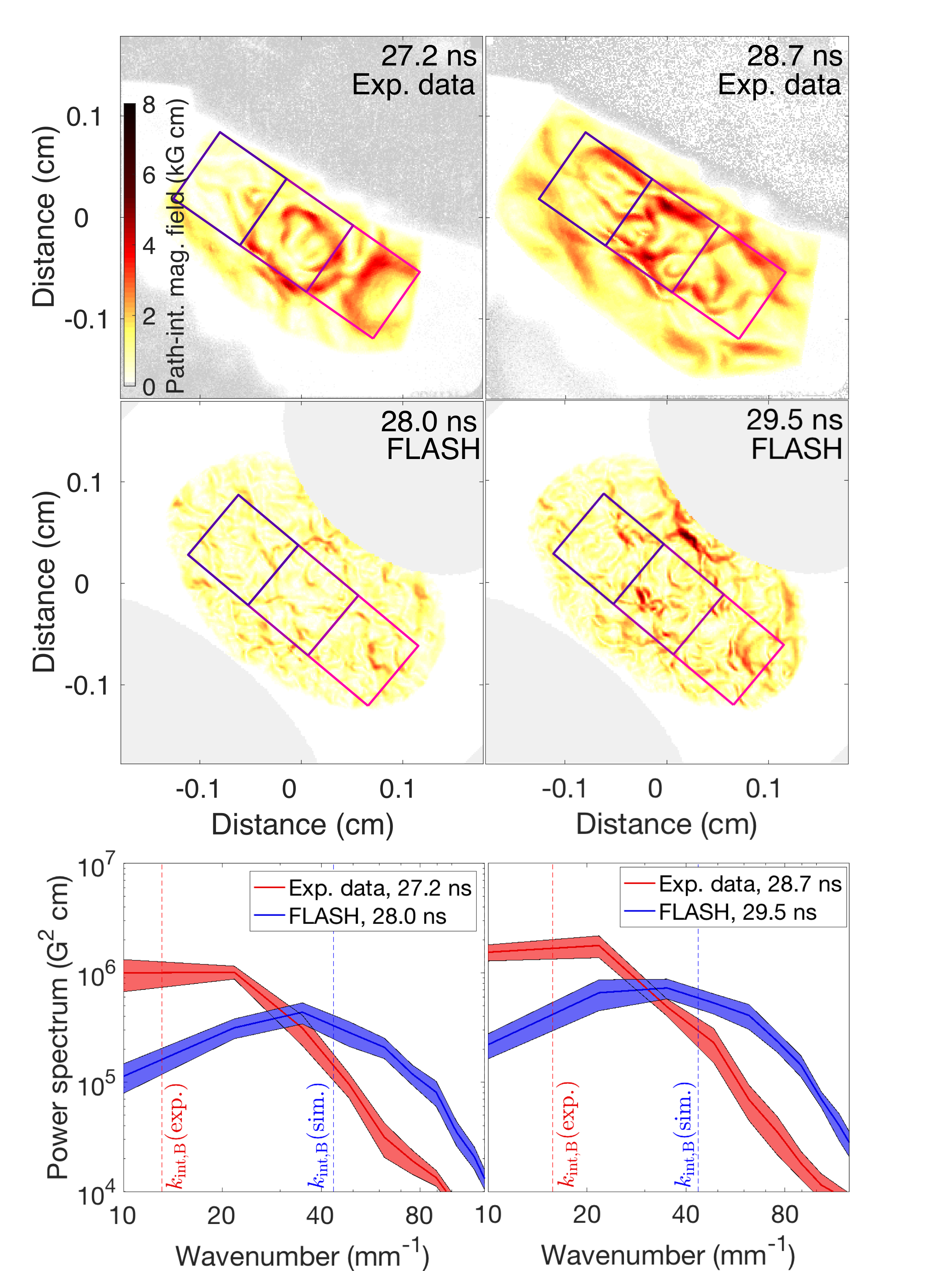}
\caption[Comparing the characteristic scale of stochastic magnetic fields between the experiment and the FLASH simulations]{\textbf{Comparing the characteristic scale of amplified magnetic fields between the experiment and the FLASH simulations.} Top row: path-integrated magnetic fields recovered from the experimental proton images at 27.2 ns (left), and 28.7 ns (right). Middle row: path-integrated magnetic fields derived from the FLASH simulations at 28.0 ns (left) and 29.6 ns (right). Bottom row: magnetic-energy spectra determined from equivalent experimental and simulated path-integrated magnetic field maps using Eq. [\ref{Especform}]. The spectra in each plot are calculated for the experimental and simulation data from the three regions depicted in the images directly above that plot. The uncertainty in the spectra is derived from the uncertainty in the spectra across the three regions. The integral-scale wavenumber $k_{\rm int,B} \equiv 2 \pi/L_{\rm int,B}$ for the experimental and simulation data is also plotted.}
\label{fig:protonimaging_corrlength}
\end{figure}
Intriguingly, the characteristic value of the integral scale obtained in the FLASH simulations of our 
experiment is much closer to the previous periodic-box simulations than to the experimentally measured value. 
This can be seen qualitatively by comparing 
simulated path-integrated magnetic field maps from the FLASH simulations with 
those reconstructed from the experimental proton images (see Figure 
\ref{fig:protonimaging_corrlength}). {\color{black} The path-integrated field structures differ noticeably in their scale}. 
More quantitatively, we can compare the magnetic-energy spectra recovered
from both experimental and simulated path-integrated field maps (using Eq. 
[\ref{Especform}]). We find that there is significantly more spectral power at wavenumbers 
$k \lesssim 3 \pi/L_{\rm int,B} \approx 25 \, \mathrm{mm}^{-1}$ in the former 
than the latter. There exist certain issues that can inhibit accurate 
determination of the magnetic-energy spectra from proton-imaging data at high wavenumbers~\citep{AGGS16}, but direct 
analysis of simulated proton images of the FLASH simulations confirms that this finding is not a result of these issues 
(see Supplementary Information). 
\color{black}
Possible explanations for why the characteristic scale of the magnetic fields in our experiment is larger than anticipated from resistive MHD simulations include additional physical processes which could arise due to the order-unity Hall parameter being attained subsequent to the seed field's amplification, or differences in the mechanism of resistive dissipation between the experiments and the simulations. 
The result is tantalizing given 
the long-standing problem of explaining the observed scale of tangled magnetic fields 
present in the ICM~\citep{SC06}: current ICM simulations tend to predict magnetic fields at smaller scales than observed~\citep{Roh19,Vaz18}.
\color{black}

Finally, we note that the maximum measured ratio of $\varepsilon_B$ to
$\varepsilon_{\rm turb}$ is $\varepsilon_{B}/\varepsilon_{\rm turb} \approx 0.03$, a value that is also obtained in the FLASH simulations. This ratio is a factor of a few smaller than that obtained for $\mathrm{Pm} \approx 1$ MHD simulation at saturation with comparable Reynolds 
numbers ($\varepsilon_{B}/\varepsilon_{\rm turb} \approx 0.08$)~\citep{SBSW20}. 
There are two possible explanations for the lower measured values of $\varepsilon_{B}/\varepsilon_{\rm turb}$ 
in the experiment. First, the time at which this measurement of 
the energy ratio is taken is less than a single driving-scale eddy turnover time after the turbulent plasma is 
formed; thus, it may be that insufficient time has passed for the saturated state of the fluctuation dynamo 
to be obtained in the experiment. Second, due to conductive losses, the plasma cools significantly 
for times $\gtrsim 30$ ns, attaining characteristic temperatures $T_e \approx T_i \approx 80 \, \mathrm{eV}$
at $t = 37.5$ ns (in the absence of heating by the Thomson-scattering probe beam -- see Supplementary Information).
Since both $\mathrm{Rm} \propto T_e^{3/2}$ and $\mathrm{Pm} \propto T_e^{3/2} T_i^{5/2}$
are sensitive functions of temperature, this cooling results in a transition to 
a different parameter regime: $\mathrm{Rm} \approx 20$, and $\mathrm{Pm} \approx 10^{-3}$. 
This transition should inhibit dynamo action, although to 
our knowledge, such a transition occurring during the nonlinear phase of the fluctuation dynamo has not been studied previously.
 
 In summary, our experiment supports the notion that turbulent plasma with $\mathrm{Pm} \gtrsim 1$ and sufficiently 
large $\mathrm{Rm}$ is capable of amplifying magnetic fields up to dynamical strengths.
 Furthermore, the time-resolved characterization provided by the experiment has demonstrated that magnetic-field amplification 
 in the plasma occurs at a much larger rate than the stretching rate associated with the outer scale of the turbulent motions. This rate of growth 
 is greater than is typically obtained in periodic-box MHD simulations with equivalent Mach number, $\mathrm{Rm}$, and $\mathrm{Pm}$, \color{black}{a finding that we attribute to the presence of strong directed shears in the interaction-region plasma.} \color{black} 
 The characteristic scale of these fields is found to be larger than anticipated by resistive-MHD simulations,\color{black}{ including our MHD FLASH simulations of the experiment, which otherwise faithfully reproduce the plasma's evolution. 
  Both findings suggest that the fluctuation dynamo -- when operating in realistic plasma -- 
 may be capable of generating large-scale magnetic 
 fields more efficiency than currently expected by analytic theory or MHD simulations.}\color{black} 

\matmethods{\subsection*{X-ray framing camera specifications}

Images of self-emitted soft X-rays from the interaction-region plasma were recorded using a 
framing camera~\citep{KBH88,BBLKO95} configured with a two-strip microchannel plate (MCP)~\citep{RB06} 
and a 50 $\mu$m pinhole array. The pinhole array was situated 9.14 cm away from the center of 
the target and the main detector at 27.4 cm, giving rise to a $\times 2$ image magnification. 
A thin filter composed of 0.5 $\mu$m polypropene and 150 nm of 
aluminum was placed in front of the MCP, removing radiation with photon energy $\lesssim 100 \, \mathrm{eV}$. 
The MCP itself was operated with a 1 ns pulse-forming module at a constant 400 V bias, 
and the two strips sequentially gated: this allowed for two images (time-integrated over a 1 ns interval) 
of the plasma at pre-specified times to be detected for each experimental shot. Electrons 
exiting the MCP struck a phosphor plate, producing an optical image, which 
was recorded using a 4096 $\times$ 4096 9-$\mu$m pixel charge-coupled-device (CCD) 
camera. The chosen voltage bias was such that the response of the CCD camera was 
linear and thus the relative counts of two given pixels provided a measure of 
the relative (optical) intensity incident on the CCD. To allow comparison between the X-ray images 
of the interaction-region plasma at different
stages of its evolution, the framing-camera bias was fixed 
throughout the experiment and its value optimized for probing the interaction-region plasma at 
peak emission. Given this normalization and the measured signal-to-noise ratio, the  
effective dynamic range of the camera was $\sim$100. The frequency-response curves of various components of the X-ray framing camera, along with the combined response, are shown in Figure S1a of the Supplementary Information.  

\subsection*{Thomson-scattering diagnostic specifications} 

The Thomson-scattering diagnostic employed a 30 J, frequency-doubled (526.5 nm)
laser, which probed the plasma in a cylindrical volume with cross-sectional area 50 $\mu$m$^2$
and length 1.5 mm centered on the target's center, which coincided with the target-chamber centre (TCC). 
The orientation of the scattering volume is shown in 
Figure \ref{fig:expsetup}. The 
scattered light was collected at scattering angle 63$^\circ$. As 
mentioned in Experimental Design, the Thomson-scattering signal was resolved spatially along the 
cylindrical scattering volume and integrated over the 1 ns duration of the laser pulse. 
The high- and low-frequency components of the spectrum were recorded separately using two distinct
spectrometers; the separation was performed using a beam splitter.  

\subsection*{Thomson-scattering data analysis}

To interpret the IAW and EPW features, a theory relating the scattered laser light detected at a particular wavelength -- or, equivalently, frequency -- to fundamental properties of the plasma is 
needed. For a given scattering vector $\boldsymbol{k}$, it can be shown~\citep{E69} that the spectrum $I(\boldsymbol{k},\omega)$
of the laser light scattered by the plasma at frequency $\omega$ is given by %SILLYREF EVANS
\begin{equation}
I(\boldsymbol{k},\omega) = N_e I_{0} \sigma_\mathrm{T} S(\boldsymbol{k},\omega) \, ,
\end{equation}
where $N$ is the total number of scattering electrons, $I_0$ the 
intensity of the incident laser, $\sigma_T \equiv (q_e^2/m_e c)^2 \sin^2{\vartheta_{\rm T}}$ 
the Thomson cross-section for scattering by a free electron ($q_e$ is the elementary charge, $m_e$ the electron mass,  
$c$ the speed of light, and $\vartheta_{\rm T}$ the angle between the direction of the electric field of the incident and scattered light), and
\begin{equation}
S(\boldsymbol{k},\omega) \equiv \frac{1}{2 \pi N_e} \int \mathrm{d}t \, \exp{\left[\mathrm{i} (\omega-\omega_0) t \right]} \langle n_e(\boldsymbol{k},0) n_e(\boldsymbol{k},t)^{*} \rangle  
\end{equation}
is the dynamic form factor ($\omega_0$ being the frequency of the incident light). Assuming that
the distribution functions of the electrons and ions are close to shifted Maxwellian distributions, with electron 
number density $n_e$, electron temperature $T_e$, temperature $T_j$ of ion 
species $j$, and bulk fluid velocity $\boldsymbol{u}$, and also that the Debye length is $\lambda_\mathrm{D} \lesssim 10^{-6} \, \mathrm{cm}$ 
(assumptions justified by Table S2 of the Supplementary Information), we find that $\alpha \equiv 1/k \lambda_\mathrm{D} \gtrsim 
8 > 1$; thus, we can employ the Salpeter approximation for the dynamic form 
factor~\citep{E69}: %SILLYREF EVANS
\begin{eqnarray}
  S(\boldsymbol{k},\omega) & \approx & \frac{1}{k v_{\mathrm{th}e}} \Gamma_\alpha\!\left(\frac{\tilde{\omega}-\omega_0}{k v_{\mathrm{th}e}}\right)
  \nonumber \\ && + \sum_{j} \frac{Z_{j} }{k v_{\mathrm{th}j}}  \left(\frac{\alpha^2}{1+\alpha^2}\right)^2 \Gamma_{\bar{\alpha}_j}\!\left(\frac{\tilde{\omega}-\omega_0}{k v_{\mathrm{th}j}}\right), 
  \label{Salpeterapprox}
\end{eqnarray}
where $\tilde{\omega} \equiv \omega - \boldsymbol{k} \cdot \boldsymbol{u}$ is the Doppler-shifted frequency, 
the sum is over all ion species in the plasma, $Z_j$  is the charge of ion species $j$, 
\begin{equation}
\Gamma_\alpha(x) \equiv \frac{\exp{\left(-x^2\right)}}{\sqrt{\pi} \left|1+\alpha^2 [1 + x Z(x)]\right|^2} \, , 
\end{equation}
and $\bar{\alpha}_j = Z_{j} \alpha^2 T_e/T_j (1+\alpha^2)$.  
The complex function $Z(x)$ is the plasma dispersion function~\citep{F61}.
For low-frequency fluctuations (in particular, ion-acoustic waves), $\omega -\omega_0 \sim k 
v_{\mathrm{th}j}$ and so the first term on the right-hand side of [\ref{Salpeterapprox}] 
is small by a factor of $\mathcal{O}[Z_i (m_e T_i)^{1/2}/(m_i T_e)^{1/2}] \ll 1$ when 
compared to the second (this factor is indeed small provided the ion temperature $T_i$ -- assumed equal for all ion species -- is comparable to the electron 
temperature); thus the shape of the low-frequency spectrum is dominated by the second term. On the other hand, for high-frequency fluctuations (electron plasma 
waves) satisfying $\omega -\omega_0 \sim k v_{\mathrm{th}e}$, the second term 
is smaller than the first by an exponential factor $\mathcal{O}[\exp\left(-m_e T_i /m_i T_e \right)] \ll 
1$; thus the shape of the high-frequency spectrum is dominated by the first 
term. We conclude that we can relate physical properties of the plasma 
to the measured EPW and IAW features using fits given by
the first and second terms of [\ref{Salpeterapprox}], respectively. 

However, for our experiment, there is a complication: the 
presence of stochastic motions and density fluctuations. The presence of such fluctuations 
means that the bulk fluid velocity $\boldsymbol{u}$ and electron density $n_e$ 
are not necessarily fixed parameters inside the Thomson-scattering volume during the time-integrated measurement,
but instead possess a range of values. To account for this range, we assume 
that fluctuations of velocity and density are isotropic and normally distributed, 
with means $\bar{\boldsymbol{u}}$ and $\bar{n}_e$, and standard deviations $\Delta u$ and $\Delta n_e$, respectively. Under this 
assumption, the appropriate fit for the IAW feature is
\begin{eqnarray}
  S_{\mathrm{IAW}}(\boldsymbol{k},\omega) & \approx & \frac{\sqrt{3}}{\sqrt{\pi} \Delta u} \int \mathrm{d} \tilde{U}_{\|} \, \exp{\left[-\frac{3 (\tilde{U}_{\|}-\bar{u}_{\|})^2}{\Delta u^2}\right]} \nonumber \\
  & \times & \sum_{j} \frac{Z_{j} }{k v_{\mathrm{th}j}}  \frac{\alpha^4}{(1+\alpha^2)^2} \Gamma_{\bar{\alpha}_j}\!\left(\frac{\omega- k \bar{U}_{\|} -\omega_0}{k v_{\mathrm{th}j}}\right), \quad \;
  \label{Salpeterapprox_ion}
\end{eqnarray}
where $\bar{u}_{\|} \equiv \hat{\boldsymbol{k}} \cdot \boldsymbol{u}$. For the EPW feature, we use 
\begin{eqnarray}
  S_{\mathrm{EPW}}(\boldsymbol{k},\omega) & \approx & \frac{1}{\sqrt{\pi} \Delta n_e} \int \mathrm{d} \tilde{n}_{e} \, \exp{\left[-\frac{\left(\tilde{n}_{e}-\bar{n}_{e}\right)^2}{\Delta n_e^2}\right]} \nonumber\\
  && \qquad \times \frac{1}{k v_{\mathrm{th}e}} \Gamma_\alpha\!\left(\frac{\omega-\omega_0}{k v_{\mathrm{th}e}}\right). 
  \label{Salpeterapprox_elec}
\end{eqnarray}
In spite of the seeming complexity of these equations, for a fully ionized CH plasma the spectral 
shapes implied by [\ref{Salpeterapprox_ion}] and [\ref{Salpeterapprox_elec}] are quite 
simple: a double peak structure, where the position and width of the peaks 
depend on plasma parameters. For the IAW feature, the distance between the peaks 
provides a measure of $T_e$; the shift in the position of the double-peaked spectrum with respect to the incident probe
beam's frequency gives a measurement of the bulk velocity $\bar{u}_{\|}$; the 
width of both peaks is a function of both $T_i$ and of the small-scale stochastic velocity dispersion $\Delta u$. 
The effect of the density on the shape of the IAW feature is negligible. For the 
EPW feature, the opposite holds: the position of the peak is determined by $n_e$. 
The width of the peak is in general determined by a range of factors -- Landau damping, collisions and the range of fluctuating densities $\Delta n_e$. 
For our experiment, both collisional broadening and that by Landau damping are
small (because $k \lambda_{e} \gg 1$ and $\alpha^2 \gg 1$, respectively), but the spread of densities can be significant. The fitting procedure is described in the Supplementary Information.

\subsection*{Proton-imaging diagnostic specifications}

The proton imaging diagnostic was implemented by imploding a $\mathrm{D}^3$He capsule~\citep{L06}: the capsule (diameter 420 $\mu$m) is composed of 2 $\mu$m of Si$\mathrm{O}_2$ (coated with aluminum), and filled with 18 atm $\mathrm{D}^3$He gas (6 atm $\mathrm{D}_2$ and 12 atm 
${}^{3}\mathrm{He}$). The capsule is imploded using 17, 270 J beams, each with a 600 ps pulse 
length, and 1.82 mm defocus. This results in the generation of $\sim10^{9}$ 3.3 MeV and 15.0 MeV 
protons via nuclear fusion reactions. 
These protons rapidly travel outward from the center of the backlighter as a  
uniform spherical sheet, passing through the plasma-filled volume, 
before reaching a detector composed of interleaved metal sheets
and solid-state nuclear track detector, CR-39~\citep{SF03}
 (chemical formula $\mathrm{C}_{12}\mathrm{H}_{18}\mathrm{O}_7$).
 The specific design of the detector is as follows: 7.5 $\mu$m of tantalum, then 1.5 mm of CR-39, then 150 $\mu$m of aluminum, and finally another 1.5 mm of CR-39. This design ensures that 3.3 MeV protons are stopped in the first layer of CR-39, and 15.0 MeV protons in the second; the tantalum filter minimizes damage to the CR-39 resulting from X-rays.
 Highly charged ions deposit the majority of their 
 energy close to where they are stopped completely, leaving small tracks of 
 broken molecular bonds. The positions of these tracks is determined by 
 etching the CR-39 for two to three hours in a 6N solution of sodium hydroxide,
yielding tracks with diameters $\sim10 \, \mu \mathrm{m}$. An automated microscope 
system records the location of tracks, before removing image defects and 
counting the number of protons in preset bin sizes: the output are proton (fluence) 
images. The robust design of the detector is such that protons reaching the 
detector are recorded with close to 100\% efficiency.
The dimensions of the imaging set-up are as follows: 
the distance $r_i$ from the proton source to the center of the target is $r_i = 1 \, \mathrm{cm}$, and the distance from the proton source to the detector is 28 cm. 
The magnification of the imaging set-up is thus $\times 28$.

}

\showmatmethods{} % Display the Materials and Methods section

\acknow{The research leading to these results has received funding from the European Research Council under the European Community’s Seventh Framework Programme (FP7/2007-2013)/ERC grant agreements no. 256973 and 247039, the U.S. Department of Energy (DOE) National Nuclear Security Administration (NNSA) under Contract No. B591485 to Lawrence Livermore National Laboratory (LLNL), Field Work Proposal No. 57789 to Argonne National Laboratory (ANL), Subcontract No. 536203 with Los Alamos National Laboratory, Subcontract B632670 with LLNL, and grants No. DE-NA0002724, DE-NA0003605, and DE-NA0003934 to the University of Chicago, DE-NA0003539 to the Massachusetts Institute of Technology, and Cooperative Agreement DE-NA0003856 to the Laboratory for Laser Energetics University of Rochester. We acknowledge support from the U.S. DOE Office of Science Fusion Energy Sciences under grant No. DE-SC0016566 and the National Science Foundation under grants No. PHY-1619573, PHY-2033925, and AST-1908551. Awards of computer time were provided by the U.S. DOE ASCR Leadership Computing Challenge (ALCC) program, using resources at ANL, which is supported by the U.S. DOE Office of Science under contract No. DE-AC02-06CH11357.
We acknowledge funding from grants 2016R1A5A1013277 and 2017R1A2A1A05071429 of the National Research Foundation of Korea. Support from AWE plc., the Engineering and Physical Sciences Research Council (grant numbers EP/M022331/1, EP/N014472/1, and EP/R034737/1) and the U.K. Science and Technology Facilities Council is also acknowledged.}

\showacknow{} % Display the acknowledgments section

% Bibliography


\begin{thebibliography}{}

\end{thebibliography}


\begin{thebibliography}{99}

%\bibitem{BB96} R~Beck, A~Brandenburg, D~Moss, A~Shukurov, and D~Sokoloff,
%{Galatic Magnetism: Recent Developments and Perspectives},
%\textit{Annu. Rev. Astron. Astrophys.} \textbf{34}, 155 (1996)

%\bibitem{CT02} CL~Carilli, and GB~Taylor,
%{Cluster Magnetic Fields},
%\textit{Annu. Rev. Astron. Astrophys.} \textbf{40}, 319 (2002)

\bibitem{B15} R~Beck,
{Magnetic fields in spiral galaxies},
\textit{Astron. Astrophys. Rev.} \textbf{24}, 1 (2015)

\bibitem{V18} V~Vacca \textit{et. al.}
{Magnetic fields in galaxy clusters and in the large-scale structure of the universe},
\textit{Galaxies} \textbf{6}, 142 (2018)

\bibitem{BS51} L~Biermann, and A~Schluter,
{Cosmic radiation and cosmic magnetic fields. II. Origin of cosmic magnetic fields},
\textit{Phys. Rev.} \textbf{29}, 29 (1951)

\bibitem{KCOR97} RM~Kulsrud, R~Cen, JP~Ostriker and D~Ryu,
{The protogalactic origin for cosmic magnetic fields},
\textit{Astrophys. J.} \textbf{480}, 481 (1997)

\bibitem{K99} R~Kulsrud,
{A critical review of galactic dynamos},
\textit{Annu. Rev. Astron. Astrophys.} \textbf{37}, 37 (1999)

\bibitem{S19} K~Subramanian,
{From primordial seed magnetic fields to the galactic dynamo},
\textit{Galaxies} \textbf{7}, 47 (2019)

\bibitem{SSH06} K~Subramanian, A~Shukurov, and NEL~Haugen,
{Evolving turbulence and magnetic fields in galaxy clusters},
\textit{Mon. Not. R. Astron. Soc.} \textbf{366}, 1437 (2006) 

\bibitem{RKCD08} D~Ryu, H~Kang, J~Cho, and S~Das,
{Turbulence and magnetic fields in the large-scale structure of the universe},
\textit{Science} \textbf{320}, 909 (2008)

\bibitem{B50} GK~Batchelor,
{On the spontaneous magnetic field in a conducting liquid in turbulent motion},
\textit{Proc. R. Soc. A.} \textbf{201}, 405 (1950)

\bibitem{R19} F~Rincon,
{Dynamo theories},
\textit{J. Plasma Phys.} \textbf{85}, 205850401 (2019)

%\bibitem{TCB13} SM~Tobias, F~Cattaneo, and S~Boldyrev, in
%{Ten Chapters in Turbulence}, 
%edited by P.A~Davidson, Y~Kaneda, and K~Sreenivasan (Cambridge University Press, Cambridge, 2013), pp. 351-404.

\bibitem{K67} AP Kazentsev,
{Enhancement of a magnetic field by a conducting fluid},
\textit{Soviet-JETP} \textbf{26}, 1031 (1968)

\bibitem{VZ72} SI Vainstein, and YB Zel'dovich,
{Review of topical problems: origin of magnetic fields in astrophysics (turbulent ‘dynamo’ mechanisms)},
\textit{Sov. Phys. Usp.} \textbf{15}, 159 (1972)

\bibitem{ZRMS84} YB Zel'dovich, AA Ruzmaikin, SA Molchanov, and DD Sololov,
{Kinematic dynamo problem in a linear velocity field},
\textit{J. Fluid Mech.} \textbf{144}, 1 (1984)

\bibitem{KA92} R~Kulsrud, and SW~Anderson,  
{The spectrum of random magnetic fields in the mean field dynamo theory of the galactic magnetic field},
\textit{Astrophys. J.} {\bf 396}, 606 (1992)

\bibitem{MFP81} M~Meneguzzi, U~Frisch, and A~Pouquet,
{Helical and nonhelical turbulent dynamos},
\textit{Phys. Rev. Lett.} \textbf{47}, 1060 (1981)

\bibitem{KYM91} S Kida, S Yanase, and J Mizushima,
{Statistical properties of MHD turbulence and turbulent dynamo},
\textit{Phys. Fluids A} \textbf{3}, 457 (1991)

%\bibitem{KD95} RG Kleva, and JF Drake,
%{Nonlinear magnetohydrodynamic dynamo},
%\textit{Phys. Plasmas} \textbf{2}, 4455 (1995)

\bibitem{MMAG96} RS Miller, F Mashayek, V Adumitroaie, and P Givi, 
{Structure of homogeneous nonhelical magnetohydrodynamic turbulence},
\textit{Phys. Plasmas} \textbf{3}, 3304 (1996)

\bibitem{CV01} J~Cho, and ET~Vishniac,
{The generation of magnetic fields through driven turbulence},
\textit{Astrophys. J.} \textbf{538}, 217 (2001)

\bibitem{SCTM04} AA~Schekochihin, SC~Cowley, SF~Taylor, JL~Maron, and JC~McWilliams,
{Simulations of the small-scale turbulent dynamo},
\textit{Astrophys. J.} \textbf{612}, 276 (2004)

\bibitem{HBD04} NE~Haugen, A~Brandenburg, and W~Dobler,
{Simulations of nonhelical hydromagnetic turbulence},
\textit{Phys. Rev. E} \textbf{70}, 016308 (2004)

\bibitem{S07} AA~Schekochihin, AB~Iskakov, SC~Cowley,  JC~McWilliams, MRE~Proctor and TA~Yousef, 
{Fluctuation dynamo and turbulent induction at low magnetic Prandtl numbers}.
\textit{New J. Phys.} {\bf 9}, 300 (2007)

\bibitem{CR09} J~Cho, and D~Ryu,
{Characteristic lengths of magnetic field in magnetohydrodynamic turbulence},
\textit{Astrophys. J.} \textbf{705}, L90 (2009)

\bibitem{B12} A~Beresnyak,
{Universal nonlinear small-scale dynamo},
\textit{Phys. Rev. Lett.} \textbf{108}, 035002 (2012)

\bibitem{PJR15} DH~Porter, TW~Jones, and D~Ryu
{Vorticity, shocks, and magnetic fields in subsonic, ICM-like turbulence gas motions in the intra-cluster medium},
\textit{Astrophys. J.} {\bf 810}, 93 (2015)

\bibitem{SBSW20} A~Seta, PJ~Bushby, A~Shukurov and TS Wood, 
{On the saturation mechanism of the fluctuation dynamo at Pm > 1},
\textit{Phys. Rev. Fluids} \textbf{5}, 043702 (2020)

\bibitem{RS81} AA~Ruzmaikin, and DD~Sokolov,
{The magnetic field in mirror-invariant turbulence},
\textit{Sov. Astron. Lett.} \textbf{7}, 388 (1981)

\bibitem{BC04} S~Boldyrev, and F~Cattaneo, 
{Magnetic-field generation in Kolmogorov turbulence},
\textit{Phys. Rev. Lett.} \textbf{92}, 144501 (2004)

%\bibitem{SCMM04} AA~Schekochihin, SC~Cowley, JL~Maron, and JC~McWilliams
%{Critical magnetic Prandtl number for small-scale dynamo},
%\textit{Phys. Rev. Lett.} \textbf{92}, 054502 (2004)

\bibitem{ISCMP07} AB~Iskakov, AA~Schekochihin, SC~Cowley, JC~McWilliams, and MRE~Proctor
{Numerical demonstration of fluctuation dynamo at low magnetic Prandtl numbers},
\textit{Phys. Rev. Lett.} \textbf{98}, 208501 (2007)

%\bibitem{SHBCMM05} AA~Schekochihin, NEL Haugen, A Brandenburg, SC Cowley, JL Maron, and JC~McWilliams,
%{The onset of a small-scale turbulent dynamo at low magnetic Prandtl numbers},
%\textit{Astrophys. J.} \textbf{625}, L115 (2005)

\bibitem{CVB09} J~Cho, ET~Vishniac, A~Beresnyak, A~Lazarian, and D~Ryu,
{Growth of magnetic fields induced by turbulent motions},
\textit{Astrophys. J.} \textbf{693}, 1449 (2009)

%\bibitem{H03} NE~Haugen, A~Brandenburg, and W~Dobler,
%{Is nonhelical hydromagnetic turbulence peaked at small scales?},
%\textit{Astrophys. J.} \textbf{597}, L141 (2003)

%\bibitem{SCHMM02} AA~Schekochihin, SC~Cowley,  G.W~Hammett, J.L~Maron and J.C~McWilliams, 
%{A model of nonlinear evolution and saturation of the turbulent MHD dynamo}.
%\textit{New J. Phys.} {\bf 4}, 84 (2002)

%\bibitem{SCTHMM04} AA~Schekochihin, SC~Cowley, SF~Taylor,  G.W~Hammett, J.L~Maron and J.C~McWilliams, 
%{Saturated State of the Nonlinear Small-Scale Dynamo}.
%\textit{Phys. Rev. Lett.} {\bf 92}, 084504 (2004)

%\bibitem{KSS14} MW~Kunz, AA~Schekochihin, and JM~Stone
%{Firehose and Mirror Instabilities in a Collisionless Shearing Plasma},
%\textit{Phys. Rev. Lett.} \textbf{112}, 205003 (2014)

%\bibitem{HSS16} P~Helander, M~Strumik, and AA~Schekochihin 
%{Constraints on dynamo action in plasmas},
%\textit{J. Plasma Phys.} \textbf{82}  6(2016)

%\bibitem{HS99} P~Helander, and D.J~Sigmar
%{Collisional Transport in Magnetized Plasmas},
%(Cambridge University Press, Cambridge, 1999), pp. 59-89

%\bibitem{SK18} DA~St-Onge and M.W~Kunz, 
%{Fluctuation Dynamo in a Collisionless, Weakly Magnetized Plasma}.
%\textit{Astrophys. J. Lett.} {\bf 863}, L25 (2018)

%\bibitem{RCSV16} F~Rincon, F~Califano, AA~Schekochihin, and F~Valentini
%{Turbulent dynamo in a collisionless plasma},
%\textit{Proc. Nat. Acad. Sci.} \textbf{113}, 3950 (2016)

%\bibitem{M07} R~Monchaux \textit{et. al.},
%{Generation of a Magnetic Field by Dynamo Action in a Turbulent Flow of Liquid Sodium},
%Phys. Rev. Lett. \textbf{98}, 044502 (2007)

\bibitem{G12} G~Gregori \textit{et. al.},
{Generation of scaled protogalactic seed magnetic fields in laser-produced shock waves},
\textit{Nature} \textbf{481}, 480 (2012)

\bibitem{M14} J~Meinecke \textit{et. al.},
{Turbulent amplification of magnetic fields in laboratory laser-produced shock waves},
\textit{Nat. Phys.} \textbf{10}, 520 (2014)

\bibitem{M15} J~Meinecke \textit{et. al.}, 
{Developed turbulence and nonlinear amplification of magnetic fields in laboratory and astrophysical plasmas},
\textit{Proc. Nat. Acad. Sci.} \textbf{112}, 8211 (2015)

\bibitem{GRM15} G~Gregori, B~Reville, and F~Miniati  
{The generation and amplification of intergalactic magnetic fields in analogue laboratory experiments with high power lasers},
\textit{Phys. Reports.} \textbf{601}, 1 (2015)

\bibitem{T17} P~Tzeferacos \textit{et. al.}, 
{Numerical modeling of laser-driven experiments aiming to demonstrate magnetic field amplification via turbulent dynamo}
\textit{Phys. Plasmas} \textbf{24}, 041404 (2017)

\bibitem{T18} P~Tzeferacos \textit{et. al.}, 
{Laboratory evidence of dynamo amplification of magnetic fields in a turbulent plasma}
\textit{Nat. Commun.} \textbf{9}, 591 (2018)

\bibitem{B97} T~Boehly \textit{et. al.},
{ Initial performance results of the OMEGA laser system},
\textit{Optics Communications} \textbf{133}, 495  (1997)

\bibitem{F00} B~Fryxell \textit{et. al.},
{FLASH: An Adaptive Mesh Hydrodynamics Code for Modeling Astrophysical Thermonuclear Flashes},
\textit{Astrophys. J.} \textbf{131}, S273 (2000)

\bibitem{T15} P~Tzeferacos \textit{et. al.},
{FLASH MHD simulations of experiments that study shock generated magnetic fields.},
\textit{High Energy Dens. Phys.} \textbf{17}, 24 (2015)

\bibitem{M17} S~M\"uller \textit{et. al.}, 
{Evolution of the design and fabrication of astrophysics targets for turbulent dynamo (TDYNO) experiments on OMEGA}
\textit{Fusion Sci. Tech.} \textbf{73}, 434 (2017)

%\bibitem{R12} D.D~Ryutov \textit{et. al.},
%{Intra-jet shocks in two counter-streaming, weakly collisional plasma jets},
%Phys. Plasmas \textbf{19}, 074501 (2012)

\bibitem{R18} A~Rigby, J~Katz, AFA~Bott, TG~White, P~Tzeferacos, DQ~Lamb, DH~Froula, G~Gregori. 
{Implementation of a Faraday rotation diagnostic at the OMEGA laser facility}, 
High Power Laser Science and Engineering \textbf{6} (2018)

\bibitem{RL86} GB~Rybicki and AP~Lightman,
{Radiative processes in astrophysics}.
(Wiley-VCH, Weinheim, 2004)

\bibitem{C12} E~Churazov \textit{et. al.},
{X-ray surface brightness and gas density fluctuations in the Coma cluster},
\textit{Mon. Not. R. Astron. Soc.} \textbf{421}, 1123 (2012)

\bibitem{Z14} I~Zhuravleva \textit{et. al.},
{The relation Between gas density and velocity power spectra in galaxy clusters: qualitative treatment and cosmological simulations},
\textit{Astrophys. J.} \textbf{788}, L13 (2014)

\bibitem{W19} TG~White \textit{et. al.},
{Supersonic plasma turbulence in the laboratory},
\textit{Nature Comm.} \textbf{10}, 1758 (2019)

%\bibitem{S57} VD~Shafranov,
%{The structure of a shock wave in a plasma},
%\textit{JETP} \textbf{5}, 1183 (1957)

%\bibitem{M04} A.J~Mackinnon \textit{et. al.},
%{Proton radiography as an electromagnetic field and density perturbation diagnostic},
%Rev. Sci. Instrum. \textbf{75}, 3531 (2004)

\bibitem{K12} NL~Kugland \textit{et. al.},
{Relation between electric and magnetic field structures and their proton-beam images},
\textit{Rev. Sci. Instrum.} \textbf{83}, 101301 (2012)

\bibitem{AGGS16} AFA~Bott, C~Graziani, TG~White, P~Tzeferacos, DQ~Lamb, G~Gregori, and AA~Schekochihin 
{Proton imaging of stochastic magnetic fields},
\textit{J. Plasma Phys.} \textbf{83} 6 (2017)

\bibitem{GM96} W~Gangbo, and RJ~McCann
{The geometry of optimal transportation},
\textit{Acta Math.} \textbf{177}, 113-161 (1996)

\bibitem{S12} G~Sarri \textit{et. al.},
{Dynamics of self-generated, large amplitude magnetic fields following high-intensity laser matter interaction}, 
\textit{Phys. Rev. Lett.} \textbf{109}, 205002. (2012) 

\bibitem{S71} JA~Stamper \textit{et. al.},
{Spontaneous magnetic Fields in laser-produced plasmas},
\textit{Phys. Rev. Lett.} \textbf{26}, 1012 (1971)

\bibitem{K13} NL~Kugland \textit{et. al.},
{Visualizing electromagnetic fields in laser-produced
counter-streaming plasma experiments for collisionless shock laboratory astrophysics},
\textit{Phys. Plasmas} \textbf{20}, 056313 (2013)

%\bibitem{GT15} C~Graziani \textit{et. al.},
%{The Biermann catastrophe in numerical magnetohydrodynamics}.
%\textit{Astrophys. J.} {\bf 802}, 43 (2015)

\bibitem{B65} SI~Braginskii, 
{Transport processes in a plasma},
in: M.A. Leontovich (Ed.), Reviews of Plasma Physics, vol. 1. (1965), p. 205.

%\bibitem{CRG14} JE~Cross, B~Reville, and G~Gregori,
%{Scaling of magneto-quantum-radiative hydrodynamic equations: from laser-produced plasmas to astrophysics},
%\textit{Astrophys. J.} \textbf{795}, 59 (2014).

\bibitem{H94} JD~Huba,
{NRL plasma formulary}.
(Naval Research Laboratory, Washington DC, 1994)

\bibitem{R99} DD~Ryutov, RP~Drake, and J Kane, 
{Similarity criteria for the laboratory simulation of supernova hydrodynamics},
\textit{Astrophys. J.} \textbf{518}, 821 (1999)

%\bibitem{CL14} J~Colvin, and J~Larsen,
%{Extreme physics: properties and behaviour of matter at extreme conditions},
%(Cambridge University Press, Cambridge, 2014)

\bibitem{SM14} AN~Simakov and K~Molvig,
{Electron transport in a collisional plasma with multiple ion species}.
\textit{Phys. Plasmas} \textbf{21}, 024503 (2014)

\bibitem{SM16} AN~Simakov and K~Molvig,
{Hydrodynamic description of an unmagnetized plasma with multiple ion species. II. Two and three ion species plasmas}.
\textit{Phys. Plasmas} \textbf{23}, 032116 (2016)

\bibitem{Si19} A~Simionescu \textit{et. al.},
{Constraining gas motions in the intra-cluster medium}.
\textit{Space Sci. Rev.} \textbf{215}, 24 (2019)

\bibitem{SC06} AA~Schekochihin, and SC~Cowley 
{Turbulence, magnetic fields, and plasma physics in clusters of galaxies},
\textit{Phys. Plas.} \textbf{13}, 056501 (2006)

\bibitem{Roh19} S~Roh, D~Ryu, H~Kang, S~Ha1, and H~Jang,
{Turbulence dynamo in the stratified medium of galaxy clusters}
\textit{Astrophys. J.} \textbf{883}, 138 (2019) 

\bibitem{Vaz18} F~Vazza, G~Brunetti, M~Bruggen, and A~Bonafede,
{Resolved magnetic dynamo action in the simulated intracluster medium}
\textit{Mon. Not. R. Astron. Soc.} \textbf{472}, 1672 (2018) 

%\bibitem{H15} CM~Huntington \textit{et. al.},
%{Observation of magnetic field generation via the Weibel instability in interpenetrating plasma flows},
%\textit{Nat. Phys} \textbf{11}, 173 (2015)

\bibitem{KBH88} JD~Kilkenny, P~Bell, R~Hanks, G~Power, RE~Turner, and J~Wiedwald,
{High‐speed gated x‐ray imagers}.
\textit{Rev. Sci. Instrum.} \textbf{59}, 1793 (1988)

\bibitem{BBLKO95} DK~Bradley, PM~Bell, OL~Landen, JD~Kilkenny and J~Oertel,
{Development and characterization of a pair of 30–40 ps x‐ray framing cameras}.
\textit{Rev. Sci. Instrum.} \textbf{66}, 716 (1995)

\bibitem{RB06} GA~Rochau \textit{et. al.},
{Energy dependent sensitivity of microchannel plate detectors}.
\textit{Rev. Sci. Instrum.} \textbf{802}, 323 (2006)

\bibitem{E69} DE~Evans and J~Katzenstein,
{Laser light scattering in laboratory plasmas}.
\textit{Rep. Prog. Phys.} \textbf{32}, 207 (1969).

\bibitem{F61} BD~Fried and SD~Conte, 
{The plasma dispersion function}.
(Academic Press, New York, 1961)

\bibitem{L06} C.~Li \textit{et. al.},
{Measuring E and B Fields in laser-produced plasmas with monoenergetic proton radiography},
Phys. Rev. Lett. \textbf{97}, 3 (2006)

\bibitem{SF03} FH~S\'eguin \textit{et. al.},
{Spectrometry of charged particles from inertial-confinement-fusion plasmas}.
\textit{Rev. Sci. Instrum.} \textbf{74}, 975 (2003)

\end{thebibliography}
\end{document}

%% Comment/remove this line before generating final copy for submission
%\instructionspage  

%\maketitle

%% Adds the main heading for the SI text. Comment out this line if you do not have any supporting information text.
\SItext

\section*{Additional information about the X-ray imaging diagnostic}

\subsection*{Framing camera response}
\begin{figure}
  \centering
\includegraphics[width=0.95\linewidth]{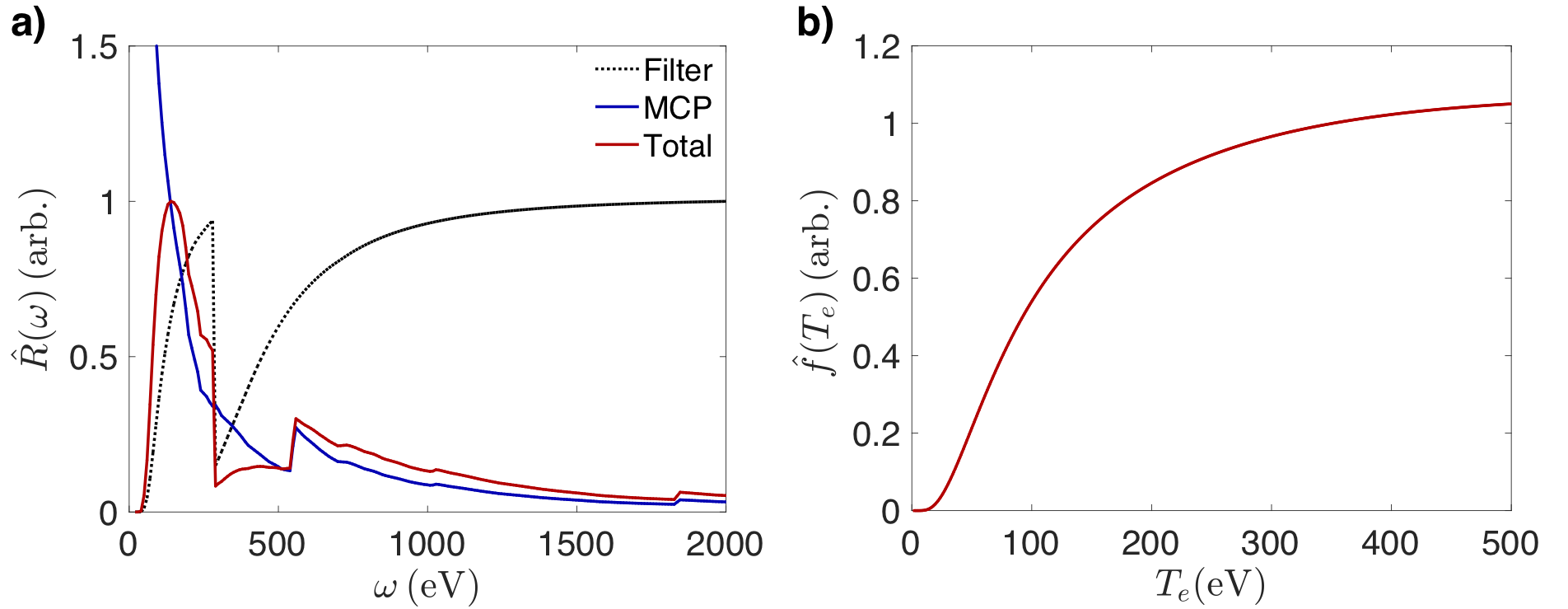}
\caption[Response of the X-ray framing camera]{\textbf{Response of the X-ray framing camera.} \textbf{a)} Response functions $\hat{R}(\omega)$ associated with the X-ray framing camera filter alone (black, dotted), 
the microchannel plate alone (blue), and the combined response (red). The total response and the filter response are normalized to their respective maximum values; the MCP response is normalized to its value at the frequency at which the total response is maximized. 
\textbf{b)} Relative temperature dependence of the measured optical intensity on the CCD camera. The temperature is normalized to its value at $T_e = 350 \, \mathrm{eV}$.} 
\label{fig:Xrays_cameraresponse}
\end{figure}

The responses of the various components of the X-ray framing camera, along with the combined temperature response, is shown in Figure \ref{fig:Xrays_cameraresponse}.
\subsection*{Calculation of mean X-ray emission profiles}
The mean emission profiles shown in Figure 2 of the main text are calculated via direct application to the images of a 57 $\times$ 57 pixel mean filter. With a pixel size of $9 \, \mu \mathrm{m}$, this corresponds to assuming that the mean emission profile 
 varies smoothly on scales $\gtrsim 500 \, \mu \mathrm{m}$; 
 the largest relative fluctuations inside the interaction region have typical size $L \lesssim 200-400 \, \mu \mathrm{m}$, 
 providing a modest scale separation. 
 However, applying only a mean filter to the images is inadequate for
 determining reasonable mean emission profiles; the presence of shocks on either side of the interaction-region plasma 
 implies that the global emission profile would in reality have sharp boundaries (on scales $\lesssim 500 \, \mu 
 \mathrm{m}$), a feature not adequately picked up by a linear filter.
This phenomenon is evident in Figure 3 of the main text, where the X-ray emission is observed to drop rapidly over $\sim 10-20$ pixels ($\sim 100 - 200 \, \mu$m). To account for this, a two-dimensional Gaussian window function on the scale of the boundary is combined with the  
mean emission profile to prevent the boundary region from distorting the calculated relative X-ray intensity map 
(similar techniques are also applied in spectral analysis of data with gaps~\citep{A12}).

\color{black}

\subsection*{Semi-deterministic feature of relative X-ray intensity maps}

For a few of the relative intensity maps that we extracted -- for example, that shown in Figure 4b of the main text -- we note that positively signed intensity fluctuations seem to be more concentrated near the equatorial plane of the interaction region, whereas negatively signed intensity fluctuations are more concentrated on the sides. In the raw X-ray images from which these particular relative intensity maps were derived, a sudden drop in emission between the central part of the interaction region is discernible by eye, suggesting that this feature is physical, rather than a numerical artifact of the algorithm separating the mean emission profile from the relative-intensity profile. We believe this feature is due to the interaction-region plasma not in fact being a flat disk, but  instead being `buckled'. It is therefore possible that the sharp drop in emission in the outer regions of the interaction region is due to a sudden decrease in the effective path length over which the plasma emits. Such an effect would not be captured by the algorithm that was used to determine the mean density profile. This interpretation is supported by the FLASH simulations of the experiment: the simulated interaction-region plasma is indeed buckled, and the aforementioned feature is evident in simulated X-ray images of the plasma (see below). The presence of this feature in some images but not others suggests that this distortion of the interaction-region shape may depend on the particular stochastic motions that emerge after the region forms. We emphasize that we find that all of our X-ray images at $t > 31$ ns lead to a similar density spectrum (including those in which the concentration of positively signed intensity fluctuations near the equatorial plane of the interaction region is much less prominent), and thus believe that our conclusions concerning the properties of fluctuations are not significantly affected by the presence of this feature.

\color{black}

%\subsection*{Assumption of small temperature fluctuations in the interaction-region plasma} %When relating relative emission fluctuations to relative density fluctuations, we assumed that $\delta T_e/\bar{T}_e \lesssim \delta n_e/\bar{n}_e$;

\section*{Additional information on the Thomson-scattering diagnostic}

\subsection*{Raw spatially resolved Thomson-scattering spectra}
\begin{figure}
  \centering
\includegraphics[width=0.7\linewidth]{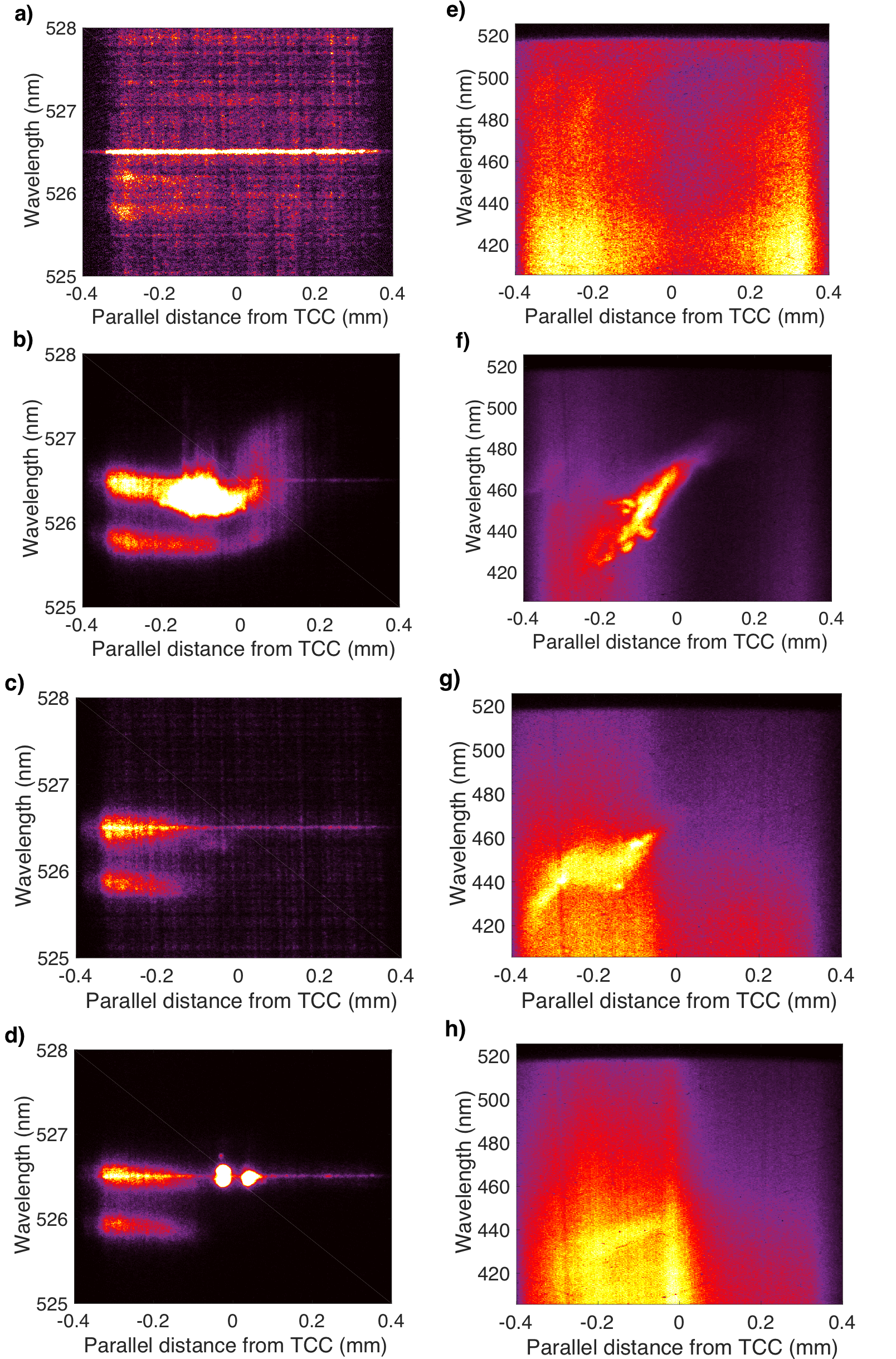}
\caption[Experimental Thomson-scattering spectra at the time of interaction-region formation]{\textbf{Experimental Thomson-scattering spectra at the time of the interaction-region formation.} The low-frequency spectra 
(IAW features) are shown in the left column (a-d) and the high-frequency spectra (EPW features) in the right column (e-h). 
The scattering volume is orientated at a 63$^{\circ}$ angle to the direction parallel to the line of centers; for convenience of reference to the X-ray images, 
we report distances projected onto the line of centers. The timings of the images are as follows: \textbf{a)}, \textbf{e)}, 24.0 ns; \textbf{e)}, \textbf{f)} 25.5 ns; \textbf{c)}, \textbf{g)} 27.0 ns; \textbf{d)}, \textbf{h)} 28.5 ns.} 
\label{fig:TS}
\end{figure}

\begin{figure}
  \centering
\includegraphics[width=0.95\linewidth]{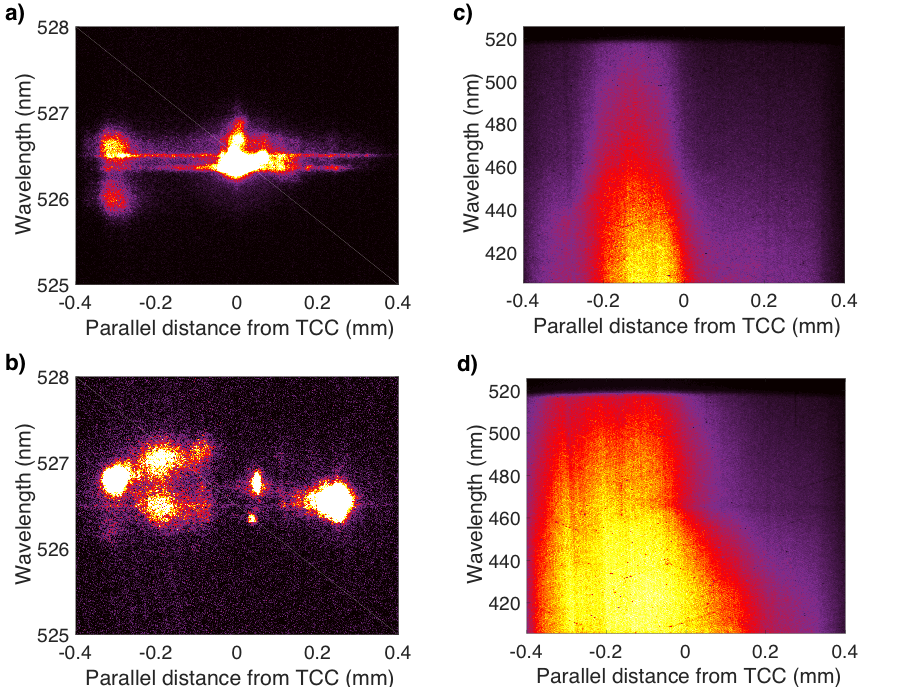}
\caption[Experimental Thomson-scattering spectra at later times]{\textbf{Experimental Thomson-scattering spectra at later times.} The low-frequency spectra 
(IAW features) are shown in the left column (a-b) and the high-frequency spectra (EPW features) in the right column (c-d). 
The timings of the images are as follows: \textbf{a)}, \textbf{c)}, 30.0 ns; \textbf{b)}, \textbf{d)} 37.5 ns.} 
\label{fig:TS_latetimedata}
\end{figure}

Figure \ref{fig:TS} shows the raw data from the spatially resolved Thomson-scattering diagnostic at times close to the interaction-region formation (Figures \ref{fig:TS}a to \ref{fig:TS}d are the IAW features and Figures \ref{fig:TS}e to \ref{fig:TS}h
the EPW features).
The Thomson-scattering data used to extract the effective ion temperature and flow velocities at late times is shown in Figure \ref{fig:TS_latetimedata}.

\subsection*{Additional information on spectral fitting procedure}

\begin{figure}
  \centering
\includegraphics[width=0.95\linewidth]{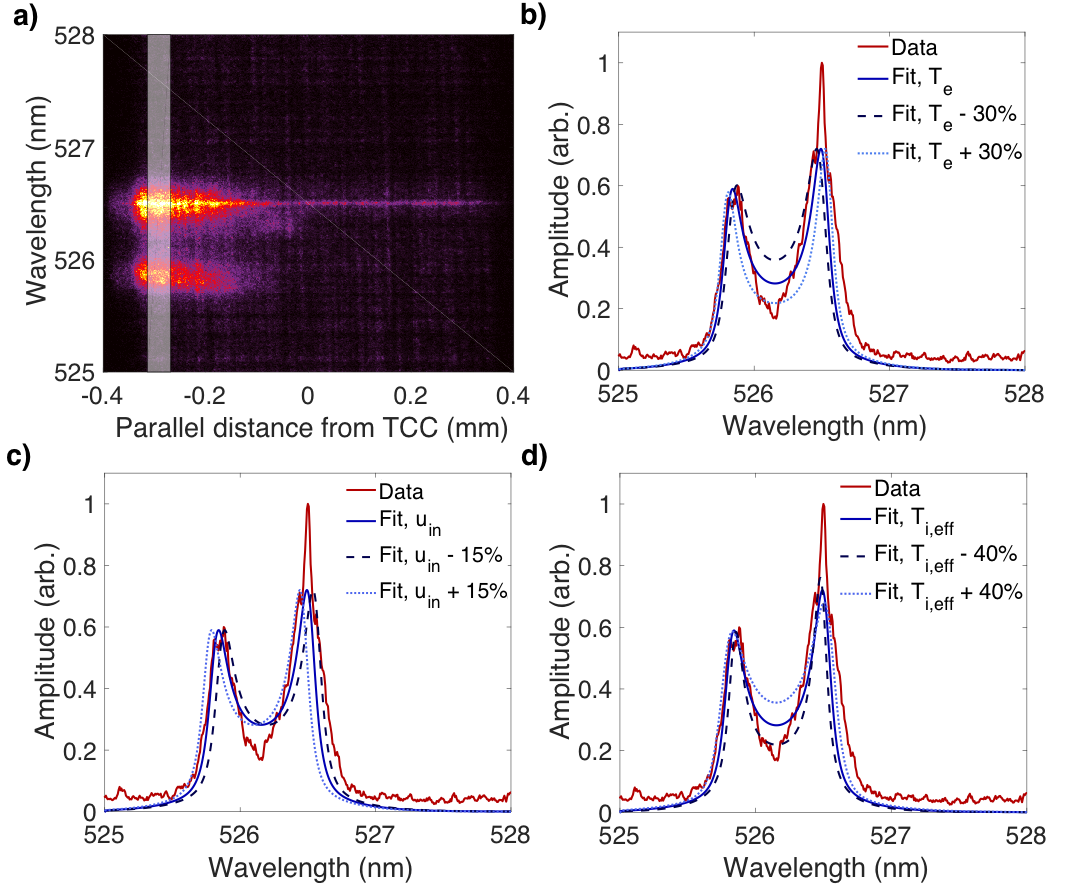}
\caption[Fitting the IAW feature]{\textbf{Fitting the IAW feature.} \textbf{a)} IAW spectrum from 27.0 ns after drive-beam pulse initiation. In (b-d), the 
experimental 1D spectrum (in red) is calculated by averaging the full spectrum horizontally over the region denoted by the translucent rectangle. 
\textbf{b)} Experimental spectrum: the best-fit spectrum ($T_e = 470 \, \mathrm{eV}$, $T_{i,\mathrm{eff}} = 620 \, \mathrm{eV}$, $\bar{u}_{\rm in} = 230 \, \mathrm{km} \; \mathrm{s}^{-1}$, $n_e = 1.2 \times 10^{20} \, \mathrm{cm}^{-3}$) in solid blue, and the
fits parametrized by $T_e$ sensitivity bounds (-30\% dashed, +30\% dotted). All fits are convolved with the (experimentally measured) instrument function, which has a characteristic width of 0.025 nm. \color{black}{The sharp peak seen at the probe-beam wavelength ($526. 5 \, \mathrm{nm}$) is the so-called `stray-light' feature, and is not part of the plasma's Thomson-scattering spectrum; it is disregarded for the purposes of 
fitting. }\color{black}
\textbf{c)} Same as b) but the fits parametrized by $\bar{u}_{\rm in}$ sensitivity bounds are plotted. \textbf{d)} Same as b) but the fits parametrized by $T_{i,\mathrm{eff}}$ sensitivity bounds are plotted.} 
\label{fig:TS_IAWfitsexample}
\end{figure}

We illustrate the spectral fitting procedure used to extract bulk plasma parameters with two examples. Figure~\ref{fig:TS_IAWfitsexample}a shows the IAW feature at 27.0 ns after the laser pulse.
To fit the experimental spectrum at a given position, we first average it over a 100 $\mu$m interval centered at that position. We then calculate a fit using
Eq. [13] of the main text, substituting the dispersion relation $\omega = \omega(k)$ for a light wave propagating through plasma, and adjusting $T_e$, $T_i$, $\bar{u}_{\|}$ and $\Delta u$. In the chosen geometry of the 
diagnostic, the bulk velocity parallel to $\hat{\boldsymbol{k}}$ is equal to the 
inflow velocity $\bar{u}_{\rm in}$ parallel to the target's line of centers, i.e., $\bar{u}_{\|} = \bar{u}_{\rm in}$. 
We do not use an absolute calibration 
of the spectrum for the fit, but 
instead normalize the height of the theoretical spectrum to the lower-wavelength experimental 
peak (the higher-wavelength experimental peak is typically distorted by a stray-light 
feature). Once a best fit is obtained, we then adjust each parameter individually to assess 
the sensitivity of the fit. Figures~\ref{fig:TS_IAWfitsexample}b and 
\ref{fig:TS_IAWfitsexample}c demonstrate this process for the electron 
temperature $T_e$ and velocity $\Delta u$, respectively: we find sensitivities of $\pm 30\%$ for $T_e$ and 
$\pm 15\%$ for $\bar{u}_{\rm in}$. Fitting $T_i$ and $\Delta u$ is more subtle because both 
quantities have a similar effect on the shape of the spectrum (they affect the width of the spectral peaks). We therefore instead choose to 
fit an effective ion temperature, $T_{i,\mathrm{eff}}$ including both thermal broadening and one due to small-scale stochastic motion. 
The sensitivity of 
the fit for $T_{i,\mathrm{eff}}$ is found to be $\pm 40\%$. To determine independently $T_i$ and $\Delta u$, we then use the fact that motions are subsonic and thus estimate $\Delta u \approx c_{\mathrm{s}} \Delta 
n_e/\bar{n}_e$, where $c_{\mathrm{s}}$ is the sound speed; this leaves $T_i$ as the only free parameter determining 
$T_{i,\mathrm{eff}}$ in the fits. It is in general true that the sensitivity of a fit can be underestimated  
systematically using our chosen methodology -- that is, fixing all physical parameters for the fits save one, and then 
varying the chosen parameter to determine the sensitivity. However, on account of 
couplings between parameters, such considerations do not apply to the IAW fits. This is because  
the parameters $\Delta u$, $T_e$, and $T_{i,\mathrm{eff}}$ each only influence 
one characteristic of the fit: the average peak position, peak separation and peak width, 
respectively. 

\begin{figure}
  \centering
\includegraphics[width=0.95\linewidth]{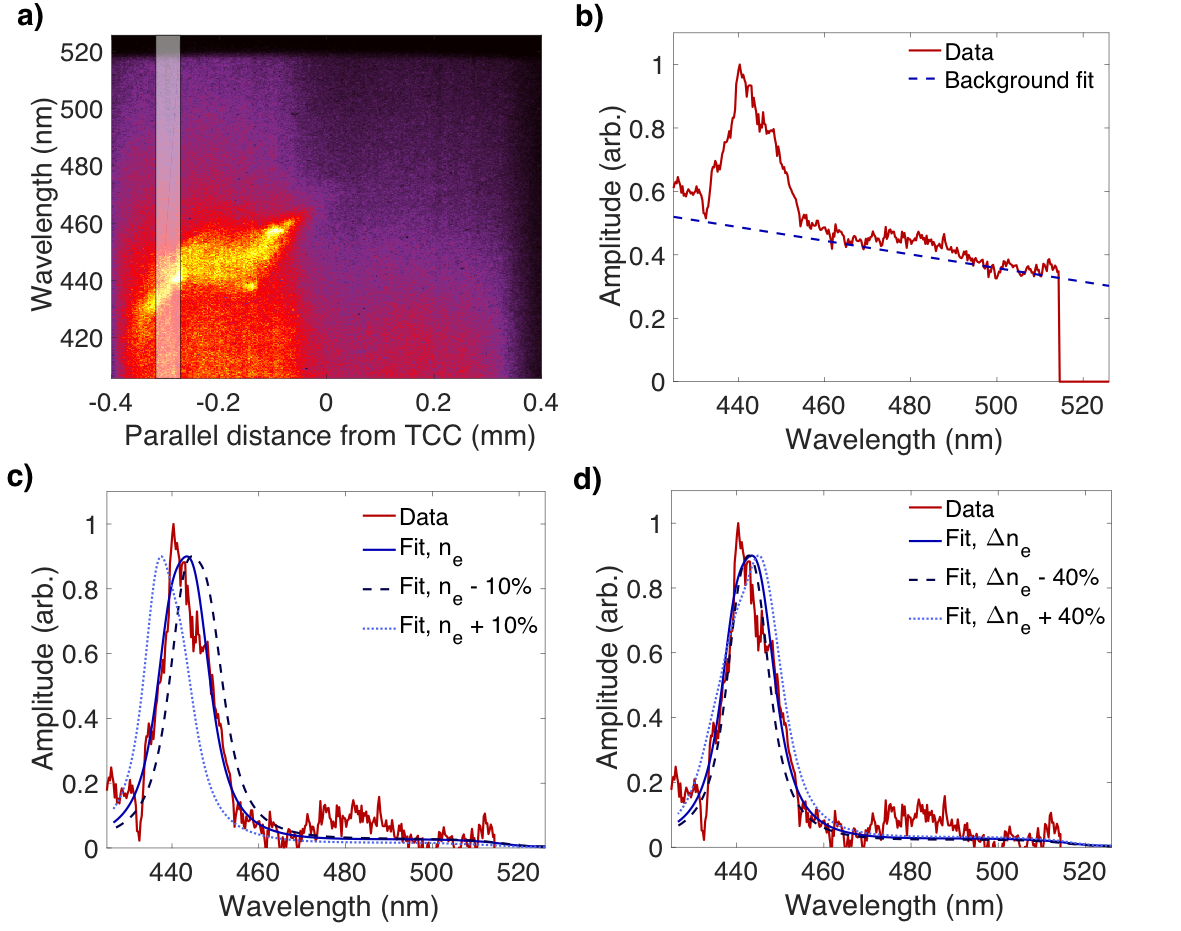}
\caption[Fitting the EPW feature]{\textbf{Fitting the EPW feature.} \textbf{a)} EPW spectrum from 27.0 ns after drive-beam pulse initiation. In (b-d), the 
experimental 1D spectrum (in red) is calculated by averaging the full spectrum horizontally over the region denoted by the translucent rectangle. 
\textbf{b)} The linear background fit to the raw experimental spectrum. 
\textbf{c)} Background-subtracted experimental spectrum: the best-fit spectrum ($T_e = 470 \, \mathrm{eV}$, $T_{i,\mathrm{eff}} = 620 \, \mathrm{eV}$, $\bar{u}_{\rm in} = 230 \, \mathrm{km} \; \mathrm{s}^{-1}$, $\bar{n}_e = 1.2 \times 10^{20} \, \mathrm{cm}^{-3}$, $\Delta n_e = 0.35 \times 10^{20} \, \mathrm{cm}^{-3}$) in solid blue, and the
fits parametrized by $\bar{n}_e$ sensitivity bounds (-10\% dashed, +10\% dotted). All fits are convolved with the (experimentally measured) instrument function, which has a characteristic width of 0.025 nm. 
\textbf{d)} Same as b) but fits parametrized by $\Delta n_{e}$ sensitivity bounds are plotted.} 
\label{fig:TS_EPWfitsexample}
\end{figure}

 Figure~\ref{fig:TS_EPWfitsexample}a shows the EPW feature for the same 
experimental shot. 
We again determine the experimental EPW spectrum at a given position by averaging over a $100 \, \mu \mathrm{m}$ 
interval. Before attempting to fit the EPW spectrum, we must first correct for the CCD/grating response, as well as subtract the 
background signal. The latter is more significant for the EPW spectrum than for the IAW spectrum -- on account of 
the former's smaller magnitude -- and is most likely to be associated with 
radiative emission from the plasma. We find that the background signal for our data is 
well characterized by a linear fit (see Figure \ref{fig:TS_EPWfitsexample}b). We then fit the experimental spectrum by varying $\bar{n}_e$ and $\Delta n_e$, before determining the sensitivity of the fits to 
variations in these quantities: we find $\pm 10\%$ for $\bar{n}_e$ (Figure \ref{fig:TS_EPWfitsexample}c) and $\pm 40\%$ 
for $\Delta n_e$ (Figure \ref{fig:TS_EPWfitsexample}d). As with the IAW fits, $\bar{n}_e$ 
and $\Delta n_e$ affect different characteristics of the EPW fit, so our 
methodology for assessing the fit sensitivity is appropriate. However, we note that the peak position of the EPW feature 
can be weakly sensitive to the assumed electron temperature as well as to $\bar{n}_e$; since the electron temperature
is held constant when fitting the EPW feature, using the value determined from the IAW feature, the quoted $\pm 10\%$ sensitivity 
for $\bar{n}_e$ could in practice be a slight underestimate. Nonetheless, we do not believe it worthwhile carrying out a multivariate assessment of the fit's sensitivity, 
because the ratio of the change in peak position arising from an order-unity change in the electron temperature to the change in peak position due to an order-unity change in the mean electron density is anticipated theoretically to be $\mathcal{O}(\alpha^{-2}) \ll 1$ (we remind the reader that $\alpha \equiv 1/k \lambda_{\rm{D}}$). Since the uncertainty in $T_e$ from the IAW fit is $\pm 30\%$, we conclude that corrections
to the sensitivity in $\bar{n}_e$ due to the uncertainty in $T_e$ will be no more than a few percent.

\subsection*{Perturbative heating effects of Thomson-scattering diagnostic}

Whilst analyzing the results of our experiment, we discovered that our Thomson-scattering diagnostic was not always non-invasive, particularly at late times. Indirect evidence of this observation is most easily obtained from our X-ray imaging diagnostic. As mentioned in Materials and Methods, for a given experimental shot, two X-ray imaging times were recorded; the earlier time was always chosen to fall before the application of the other diagnostics, and the later time after. This allowed us to assess whether invasive effects -- in particular, additional heating -- associated with our experimental diagnostics were significant.

\begin{figure}
  \centering
\includegraphics[width=0.95\linewidth]{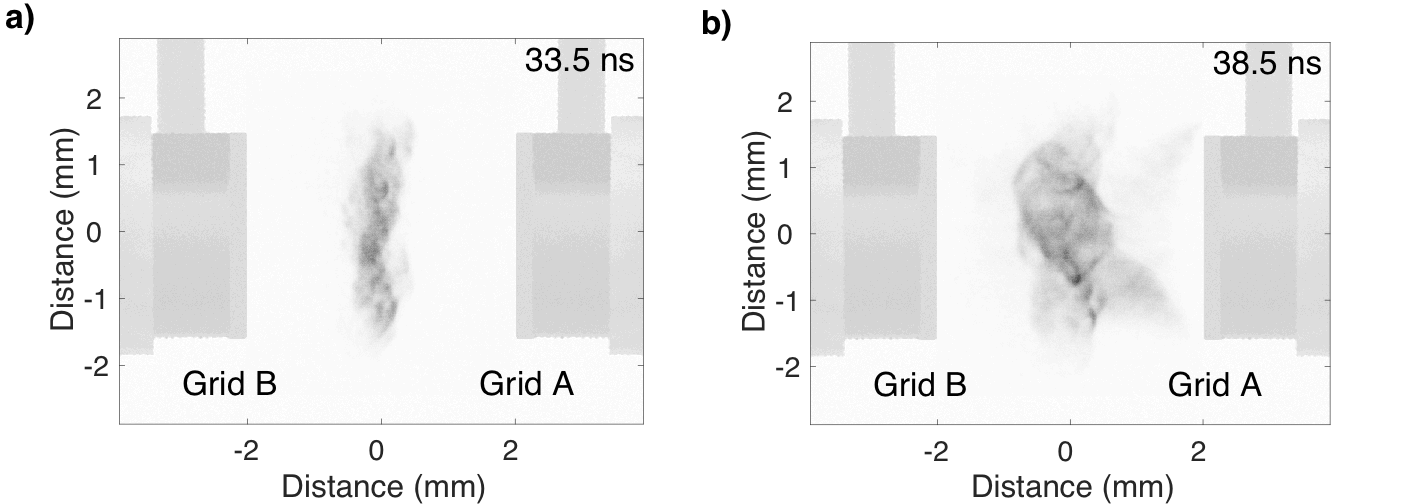}
\caption{\textbf{Imaging self-emitted X-rays for each experimental shot}. Characteristic output images from X-ray framing 
camera diagnostic in a sample experimental shot. In both images, the color map is linear and adjusted so that the maximum displayed pixel value is 1,050 
(corresponding to the maximum pixel count of all our experimental 
images). The images shown have been cropped and are both 870 $\times$ 641 pixels. To aid interpretation of the images,
an outline of the target has been superimposed onto both images. The respective designs of grid A and B are shown in Figure 1 of the main text, bottom right. 
\textbf{a)} The X-ray image recorded with the first MCP strip 33.5 ns after the drive-beams are initiated. On this experimental shot, the Thomson-scattering probe beam was initiated 
1 ns after the first MCP strip was gated and the proton backlighter capsule irradiated 2.5 ns later; thus, the self-emitted X-rays detected in the image
are purely due to the interaction-region plasma's dynamics. \textbf{b)} The X-ray image recorded with the second MCP strip 38.5 ns after the drive beams are initiated. This image was recorded after both the Thomson-scattering probe beam and the proton backlighter were used.}  
\label{fig:Xrays_overview}
\end{figure}

\begin{figure}
  \centering
\includegraphics[width=0.9\linewidth]{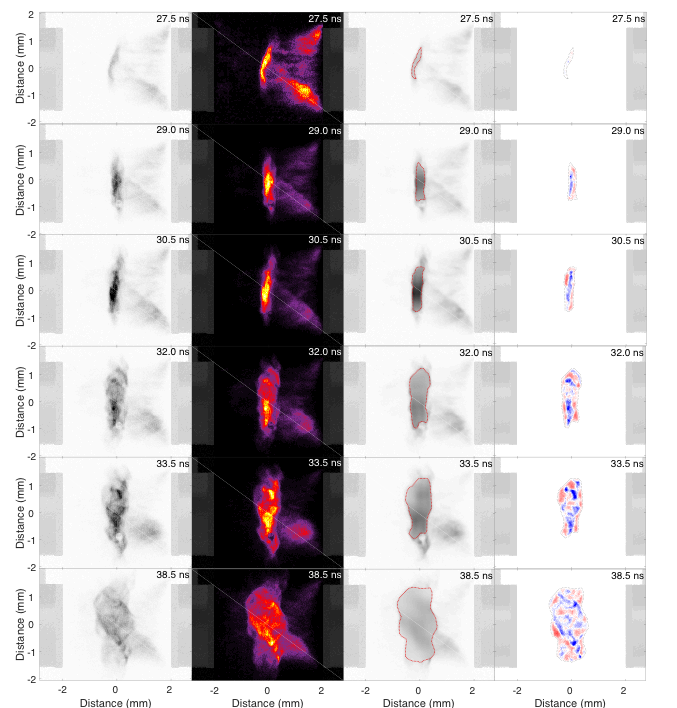}
\caption[The interaction-region plasma's evolution in the presence of invasive diagnostic heating]{\textbf{The interaction-region plasma's evolution in the presence of perturbative diagnostic heating.} Perturbed self-emission X-ray images of the interaction-region plasma.
The parameters of each column are identical to those of Figure 2 in the main text. We note that the CCD camera used to record 
this sequence of images was damaged in a small localized region slightly below the center of each image; we therefore chose our window functions so that this region was excluded from subsequent analysis.} 
\label{fig:Xrays3}
\end{figure}

Two sample X-ray images arising from one particular experimental shot timed according to these arrangements are shown in Figure
\ref{fig:Xrays_overview}. 
 Comparing Figure
\ref{fig:Xrays_overview}a and \ref{fig:Xrays_overview}b, we see that, in 
addition to the dynamical evolution of the interaction region, increased 
emission is observed in the later image from plasma between grid A and the central interaction-region plasma. This effect
is most likely due to the aforementioned heating; the diagonal feature corresponds 
to the projected trajectory of the Thomson-scattering beam and only the plasma 
between grid A and the target center is visible to the proton backlighter 
capsule. We henceforth refer to the earlier-time images as `unperturbed'
self-emission X-ray images and the later ones as `perturbed' images. All X-ray images presented in the main text are unperturbed images. For reference, the perturbed images corresponding to the unperturbed images shown in Figure 3 of the main text are presented in Figure \ref{fig:Xrays3}.
%Aside from the previously discussed characteristic differences between unperturbed and perturbed images, 
%the qualitative evolution of the perturbed images over time is broadly similar.

From our measurements of bulk plasma parameters, the claim that the Thomson-scattering probe beam heats the plasma can be justified as follows.
%explain the perturbative effects observed in the X-ray images theoretically via the following simple argument, which suggests that the Thomson scattering probe beam is predominantly responsible. More specifically, 
The fraction $f_a$ of the probe beam's energy absorbed by the plasma can be estimated as $f_a \approx 1 - \exp{(-\kappa_a l_{\mathrm{beam}})}$, where $\kappa_a$ is the 
the absorption coefficient of the beam in the plasma and $l_{\mathrm{beam}}$ 
the path length of the beam through the interaction-region plasma~\citep{CL14}. The dominant 
absorption process affecting the beam is inverse bremsstrahlung, so 
\begin{eqnarray}
\kappa_a & = & 3.6 \left[\frac{Z_{\mathrm{eff}}}{5.3}\right] \left[\frac{\log \Lambda}{7}\right] \left[\frac{n_e(\mathrm{cm}^{-3})}{1.5 \times 10^{20} \, \mathrm{cm}^{-3}} \right]^{2} \left[\frac{T_e(\mathrm{eV})}{330 \, \mathrm{eV}} \right]^{-3/2}
\left[\frac{\omega_0(\mathrm{s}^{-1})}{3.6 \times 10^{15} \, \mathrm{s}^{-1}}\right]^{-2} \left[\frac{\epsilon_{\rm D}}{0.97}\right]^{-1/2} \, \mathrm{cm}^{-1} , 
\end{eqnarray}
where $\epsilon_{\rm D} \equiv 1- \omega_{\mathrm{p}e}^2/\omega_0^2$ is the 
dielectric permittivity of the plasma. Estimating $l_{\mathrm{beam}} \approx 0.13 \, \mathrm{cm}$
from the X-ray measurements of the interaction-region width at 38.5 ns, we conclude that $\kappa_a l_{\mathrm{beam}} \approx 
0.45$ and thus $f_a \approx 0.36$. Therefore, the total energy deposited in the Thomson-scattering volume (neglecting conduction) is $E_{\mathrm{heat}} 
\approx 1.1  \times 10^8 \, \mathrm{erg}$. Noting that the internal-energy density of the plasma is 
$\varepsilon_{\rm th} \approx 3 n_e T_e \left(1+ \langle Z \rangle^{-1} \right)/2$ (assuming $T_i \approx T_e$), it follows 
that the total thermal energy $E_{\mathrm{th}}$ present in the Thomson-scattering volume is 
\begin{equation}
E_{\mathrm{th}} = 1.1 \times 10^8  \left[\frac{\langle Z \rangle +1}{4.5}\right] \left[\frac{n_e(\mathrm{cm}^{-3})}{1.5 \times 10^{20} \, \mathrm{cm}^{-3}} \right] \left[\frac{T_e(\mathrm{eV})}{330 \, \mathrm{eV}} \right] \left[\frac{\mathcal{V}_{\rm T}(\mathrm{cm}^{3})}{7.5 \times 10^{-4} \, \mathrm{cm}^{3}} \right]
\, \mathrm{erg} \, ,
\end{equation}
where $\mathcal{V}_{\rm T}$ is the volume of the Thomson-scattering collection region. 
We conclude that at the given density, $E_{\mathrm{heat}} \gtrsim E_{\mathrm{th}}$ 
for electron temperatures $T_e \lesssim 330 \, \mathrm{eV}$. It follows that the electron temperature at later times could be significantly lower than 
the value measured by the Thomson-scattering beam in the absence of the probe-beam heating.

\begin{figure}
  \centering
\includegraphics[width=0.85\linewidth]{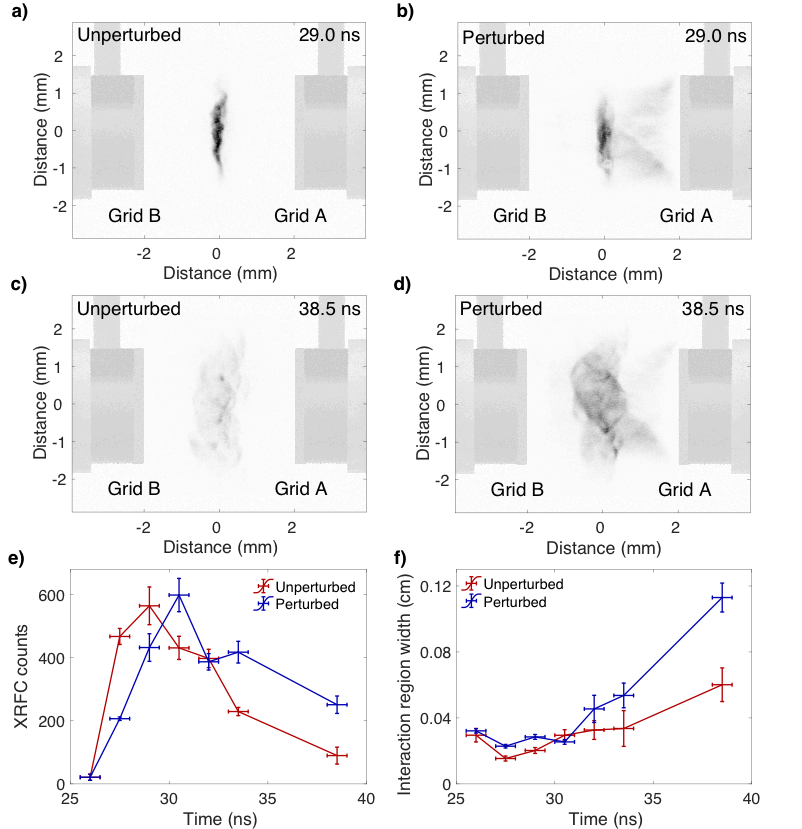}
\caption[Perturbed versus unperturbed self-emission X-ray images]{\textbf{Perturbed versus unperturbed self-emission X-ray images.} \textbf{a)} Self-emission X-ray image at 29.0 ns after the initiation of the drive-beam pulse, in the absence of perturbative diagnostics. 
The color map of the image is the same as Figure \ref{fig:Xrays_overview}. \textbf{b)} Same as a), except in the presence of perturbative diagnostics. \textbf{c)} Unperturbed X-ray image at 38.5 ns. \textbf{d)} Perturbed X-ray image at 38.5 ns. \textbf{e)} 
Evolution of mean maximum value of one-dimensional mean emission profiles over time for unperturbed (red) and perturbed (blue) X-ray images. The error in each measurement is the error 
arising from the three distinct lineouts calculated from different perpendicular positions in the interaction-region plasma -- see Figure 4a). \textbf{f)} Mean-interaction width over time for the unperturbed and perturbed X-ray images. The width and errors are calculated in the same manner as described in the caption of Figure 4 of the main text.} 
\label{fig:Xrays4}
\end{figure}

We now consider the effects of this diagnostic heating on our results. Figures~\ref{fig:Xrays4}a and \ref{fig:Xrays4}b compare the X-ray self-emission from the plasma $\sim 3 \, \mathrm{ns}$ 
after the jet collision (which is at $\sim 26 \, \mathrm{ns}$) in the absence and presence, respectively, of probe-beam heating; the equivalent comparison $\sim 12.5 \, \mathrm{ns}$ after the jet collision is given by Figures \ref{fig:Xrays4}c and 
\ref{fig:Xrays4}d. 
Qualitatively, we note that emission from the interaction-region plasma itself in the former case is not 
obviously different in the perturbed and unperturbed images: the absolute value of the emission is similar, as is the 
morphology of the region. However, at later times, emission from the perturbed X-ray images 
is noticeably higher. More quantitatively, we use the maximum pixel values of the one-dimensional mean emission 
profiles used to calculate the interaction region width (cf. Figure 4 of the main text) to compare the relative 
emission levels associated with the unperturbed and perturbed cases, 
respectively (Figure \ref{fig:Xrays4}e). Somewhat unexpectedly, we find that, 
immediately after the interaction-region formation, the emission from the unperturbed cases is greater. 
This trend is most likely explained by thermal expansion of the interaction-region plasma induced by 
the additional diagnostic heating: although higher
temperatures would result in slightly increased emission, this trend would be 
counteracted by lower mean densities in the expanded interaction region, particularly since the measured X-ray intensity
is much more strongly dependent on density than on
temperature for $T_e \sim 300-500 \, \mathrm{eV}$. 
This argument is given some weight by the observation that the 
measured interaction-region width in the perturbed images is slightly larger (Figure \ref{fig:Xrays4}f) 
immediately after collision. At later times, Figure \ref{fig:Xrays4}e illustrates the (expected) trend of significantly 
reduced X-ray intensity in the unperturbed images compared to the perturbed ones. 
The most likely explanation of this observation is a lower temperature of the interaction-region plasma  
in the unperturbed case ($T_e \sim 50 - 100$ eV).

We can, in fact, use the observed changes in self-emitted X-ray intensity (combined with the Thomson-scattering measurements) to derive a more precise estimate of the amount of heating. We recall that the intensity $I$ recorded by the CCD camera depends on temperature: $I  \propto \int \mathrm{d}s \, n_e^2 \hat{f}(T_e)$, where $\hat{f}(T_e)$ is defined by Eq. [2] of the main text and plotted in Figure \ref{fig:Xrays_cameraresponse}b. 
Thus, if the electron number density of the interaction-region 
plasma is known at two given times, any difference in total intensity evident in an X-ray image 
can be attributed to distinct temperatures. Applying this logic to unperturbed 
and perturbed X-ray images recorded at the same point in time (in different experimental 
shots) and assuming that at later times probe-beam heating has a negligible effect on 
plasma density, we can use the differences in intensity calculated in Figure
\ref{fig:Xrays4}f to estimate the unperturbed electron temperature $(T_e)_{\mathrm{up}}$. 
Namely, it follows that $(T_e)_{\mathrm{up}} \approx (T_e)_{\mathrm{p}}\hat{f}^{-1}\left({I_{\mathrm{up}}}/{I_{\mathrm{p}}}\right) \approx 75 \, \mathrm{eV}$,
where $(T_e)_{\mathrm{p}}$ is the perturbed electron temperature, $I_{\mathrm{p}}$ 
the mean X-ray intensity in the unperturbed image, and $I_{\mathrm{up}}$ the 
mean X-ray intensity in the perturbed image. This inferred bound is 
significantly below the temperature measured by the Thomson-scattering 
diagnostic. 

We confirm the importance of the Thomson-scattering probe beam heating using the FLASH simulations -- see `Overview of FLASH simulations of experiment'. 

\section*{Additional information on the proton-imaging diagnostic}

\subsection*{Parameterized one-dimensional model for the path-integrated magnetic field arising from the double-cocoon configuration}

In the main text, we employ a double-cocoon configuration to describe the magnetic field at the time of interaction-region formation; we further claim that if the parallel scale of the cocoons is much smaller than their perpendicular scale, then the path-integrated field experienced by imaging protons is quasi-one-dimensional, and is oriented in the (unique) direction that is perpendicular both to the imaging protons' direction and the axis of symmetry of the cocoons. We validate this claim here. 

The double-cocoon configuration is constructed from the simple model of a single cocoon provided by~\citep{K12}, in which the magnetic field $\boldsymbol{B}$ is given by
\begin{equation}
\boldsymbol{B} = \sqrt{2} B_{\rm max} \frac{r}{a} e^{\frac{a^2 -2 r^2}{2a^2}-\frac{z^2}{b^2}} \boldsymbol{e}_\phi \, 
,
\end{equation}
where $(r,\phi,z)$ is a cylindrical coordinate system with symmetry axis $z$, $B_{\rm max}$ is the maximum magnetic-field strength, $a$ the characteristic 
perpendicular size of the cocoon, $b$ the characteristic parallel size of 
the cocoon, and $\boldsymbol{e}_\phi$ the azimuthal unit vector. It is shown in \citep{K12} that the path-integrated magnetic field associated with such a structure when viewed 
at an angle $\theta$ with respect to the $z$ axis is
\begin{eqnarray}
  \int \mathrm{d}s \, B_1 & = & \frac{B_{\rm max} a^2 b \sqrt{2\pi} \cos \theta}{d^2}  \frac{u_2}{d} e^{\frac{1-2u_1^2}{2a^2}-\frac{u_2^2}{d^2}}  \, ,  \label{full_path_intfield_A} \\
  \int \mathrm{d}s \, B_2 & = & -\frac{B_{\rm max} a b \sqrt{2\pi} \cos \theta}{d} \frac{u_1}{a} e^{\frac{1-2 u_1^2}{2 a^2}-\frac{u_2^2}{d^2}}  \, ,  
  \label{full_path_intfield_B}
\end{eqnarray}
where 
\begin{equation}
 d \equiv \sqrt{a^2 \cos^2 \theta + b^2 \sin^2 \theta} \, .
\end{equation}
Here the two-dimensional Cartesian coordinate system $(u_1,u_2)$ is chosen to be perpendicular to the viewing 
direction, with basis vectors satisfying $\hat{\boldsymbol{e}}_1 \cdot \hat{\boldsymbol{z}} = 
0$, $\hat{\boldsymbol{e}}_2 \cdot \hat{\boldsymbol{z}} \neq 0$, $B_1 \equiv \boldsymbol{B} \cdot \hat{\boldsymbol{e}}_1$ and $B_2 \equiv \boldsymbol{B} \cdot \hat{\boldsymbol{e}}_2$. If we assume that the perpendicular extent of the cocoon 
structures is much greater than the parallel extent, viz., $a \gg b$, it follows that
for angles $\theta$ such that $\pi/2 - \theta \gg b/a$, $d \approx a \cos 
\theta$. For our experiment, $\theta \approx 
55^{\circ}$, so $d \approx a/\sqrt{3}$. Under these assumptions, Eqs. [\ref{full_path_intfield_A}] and [\ref{full_path_intfield_B}] 
become
\begin{eqnarray}
  \int \mathrm{d}s \, B_1 & \approx & 3 B_{\rm max} b \sqrt{2 \pi} \frac{u_2}{a} e^{\frac{1-2u_1^2}{2a^2}-3 \frac{u_2^2}{a^2}}  \, , \\
  \int \mathrm{d}s \, B_2 & \approx & - B_{\rm max}  b \sqrt{2 \pi} \frac{u_1}{a} e^{\frac{1-2u_1^2}{2a^2}-3 \frac{u_2^2}{a^2}}  \, .  
  \label{full_path_intfield_simp}
\end{eqnarray}
We conclude that in such a model, the path-integrated field is indeed elongated 
in the $u_1$ direction (which by definition is precisely the direction perpendicular to the 
projected line of centers), and, for $u_1 \lesssim u_2$, the path-integrated perpendicular magnetic field is 
also predominantly in the $u_1$ direction. 

For the double-cocoon configuration, we therefore obtain
\begin{eqnarray}
\int \mathrm{d}s \, \boldsymbol{B} & \approx & 3 \sqrt{2 \pi} \mathrm{e}^{\frac{u_1^2-2a^2}{2 a^2}} \Bigg[b B_{\rm max}^{+} \frac{u_2+\tilde{\ell}_c}{a} \mathrm{e}^{-\frac{3 (u_2+\tilde{\ell}_c)^2}{a^2}} + b B_{\rm max}^{-} \frac{u_2-\tilde{\ell}_c}{a} \mathrm{e}^{-\frac{3 (u_2-\tilde{\ell}_c)^2}{a^2}} \Bigg] \hat{\boldsymbol{e}}_1 \, ,
\label{pathintdoublecocoon}
\end{eqnarray} 
where $\tilde{\ell}_c \approx \sqrt{2} \ell_c/\sqrt{3}$. Eq. [\ref{pathintdoublecocoon}] is the quasi-one-dimensional model for the path-integrated magnetic field employed in the main text. 

\subsection*{Validating the double-cocoon configuration}

\begin{figure}
  \centering
\includegraphics[width=\linewidth]{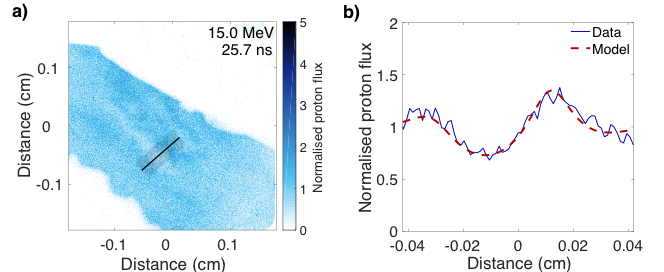}
\caption[Validating double-cocoon model for magnetic field at the time of interaction-region formation]{\textbf{Validating the double-cocoon model for the magnetic field at the time of interaction-region formation.} 
\textbf{a)} The 15.0 MeV proton image from which Figure 6a of the main text is calculated, portrayed with the translucent rectangular region.
\textbf{b)} One-dimensional lineout of the normalized proton flux (blue, solid line) calculated by averaging translucent rectangular region 
denoted in a) across its width, along with one-dimensional prediction from model [4] of the main text using the optimized parameters given in the caption of Figure 6.}
\label{fig:protoninitfieldsB}
\end{figure}

As a further check of the accuracy of our double-cocoon model, we 
calculate the predicted one-dimensional profile of proton flux associated with this model, and compare the result to a one-dimensional lineout calculated directly from the relevant proton image in the same region as was used to determine the experimental path-integrated-field profile (see Figures \ref{fig:protoninitfieldsB}a and
\ref{fig:protoninitfieldsB}b). We find that the predicted profile is a 
close match to the experimental one. 

\subsection*{Heuristic estimate of magnetic-field strength at the time of interaction-region formation}

An alternative, simple way of estimating the average strength of the
magnetic field in the interaction region is to calculate the mean value of the path-integrated perpendicular magnetic field and divide it by the path length $l_p$ of 
the protons through the interaction region. We do this for our data by 
first evaluating $\langle|\int \mathrm{d}^2 \boldsymbol{x} \, \boldsymbol{B}_{\perp}'|\rangle$
in three square regions (see Figure 7a of the main text and the caption for their precise 
dimensions), and determining the mean value (and errors) across the three regions. 
We then determine $l_p$ as described in the main text. Finally, we estimate $\langle B \rangle $ via $\langle B \rangle \approx \langle|\int \mathrm{d}^2 \boldsymbol{x} \, \boldsymbol{B}_{\perp}'|\rangle/l_{p}$. 
We obtain $\langle B \rangle  \approx 6 \, \mathrm{kG}$, in agreement with the parameterized double-cocoon configuration.  

However, such an estimate implicitly makes a number of assumptions about the nature of 
the underlying structure of the (non-stochastic) magnetic field. First, 
components of the magnetic field parallel to the path of the proton-imaging beam
are assumed to be negligible compared to perpendicular components. Secondly, the estimate 
presupposes that the path-integrated field does not change sign along the path.  We can, however, assess the validity of these two assumptions using our parameterized double-cocoon model of the magnetic field at the time of interaction-region formation. Inside the rectangular region 
to which we have applied our model, the magnetic field is predominantly in the $x$ direction, so the component parallel to the protons' path is indeed small. Due
to the $55^{\circ}$ angle of imaging, the $B_x$ components of the opposing cocoon structures add 
constructively rather than destructively at all positions. Thus, we conclude 
that the stated assumptions are reasonable at least for the strongest path-integrated structure.

\subsection*{Testing statistical assumptions about stochastic magnetic fields in our experiment}

\begin{figure}
  \centering
\includegraphics[width=\linewidth]{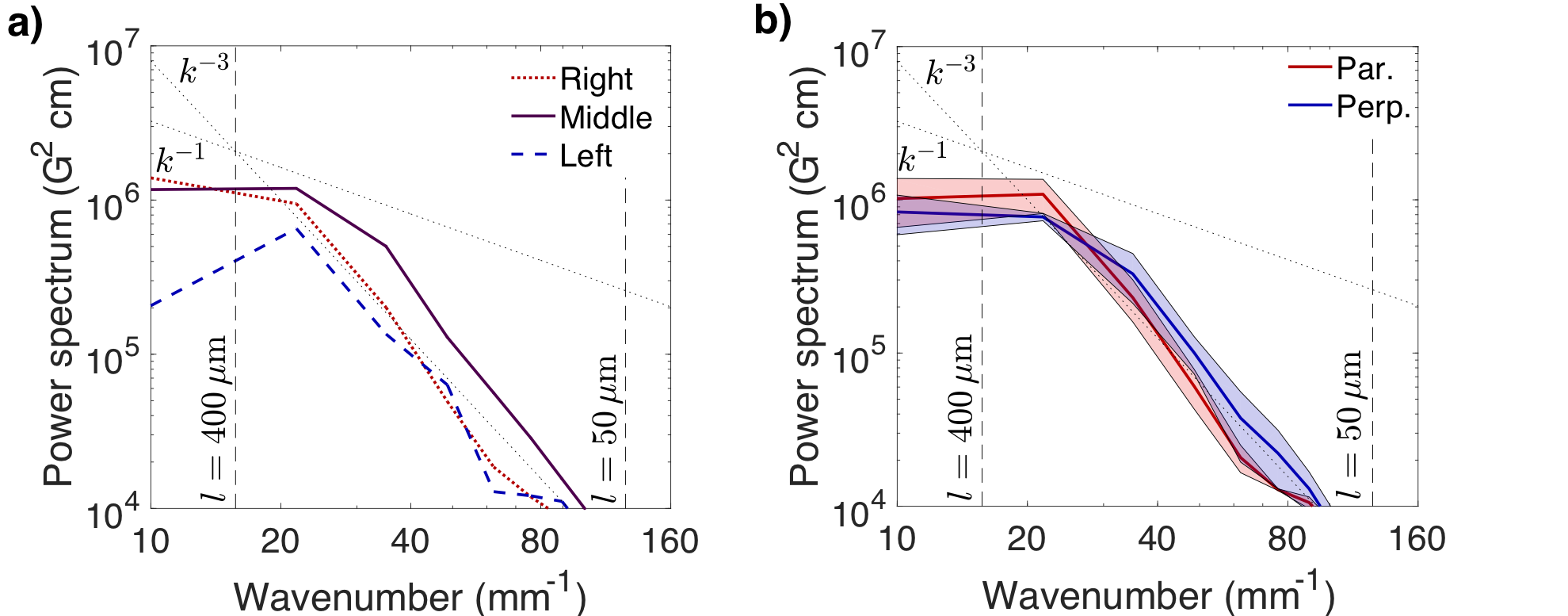}
\caption[Validating assumed statistical symmetries of stochastic magnetic fields subsequent to interaction-region formation]{\textbf{Validating assumed statistical symmetries of stochastic magnetic fields subsequent to interaction-region formation.} 
\textbf{a)} Test of statistical homogeneity of the stochastic magnetic field. 
The magnetic-energy spectra associated with the left, middle and right square regions shown in Figure 7 of the main text (which are calculated using Eq. [7]) are plotted, along with the nominal resolution of the proton-imaging diagnostic and the grid scale. The slope $k^{-3}$ is plotted for reference, 
but does not have physical significance. \textbf{b)} 
Test of statistical isotropy of the stochastic magnetic field. For each square region, the magnetic-energy 
spectrum associated with wavenumbers predominantly parallel to the line of centers
is calculated by determining 
the spectral slope from magnetic-field Fourier components with $k_{\|} > k_{\perp}$, and 
vice versa for the magnetic-energy spectrum associated with predominantly
perpendicular wavenumbers. Mean parallel and perpendicular spectra are 
subsequently obtained by averaging the results from each region (the errors are determined 
similarly).}
\label{fig:protonimaging_stocmagfields}
\end{figure}

In the main text, we claim that the stochastic magnetic fields produced after the interaction-region formation can be assumed to be statistically homogeneous and isotropic within the plasma. Here we test that assumption. Figure~\ref{fig:protonimaging_stocmagfields}a
shows the magnetic-energy spectra calculated using Eq. [7] of the main text for the three chosen regions 
in the case of the path-integrated fields calculated at 27.2 ns since the initiation of the drive beams.
Except at the largest scales, we find that the three spectra match closely,
supporting the homogeneity assumption. The isotropy assumption is tested in Figure \ref{fig:protonimaging_stocmagfields}b. Given the symmetry of our experiment, any anisotropy would 
manifest itself with respect to the line of centers. So, in each box, we calculate the 
magnetic-energy spectra both for wavenumbers predominantly parallel to the projected 
line of centers, and for those predominately perpendicular to it. The results are then combined to obtain averaged parallel and perpendicular spectra. 
We find that the two spectra are the same within the uncertainty of the measurement.

\section*{Plasma characterization}

\subsection*{Plasma parameter tables}

In the main text, we quote some calculated parameters in the plasma, which are of particular theoretical interest. These values (and the bulk plasma parameters from which they are calculated) are shown in Tables \ref{table:TableTheoParam} and \ref{TableParam}, respectively.
\begin{table}[p]
\centering
\caption[Summary of measured plasma parameters related to the experiment at $\approx27 \, \mathrm{ns}$ after the drive-beam laser pulse is initiated]{Summary of measured plasma parameters related to the experiment at $\approx27 \, \mathrm{ns}$ after the drive-beam laser pulse is initiated. The effective ion charge (which appears in various physical parameters)
is given by $Z_{\mathrm{eff}} = (Z_\mathrm{C}^2+Z_\mathrm{H}^2)/(Z_\mathrm{C}+Z_\mathrm{H})$.} 
{\renewcommand{\arraystretch}{1.1}
\renewcommand{\tabcolsep}{0.18cm}
\begin{tabular}[c]{c|c}
Quantity & Value \\
\hline
Carbon/hydrogen masses ($M_\mathrm{C}$,$M_\mathrm{H}$) & 12, 1 \\
Average atomic weight ($\langle M \rangle$) &\textcolor{black}{$6.5$} \\
Carbon/hydrogen charges ($Z_\mathrm{C}$,$Z_\mathrm{H}$) & 6, 1 \\
Mean ion charge ($\langle Z \rangle $)  & \textcolor{black}{$3.5$} \\
Effective ion charge ($Z_{\mathrm{eff}}$)  & \textcolor{black}{$5.3$} \\
Electron temperature ($T_{e}$) & $ 460$ eV \\
Ion temperature ($T_i$) & $ 580$ eV \\
Electron number density ($n_{e}$) & $1.2 \times 10^{20}$ cm$^{-3}$ \\
Carbon number density ($n_{\mathrm{C}}$) & $1.7 \times 10^{19}$ cm$^{-3}$ \\
Hydrogen number density ($n_{\mathrm{H}}$) & $1.7 \times 10^{19}$ cm$^{-3}$ \\
Bulk velocity ($\bar{u}_{\rm in}$) & 1.9 $\times 10^7$ cm $\mathrm{s}^{-1}$ \\
Turbulent velocity ($u_{\rm rms}$) & 1.1 $\times 10^{7}$ cm $\mathrm{s}^{-1}$  \\
Outer scale ($L$) & 0.04 cm \\
RMS magnetic field ($B$) & 90 kG \\
Maximum magnetic field ($B$) & 250 kG \\
Adiabatic index ($\gamma_{I}$) & 5/3 \\
\end{tabular}}
\label{TableParam}
\end{table}
\begin{table}[htbp]
\centering
\caption{Summary of relevant theoretical plasma parameters for the experiment at $\approx27 \, \mathrm{ns}$ after the drive-beam laser pulse is initiated. 
The unit system used for all these physical quantities in the above formulas is again Gaussian CGS, except for the temperature, which is expressed in eV. We assume the conventions given
in~\citep{SCTM04} for the resistive and viscous dissipation scales; the pre-factor is included in order to match with the dissipation scales typically found in simulations of the fluctuation dynamo. }
{\renewcommand{\arraystretch}{1.12}
\renewcommand{\tabcolsep}{0.18cm}
\begin{tabular}[c]{c|c|c}
Quantity & Formula & Value \\
\hline
Coulomb logarithm $(\log{\Lambda})$ & $23.5 - \log{n_{e}^{1/2} T_{e}^{-5/4}}-\sqrt{10^{-5}+{\left(\log{T_{e}}-2\right)^2}/{16}}$ & $\sim 7$ \\
Mass density $(\rho)$ & $1.7 \times 10^{-24} \left(M_\mathrm{C} n_\mathrm{C} + M_\mathrm{H} n_\mathrm{H}\right)$ & $3.7 \times 10^{-4}$ g cm$^{-3}$ \\
Debye Length ($\lambda_{\mathrm{D}}$) & $ 7.4 \times 10^2 \,{T_e^{1/2}}\left[1+ {T_e}Z_{\mathrm{eff}}/{T_i}\right]^{-1/2} {n_{\mathrm{e}}^{-1/2}} $ & \textcolor{black}{$6.3 \times 10^{-7}$ cm} \\
Sound speed ($c_{\mathrm{s}}$) & $ 9.8 \times 10^5 \, {\left[(\langle Z \rangle +1) \gamma_I T_{e}\right]^{1/2}}{\langle M \rangle^{-1/2}}$  & \textcolor{black}{$2.3 \times 10^{7}$ cm s$^{-1}$} \\
Mach number & $u_{\rm rms}/c_{\mathrm{s}}$  & $0.5$ \\
Plasma $\beta$ & $4.0 \times 10^{-11} \, \left[{n_e T_{e} + \left( n_\mathrm{C}+  n_\mathrm{H}\right) T_i}\right]/{B^2} $ & $370$\\
Carbon-carbon mean free path ($\lambda_{\mathrm{CC}}$) & $2.9 \times 10^{13} \, {T_{i}^2}/{Z_{\rm C}^4 n_{\mathrm{C}} \log{\Lambda}}$ & \textcolor{black}{$6.2 \times 10^{-5}$ cm}\\
Hydrogen-carbon mean free path ($\lambda_{\mathrm{HC}}$) & $2.1 \times 10^{13} \, {T_{i}^2}/{Z_{\rm H}^2 Z_{\rm C}^2 n_{\mathrm{C}} \log{\Lambda}}$ & \textcolor{black}{$1.6 \times 10^{-3}$ cm}\\
Electron-ion mean free path ($\lambda_{\mathrm{e}}$) & $2.1 \times 10^{13} \, {T_{e}^2}/{Z_{\mathrm{eff}} n_{\mathrm{e}} \log{\Lambda}}$ & \textcolor{black}{$1.0 \times 10^{-3}$ cm}\\
Carbon-electron equilibration time ($\tau_{\mathrm{C}e}^{\epsilon}$) & $3.2 \times 10^8  \, {M_C T_{e}^{3/2}}/{Z_{\rm C}^2 n_{\mathrm{e}}  \log{\Lambda} }$ & \textcolor{black}{$1.4 \times 10^{-9}$ s} \\
Electron Larmor radius ($\rho_{\mathrm{e}}$) &\textcolor{black}{ $2.4  \, {T_{e}^{1/2}}/{B}$ }& \textcolor{black}{$5.7 \times 10^{-4}$ cm}\\
Carbon Larmor radius ($\rho_{\mathrm{C}}$) & \textcolor{black}{ $1.0  \times 10^2 \, {M_\mathrm{C}^{1/2} T_{i}^{1/2}}/{Z_\mathrm{C} B}$} & \textcolor{black}{$1.5 \times 10^{-2}$ cm} \\
Hydrogen Larmor radius ($\rho_{\mathrm{H}}$) & \textcolor{black}{ $1.0  \times 10^2 \, {M_{\rm H}^{1/2} T_{i}^{1/2}}/{Z_\mathrm{H} B}$} & \textcolor{black}{$2.6 \times 10^{-2}$ cm} \\
Thermal diffusivity ($\chi$) & $3.1 \times 10^{21} {T_{e}^{5/2}}/{Z_{\mathrm{eff}} n_{\mathrm{e}} \log{\Lambda}}$& $3.0 \times 10^{6}$ cm$^2$ s$^{-1}$\\
Turbulent Peclet number ($\mathrm{Pe}$) & $u_{\rm rms} L/\chi$ & 0.2 \\
Dynamic viscosity ($\mu$) & $3.7 \times 10^{-5} \, {M_{\mathrm{H}}^{1/2} T_{i}^{5/2}}/{Z_{\rm C}^{2} \log{\Lambda} }$  & \textcolor{black}{$1.2 \times 10^{2}$ g cm$^{-1}$ s$^{-1}$} \\
Kinematic viscosity ($\nu$) & $\mu/\rho$  & \textcolor{black}{$3.1 \times 10^{3}$ cm$^2$ s$^{-1}$} \\
Turbulent Reynolds number ($\mathrm{Re}$) & $u_{\rm rms} L/\nu $ & 140 \\
Viscous dissipation scale ($l_{\nu}$) & $5 L/{\mathrm{Re}^{3/4}}$ & \textcolor{black}{$4.9 \times 10^{-3}$ cm} \\
Resistivity ($\eta$) & $3.1 \times 10^5 \,  {Z_{\mathrm{eff}} \log{\Lambda}}/{T_{e}^{3/2}}$  & \textcolor{black}{$1.0 \times 10^{3}$ cm$^2$ s$^{-1}$} \\
Magnetic Reynolds number ($\mathrm{Rm}$) & $u_{\rm rms} L/\eta$ & \textcolor{black}{$380$} \\
Magnetic Prandtl number ($\mathrm{Pm}$) & $\mathrm{Rm}/\mathrm{Re}$ & $ \sim 1$ \\
Resistive dissipation scale ($l_{\eta}$) & \textcolor{black}{$5 \ell_{\nu}/{\mathrm{Pm}^{1/2}}$} & \textcolor{black}{$3.0 \times 10^{-3}$ cm} \\
\end{tabular}}
\label{table:TableTheoParam}
\end{table} 

\subsection*{Collisionality assumption}

In the main text, we claim that during the formation of the interaction region (at $\sim25 \, \mathrm{ns}$ after the initiation of the drive beams), the interaction between the two plasma jets is predominately collisional. To verify this claim, we 
estimate the characteristic linear growth rate of the ion Weibel instability associated 
with the counter-propagating jets to
\begin{equation}
\gamma_{\mathrm{W}} \sim \frac{\bar{u}_{\rm in}}{c} \omega_{\mathrm{p}i} \approx 2.8 \times  10^{9} \alpha_\mathrm{W} \left[\frac{Z_\mathrm{C}}{6}\right] \left[\frac{M_{\mathrm{C}}}{12}\right]^{-1/2} \left[\frac{\bar{u}_{\rm in}\left(\mathrm{cm} \; \mathrm{s}^{-1}\right)}{2.2 \times 10^{7} \, \mathrm{cm} \; \mathrm{s}^{-1}}\right] \left[\frac{n_e\left(\mathrm{cm}^{-3}\right)}{8 \times 10^{19} \, \mathrm{cm}^{-3}}\right]^{1/2}\, \mathrm{s}^{-1} ,
\end{equation}
at the wavenumber $k \gtrsim d_i^{-1} \equiv \omega_{\mathrm{p}i}/c$ ( $\omega_{\mathrm{p}i}$ being the plasma frequency). 
% Check this - different to Petros by factor of 2!
Here $\alpha_\mathrm{W}$ is a growth-rate reduction factor associated with the stabilizing effects of intra-jet collisions~\citep{R12b}.  By comparison, the
collisional slowing-down rates $\nu_{\mathrm{S}}^{\mathrm{C}|\mathrm{C}'}$ and $\nu_{\mathrm{S}}^{\mathrm{C}|e}$ (due to interaction of the carbon ions in one jet with carbon ions and electrons, respectively, in the other jet), are given by
\begin{eqnarray}
\nu_{\mathrm{S}}^{\mathrm{C}|\mathrm{C}'} & \approx & 4.9 \times  10^{9}  \left[\frac{Z_\mathrm{C}}{6}\right]^{3} \left[\frac{M_{\mathrm{C}}}{12}\right]^{-1/2} \left[\frac{\bar{u}_{\rm in}\left(\mathrm{cm} \; \mathrm{s}^{-1}\right)}{2.2 \times 10^{7} \, \mathrm{cm} \; \mathrm{s}^{-1}}\right]^{-3} \left[\frac{n_e\left(\mathrm{cm}^{-3}\right)}{8 \times 10^{19} \, \mathrm{cm}^{-3}}\right] \, \mathrm{s}^{-1} \\
\nu_{\mathrm{S}}^{\mathrm{C}|e} & \approx & 1.0 \times  10^{9}  \left[\frac{Z_\mathrm{C}}{6}\right]^{2} \left[\frac{M_{\mathrm{C}}}{12}\right]^{-1} \left[\frac{T_{e,\mathrm{jet}}\left(\mathrm{eV}\right)}{180 \, \mathrm{eV}}\right]^{-3/2} \left[\frac{n_e\left(\mathrm{cm}^{-3}\right)}{8 \times 10^{19} \, \mathrm{cm}^{-3}}\right] \, \mathrm{s}^{-1}
, \label{collslowtime}
\end{eqnarray}
where $T_{e,\mathrm{jet}}$ is the jets' electron temperature.
To obtain these expressions, we have used the fact that, with respect to the carbon-ion and electron populations in one jet, the distribution 
of carbon ions in the other jet is effectively a beam traveling with velocity $2 \bar{u}_{\rm in}$, which is fast with respect to the opposing carbon-ion population, but slow with respect to the electron population~\citep{H94}.
Assuming $\alpha_\mathrm{W} \lesssim 0.5$ [in fact, a conservative choice of the upper bound; 
see~\citep{R12b}], we conclude that collisional relaxation prevents the Weibel 
instability from being present. Once the interaction region has formed, the 
collisional slowing-down rate of carbon ions arriving into the interaction region
increases approximately eight-fold, on account of the interaction region itself 
being approximately stationary in the laboratory frame. Thus, the Weibel 
instability is inhibited in this experiment, perhaps apart from at very early times (that is, before $25$ ns), when the densities of the extended fronts of the counter-propagating jets are lower and flow velocities higher than those given in Eq. [\ref{collslowtime}]. Note that this finding is completely consistent with the observation of the Weibel instability found in other laser-plasma experiments on the OMEGA laser facility [e.g.,~\citep{H15}]. Those other experiments, which involved front-side rather than rear-side blow-off plasma, typically achieve jet velocities nearly an order-of-magnitude greater than our experiments. Since the ion-ion collisional slowing-down rate has a strong power-law dependence on the jet velocities, the collisional slowing-down rate is significantly suppressed compared to the estimate given by Eq. [\ref{collslowtime}]. 

\section*{FLASH simulations}

\subsection*{Overview of FLASH code}

%The experiments were designed, executed, and interpreted using three-dimensional radiation magnetohydrodynamic (MHD) simulations with the FLASH code.
FLASH is a parallel, multi-physics, adaptive-mesh-refinement, finite-volume Eulerian hydrodynamics and radiation MHD
code~\citep{Fryxell2000, Tzeferacos2015}.
The code scales to over 100,000 processors, and
uses a variety of parallelization techniques including domain
decomposition, mesh replication, and threading to utilize
hardware resources in an optimal fashion. FLASH is professionally managed, with
version control, coding standards, extensive documentation, user
support, and integration of code contributions from external users;
and is subject to daily, automated regression testing on a variety of
platforms.  
%To date, FLASH has been downloaded more than 3,500 times, and more than 1,200 papers have been published that directly use the code to model problems in a  wide range of disciplines. These include astrophysics, cosmology, combustion, fluid dynamics, turbulence, and high energy density physics (HEDP). 
HEDP capabilities crucial for the accurate numerical modeling of the physical processes present in laser-driven experiments~\citep{Tzeferacos2015} have been added to the FLASH code over the
past eight years as part of the U.S. Department of Energy (DOE) National Nuclear Security Administration (NNSA)-funded FLASH HEDP Initiative and U.S. DOE NNSA support from Los Alamos National Laboratory and Lawrence Livermore
National Laboratory to the Flash Center for Computational Science.  
%These capabilities are  
The FLASH code and its capabilities have been validated through benchmarks and
code-to-code comparisons~\citep{fatenejad2013b, orban2013}, as well as through
direct application to laboratory experiments~\citep{meinecke2014, falk2014, yurchak2014, meinecke2015, Tzeferacos2015, Li2016, Tzeferacos2017, Tzeferacos2018, Rigby2018, Lu2019, Gao2019, White2019, Chen2020}.

\subsection*{Overview of FLASH simulations of experiment} \label{FLASHsim_overview}

We initialize three-dimensional (3D) FLASH simulations with the target specifications outlined in Figure 1 of the main text. The target specification was
tailored -- via the removal of chlorine doping in the target foils and the
larger opening fraction in the grid~\citep{Tzeferacos2017, Tzeferacos2018} -- to achieve subsonic magnetized turbulence at supercritical magnetic
Reynolds numbers and order-unity magnetic Prandtl numbers.
The FLASH simulations were then validated using the experimental data. They show that the changes that we made to the platform following our previous
experiments~\cite{Tzeferacos2018} do indeed enable us to study turbulent dynamo at order-unity magnetic Prandtl numbers in the experiments described in this paper (as we demonstrate below).

\begin{figure}
  \centering
\includegraphics[width=\linewidth]{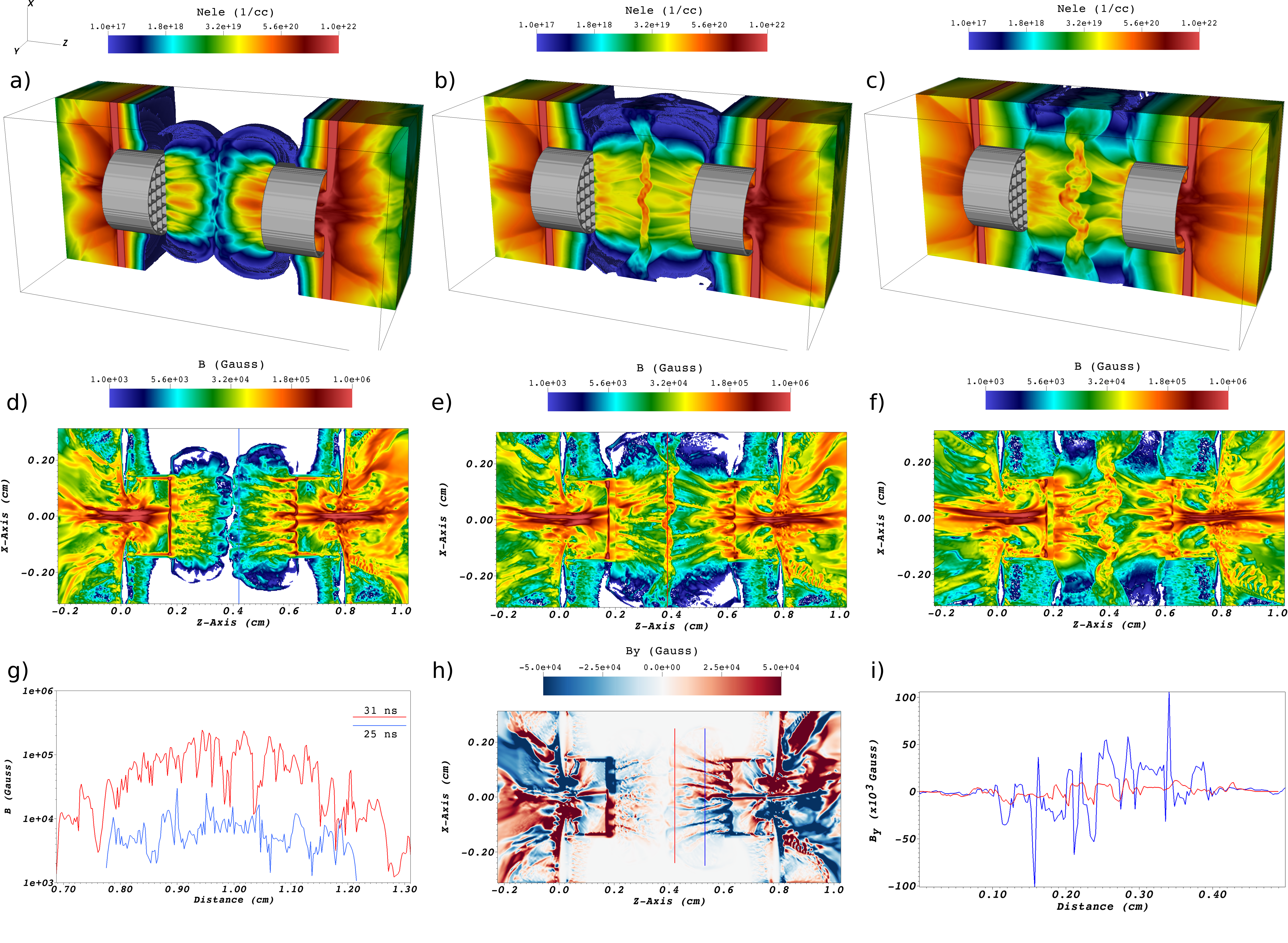}
\caption[3D FLASH simulations of the experiment.]
{\textbf{3D FLASH simulations of the experiment.}
The temporal evolution of electron density and magnetic field strength
in the FLASH simulations of the experiment, demonstrating the onset of turbulence
and the turbulent dynamo amplification of the magnetic fields.
\textbf{a)} Pseudo-color plot of electron density at $t=25 \rm{ns}$, when the
counter-propagating plasma flows collide. The gray contours denote the grids and
cans of the assembly.
\textbf{b)} Same as a) at $t=31 \rm{ns}$, when the turbulent region is well
formed.
\textbf{c)} Same as a) at $t=37 \rm{ns}$.
\textbf{d)} Pseudo-color plot of the magnetic-field magnitude at $t=25 \rm{ns}$,
showing the seed fields being advected into the turbulent region.
\textbf{e)} Same as d) at $t=31 \rm{ns}$. The magnetic fields are amplified by
turbulent dynamo.
\textbf{f)} Same as d) at $t=37 \rm{ns}$, i.e., at late times.
\textbf{g)} Lineouts of magnetic field magnitude from d) and e), showing the
size and fluctuations of the seed field and the amplified magnetic field.
\textbf{h)} Pseudo-color plot of the y-component of magnetic field in the x-z
plane at $t=25 \rm{ns}$, showing the helical nature of the seed field.
\textbf{i)} Lineouts of the y-component of the magnetic field from h) at different
values of z, showing a decrease in the seed-field magnitude due to lateral expansion of the
plasma flows.}
\label{fig:flashevolution}
\end{figure}

%combined with the heating produced by the Thomson-scattering probe beam,
The temporal evolution of the system obtained in the FLASH
simulations is shown in Figure~\ref{fig:flashevolution}. The laser beams ablate the back
of the foil targets and a pair of hot plasma plumes are created and expand
outwards. The ablation results in a pair of shocks -- driven inside the
polystyrene foils -- that break out and propagate supersonically towards the
grids. The lateral expansion of the inward  plasma flows is inhibited by
the collimating effect of the washers. Subsequently, the flows traverse the
grids creating ``finger''-like features and corrugated fronts with
characteristic length scale $L \approx 400 \mu\rm{m}$ – the sum of the grid holes
and the grid hole spacing – and continue towards the center of the domain
(Figure~\ref{fig:flashevolution}a). The flows then collide and the ``finger''-like 
features shear to form an interaction region of hot, subsonic turbulent plasma
(Figure~\ref{fig:flashevolution}b), which then evolves for multiple eddy turnover times
(Figure~\ref{fig:flashevolution}c).

The FLASH simulations replicate most aspects of the evolutionary history of the magnetic field that was described in the main text. More specifically, the simulated laser drive results in the generation of Biermann-battery magnetic fields, which the counter-propagating plasma flows then advect towards the center
(Figure~\ref{fig:flashevolution}d). The topology of the advected fields is
largely helical (Figure~\ref{fig:flashevolution}h), because the laser drive on
each foil generates magnetic fields that are toroidal~\citep{Li2013} in the plane perpendicular to the line of centers. The strength of these magnetic fields declines considerably as they are advected
(Figure~\ref{fig:flashevolution}i) because the counter-propagating plasma flows expand laterally. These weak fields then serve as the seed for the fluctuation dynamo that operates in the turbulent interaction region (as claimed in the main text). The FLASH
simulations show an amplification of the seed field
(Figure~\ref{fig:flashevolution}e-f) from the RMS value of a few kG (blue
line in Figure~\ref{fig:flashevolution}g) to RMS value of $\sim$80-130 kG and
peak value of $\sim$ 300-500 kG (blue line in Figure~\ref{fig:flashevolution}g,
see also Figure 8 in the main text), consistent with the experimental results.

\begin{figure}
  \centering
\includegraphics[width=\linewidth]{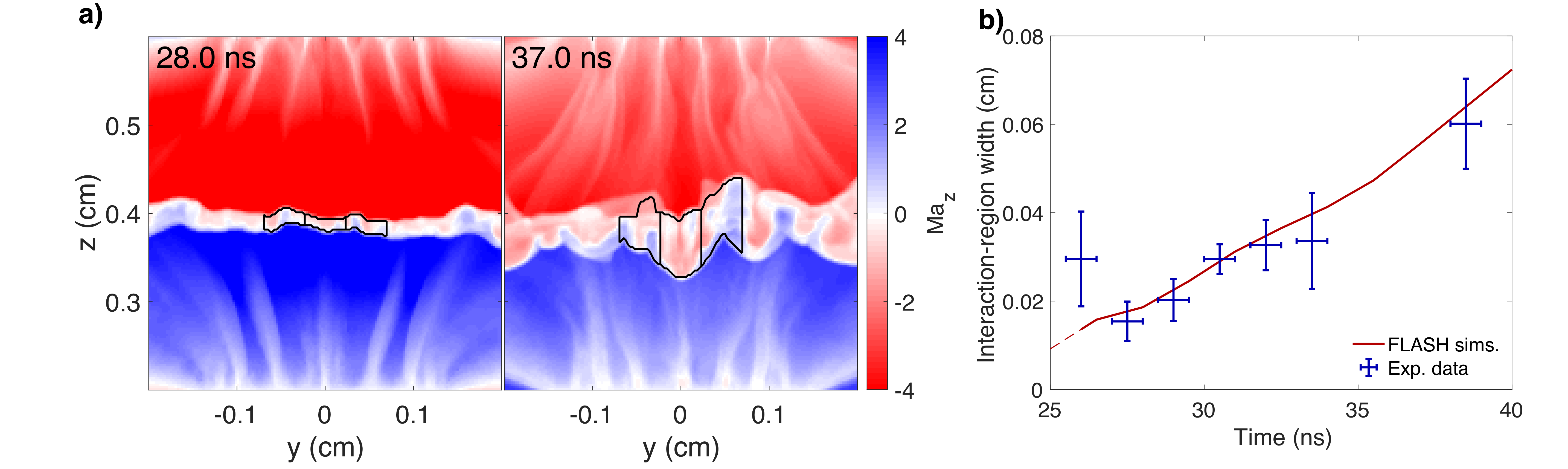}
\caption[Calibration and validation of FLASH simulations using the width of the
interaction region.]
{\textbf{Calibration and validation of FLASH simulations using the width of the
interaction region.}
We use the width of the interaction region as a function of time to calibrate
the offset in time between the FLASH simulations and the experiment.
\textbf{a)} Signed Mach number, $\mathrm{Ma}_z$ (defined using the $z$-component of the bulk velocity, viz., $\mathrm{Ma}_z \equiv \boldsymbol{u} \cdot \hat{\boldsymbol{z}}/c_s$) at 28.0 ns in the FLASH simulations. Examples of the adaptive fiducial volumes (central and full) at this time are also shown.
\textbf{b)} Signed Mach number, $\mathrm{Ma}_z$, at 37.0 ns, in the FLASH simulations. Examples of the adaptive fiducial volumes (central and full) at this time are also shown.
\textbf{c)} Comparison of the width of the interaction region as a function of
time as given by the FLASH simulations and the experimental data.
Good agreement between the results of the FLASH simulations and the experimental
data is obtained by shifting the results of
the simulations 4.0 ns later in time.}
\label{fig:interactionregionwidth}
\end{figure}

Analysis reveals that there is an offset in time between the results of the
FLASH simulations and the experimental data. We use the width of the interaction region as a function of time to determine the offset. In the FLASH simulations, we use the width of the interaction region as the distance between the
locations at which the signed Mach number of the flows decreases below its signed half-maximum.  For the experiment, we use the width of the interaction
region determined from the X-ray self emission images, as was described previously
in~\citep{Chen2020}. Figure~\ref{fig:interactionregionwidth} shows that good
agreement is obtained between the results of the FLASH simulations and the
experimental data by shifting the simulations 4 ns later in time.

\begin{figure}
  \centering
\includegraphics[width=\linewidth]{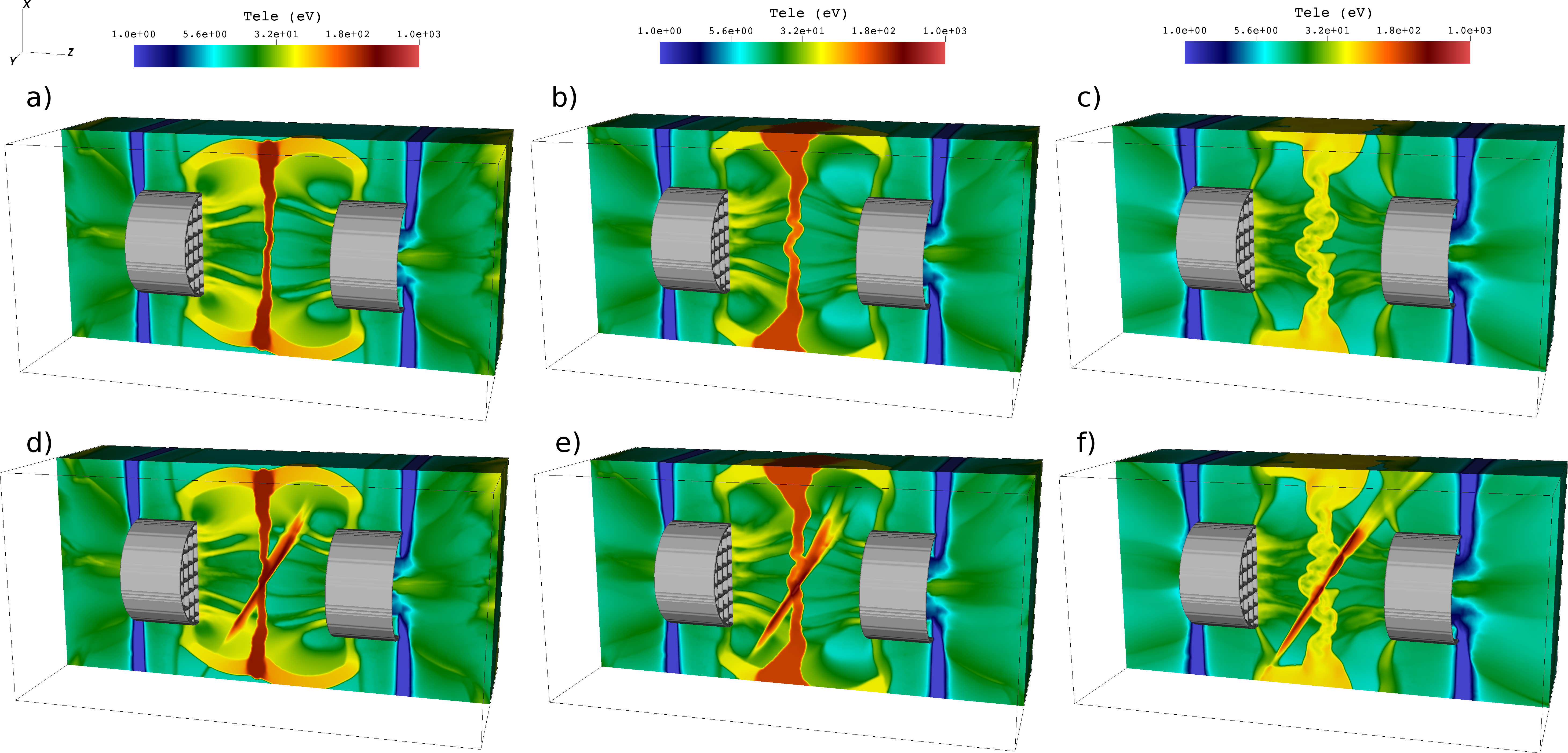}
\caption[Modeling of the Thomson-scattering probe beam heating in the FLASH simulations.]
{\textbf{Modeling of the Thomson-scattering probe beam heating in the FLASH simulations.}
We use the FLASH code's laser-deposition module to incorporate the effects of the
Thomson-scattering probe on the turbulent plasma. The main effect of the probe is
the increase of the electron temperature, predominantly at later times in the
evolution of the turbulent plasma.
\textbf{a)} Pseudo-color plot of the electron temperature at $t=27 \rm{ns}$.
The gray contours denote the grids and cans of the assembly.
\textbf{b)} Same as a) at $t=31 \rm{ns}$.
\textbf{c)} Same as a) at $t=37 \rm{ns}$.
\textbf{d)} Same as a) but with the Thomson-scattering laser centered at
$t=27 \rm{ns}$.
\textbf{e)} Same as b) but with the Thomson-scattering laser fired as at
$t=30 \rm{ns}$.
\textbf{f)} Same as a) but with the Thomson-scattering laser fired as at
$t=37 \rm{ns}$.
}
\label{fig:flashtemperature}
\end{figure}

To enhance further the fidelity of the FLASH simulations, we incorporated
the effects of the Thomson-scattering laser probe (which, according to the analysis presented in `Perturbative heating effects of Thomson-scattering diagnostic', we believe to be significant).  Taking into account the
temporal offset, we illuminate the turbulent region at selected times of the
evolution that correspond to the experimental timing of the Thomson-scattering
diagnostic. The $2\omega$, $1\,\mathrm{ns}$ probe laser is configured to match the
experimental configuration in terms of energy, spatial profile, and pointing
properties. The Thomson-scattering beam refracts on the turbulent plasma and
deposits its energy via inverse bremsstrahlung. The main effect is an increase in electron temperature (Figure~\ref{fig:flashtemperature}), which is more pronounced at later times of the evolution.

\subsection*{Validation of FLASH simulations}

An extensive effort was made to validate the FLASH simulations using the
experimental data. To do so, we use the results of the FLASH simulations for 
plasma-state physical quantities in two adaptive fiducial volumes (AFVs) 
defined as follows. 
We center both AFVs on the line of centers passing through the TCC and take the 
width of the interaction region defined above as their dimension parallel to the 
line of centers. We take $500\, \mu \rm{m}$ as the dimension 
in the directions perpendicular to the line of centers for the first AFV (which we will refer to as the 
central AFV). The resulting volume is comparable to the volume sampled by the Thomson-scattering diagnostic. This volume is a cylinder with a diameter 
$\sim 70$-$100 \, \mu \rm{m}$, and, depending on the time at which the measurement was made, a length of $\sim 300$-$1200 \, \mu \rm{m}$, due to the oblique angle of the Thomson probe-beam relative to the line of centers. For the second AFV (which we will refer to as the full AFV), we take $1500\, \mu \rm{m}$ as the dimension in the directions perpendicular to the line of centers. These dimensions are more than triple the $400 \, \mu \rm{m}$ grid spacing, which ensures that this AFV contains a relatively fair sample of the interaction region, making it appropriate to compare the FLASH simulation results for the plasma state properties for this AFV with the experimental data obtained using the X-ray framing camera and proton-imaging diagnostics.  

\begin{figure}
  \centering
\includegraphics[width=\linewidth]{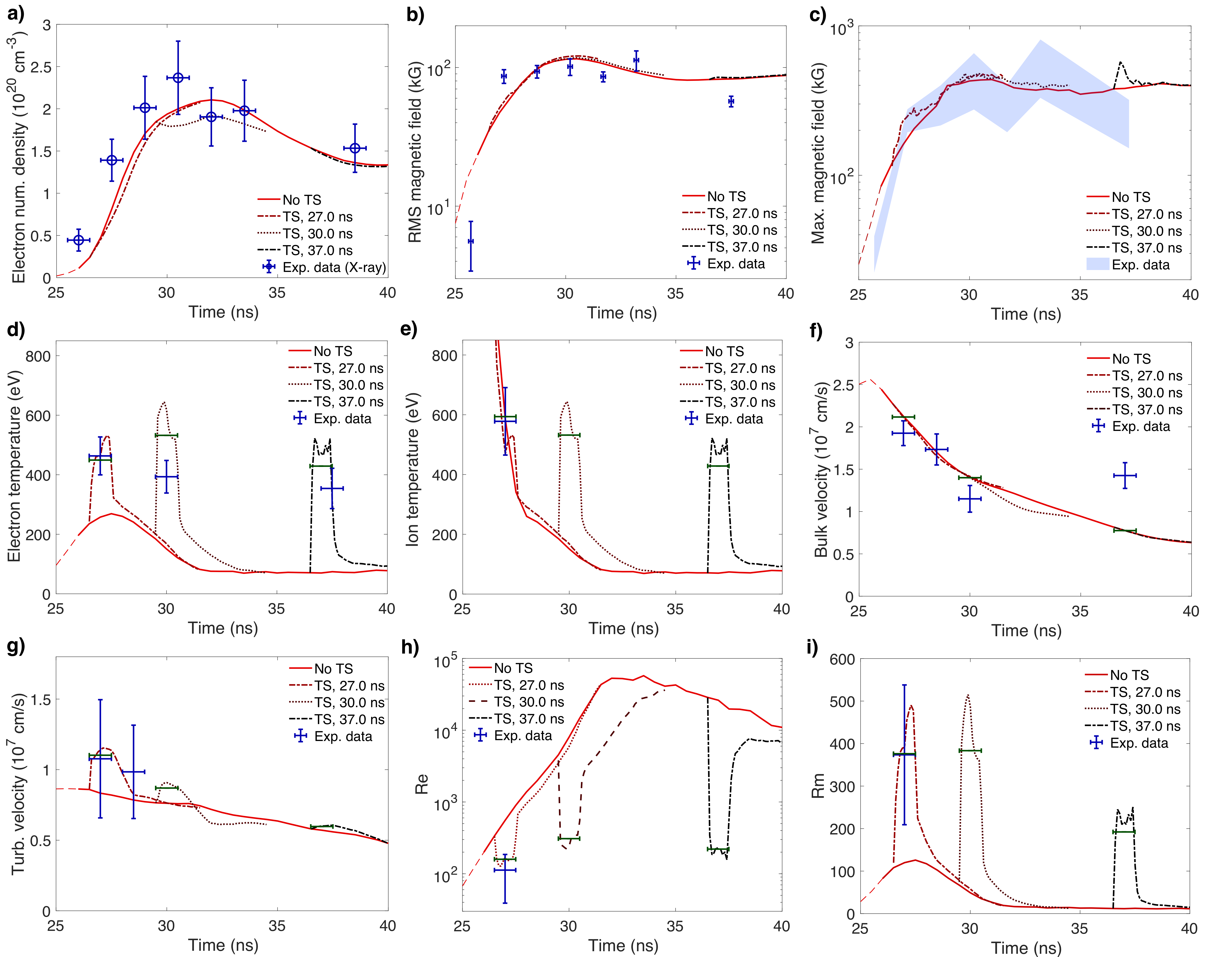}
\caption[Further validation of the FLASH simulations using experimental data.]
{\textbf{Further validation of the FLASH simulations using experimental data.}
FLASH simulation results for nine physical quantities in the interaction
region are shown as a function of time with and without the heating 
due to the Thomson-scattering diagnostic as a result of 1-ns-long pulses
centered at 27, 30 and 37 ns after the start of the laser drive. 
For the physical quantities measured with the Thomson-scattering diagnostic 
(temperatures, flow velocities, and Reynolds numbers), 
the results of the FLASH simulations are presented for the central adaptive 
fiducial volume (AFV), 
which is comparable to the volume sampled by the Thomson-scattering diagnostic. 
The results are shown as horizontal green lines whose length is equal to the 
duration of the 1-ns pulse of the Thomson-scattering probe beam and whose value 
corresponds to the mean value of the physical quantity in that time interval. 
The experimental data points are shown with horizontal error bars equal to
the duration of the 1-ns pulse and vertical error bars corresponding to 1$\sigma$ 
in the physical quantity.
For the other physical quantities (the electron number density measured using the 
X-ray imaging diagnostic and the magnetic field measured using proton imaging), 
the results of the FLASH simulations are presented for the full AFV. 
\textbf{a)} Electron number density. 
\textbf{b)} RMS magnetic field.
\textbf{c)} Maximum magnetic field.
\textbf{d)} Electron temperature.
\textbf{e)} Ion temperature.
\textbf{f)} Bulk velocity.
\textbf{g)} Turbulent velocity.
\textbf{h)} Fluid Reynolds number.
\textbf{i)} Magnetic Reynolds number.
Excellent agreement is found between the results of the FLASH simulations and the
experimental data for all nine quantities, which provides further support for a 
4-ns offset in time between the results of the FLASH simulations and the 
experimental data, and further validates the simulations.}
\label{fig:validationFLASHsimulations}
\end{figure}

Figure~\ref{fig:validationFLASHsimulations} compares the time histories obtained in
the FLASH simulations for nine physical quantities with 
the experimental data for these quantities. 
Panels a)-c) compare the time histories of the electron number density, the RMS magnetic field and the maximum magnetic field for the full AFV (which represents a relatively fair sample of the interaction region), while panels d)-i) compare the time histories of the electron temperature, ion temperature, bulk velocity, turbulent velocity, fluid and magnetic Reynolds numbers for the central AFV (which is comparable to the Thomson-scattering diagnostic's sampling volume).
Without the heating by the Thomson-scattering probe beam, there is good agreement for all quantities save the electron and ion temperatures, which are smaller than the reported experimental values (particularly at late times). When its heating effect is incorporated, the Thomson-scattering probe beam raises the electron temperature in the 
Thomson-scattering volume to $500$-$600\,\rm{eV}$ for the duration of the
probe beam (1 ns). 
This heating of the electrons has only a modest effect on the electron 
density, the bulk velocity and the turbulent velocity [see panels a), f) and 
g)] and virtually no effect on the RMS magnetic field and the maximum magnetic 
field [see panels b) and c)]; it does, however, have a significant effect on the fluid Reynolds 
number $\mathrm{Re}$, decreasing it from $\mathrm{Re} \approx 5 \times 10^3$ - $5 \times 10^5$ to 
$\mathrm{Re} \approx 200$-$300$, and on the magnetic Reynolds number $\mathrm{Rm}$ of the turbulent 
magnetized fluid, increasing it from $\mathrm{Rm} \approx 25$-$120$ to $\mathrm{Rm} \approx 200$-$500$, as panels h) and i) 
show. When the heating by the Thomson-scattering probe beam is included, the agreement is excellent in all cases, validating the simulations. Furthermore, the agreement for the electron density and the RMS magnetic field, which are physical quantities of particular 
interest and both of which increase rapidly, confirms that the 4 ns time offset between the FLASH simulations and the experimental data inferred from the width of the interaction as a function of time is correct.

\color{black}

\begin{figure}
  \centering
\includegraphics[width=0.95\linewidth]{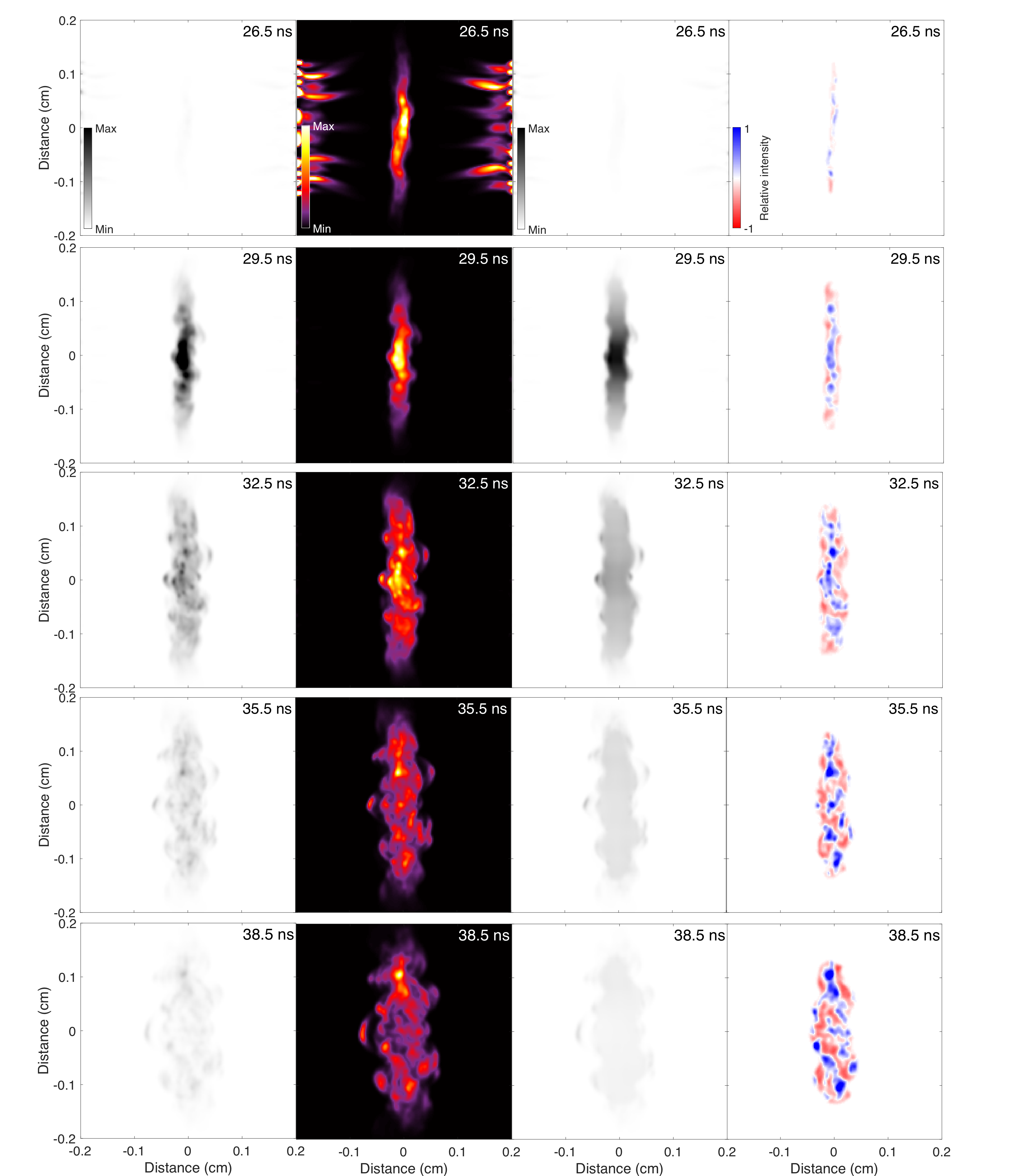}
\caption[Self-emission X-ray images of the FLASH-simulated interaction-region plasma]
{\textbf{Self-emission X-ray images of the FLASH-simulated interaction-region plasma.}
First column: simulated `absolute' X-ray intensity images, 
normalized to the maximum count value associated with the interaction-region plasma across all of the images (which is realized at $t = 29.5$ ns). The 50 $\mu$m resolution limit of the experimental X-ray images is incorporated into the simulated images via a convolution with a two-dimensional Gaussian filter of the appropriate width. Second column: X-ray intensity images normalized
by the maximum pixel value in each image. Third column: mean emission profiles calculated from the far-left column; the gray-scale map is the same as in the far-left images. Fourth column: 
relative X-ray intensity map calculated from the mean emission profile. Fluctuations with a positive value with respect to the mean intensity are denoted in blue, negative in red, with maximum 
and minimum values set at $\pm 100\%$ of the mean value.}
\label{fig:FLASHXray}
\end{figure}

We can also use the self-emission X-ray imaging diagnostic to provide further confirmation that the FLASH simulations are modeling the hydrodynamic physical variables of the interaction-region plasma correctly. In the main text, we establish a simple relationship between the electron density $n_e$, electron temperature $T_e$, and the measured (optical) intensity $I$ on the CCD camera associated with X-rays emitted by the plasma: $I \propto \int \mathrm{d}s \ n_e^2 \hat{f}(T_e)$, where $\hat{f}(T_e)$ is a function defined by Eq. [2] of the main text and plotted in Figure S1b. This relationship can be used to simulate artificial X-ray images of the FLASH-simulated plasma; the resulting images are shown in Figure \ref{fig:FLASHXray}.
The agreement (both qualitative and quantitative) between the evolution of the FLASH images and those from the experiment is excellent: at $t = 26.5$ ns, the emission from the interaction region is comparable to that of the grid jets, and is considerably weaker than 1.5 ns later; peak emission is at 29.5 ns; then the total emission drops off considerably by $t = 38.5$ ns. As a further validation, the synthetic X-ray images derived from the FLASH simulations were analyzed using the same procedure that was applied to the experimental images: we extract mean-emission and relative-intensity maps, and then evaluate the RMS of the latter. As shown in Figure 4c of the main text, the RMS of relative fluctuations derived from the synthetic images evolves similarly to the RMS derived from the experimental images (save perhaps at late times). In summary, the results of the simulated X-ray self-emission images provide additional validation of the accuracy of the hydrodynamic modeling of the FLASH simulations. 

\color{black}

\subsection*{Evolution of the interaction-region plasma in the FLASH simulation} 

\begin{figure}
  \centering
\includegraphics[width=\linewidth]{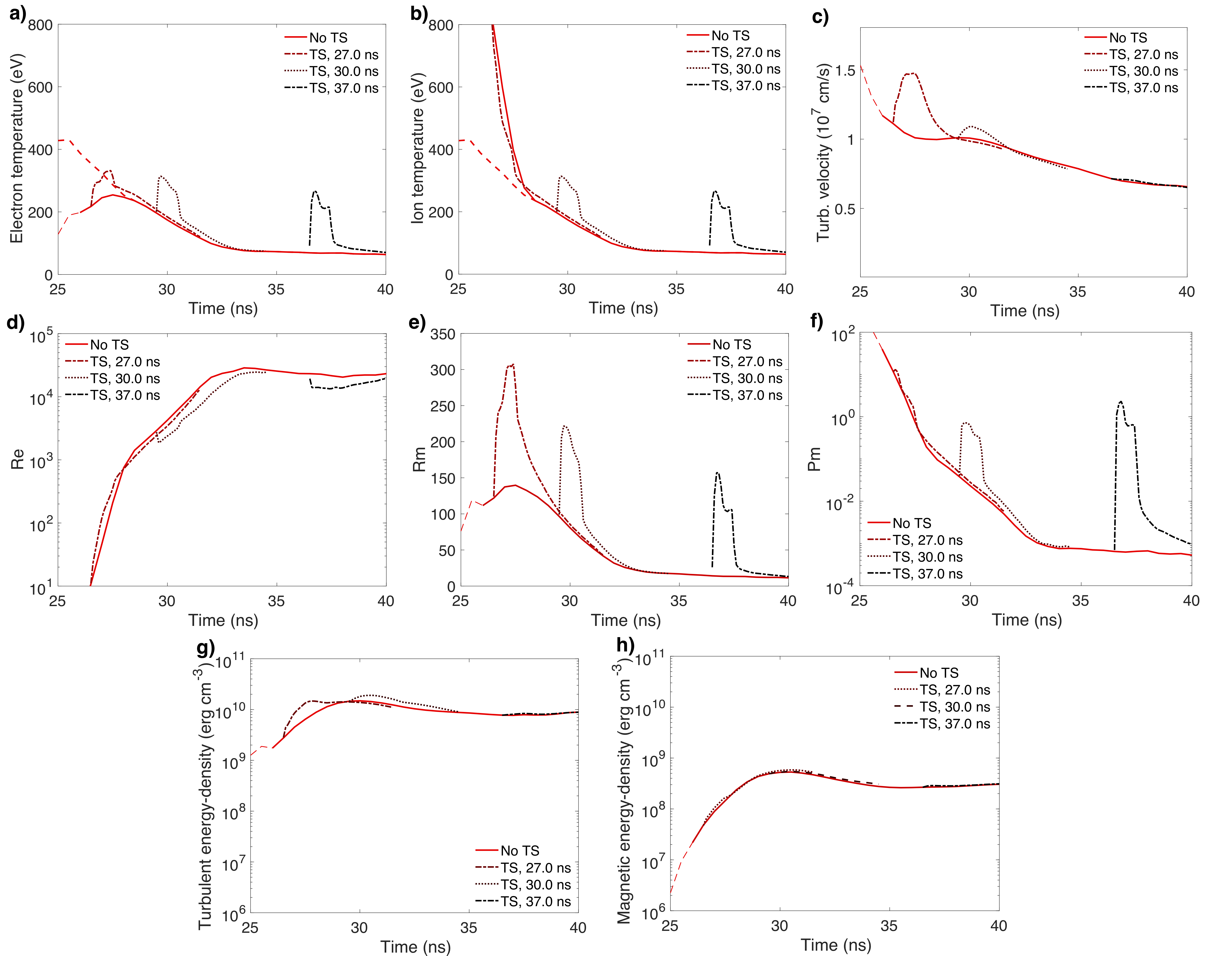}
\caption[Evolution of physical quantities in the FLASH simulations]
{\textbf{Evolution of physical quantities in the FLASH simulations.}
FLASH simulation results are shown as a function of time for eight physical 
quantities, averaged over the full adaptive fiducial volume for no heating and for
heating by the 1-ns-long Thomson-scattering probe beam at times centered at 
27, 30 and 37 ns after the start of the laser drive. 
\textbf{a)} Electron temperature. The thick dashed red line depicts the 
electron-temperature history under the assumption of equal and ion 
temperatures in the FLASH simulations, but conserving total thermal energy 
of the plasma. 
\textbf{b)} Ion temperature.
\textbf{c)} Turbulent velocity.
\textbf{d)} Turbulent fluid Reynolds number.
\textbf{e)} Turbulent magnetic Reynolds number.
\textbf{f)} Magnetic Prandtl number.
\textbf{g)} Turbulent-energy density.
\textbf{h)} Magnetic-energy density.}
\label{fig:FLASH3b3evolve}
\end{figure}

Figure~\ref{fig:FLASH3b3evolve} shows the time histories of eight physical
quantities in the FLASH simulations for the full AFV (which represents 
a relatively fair sample of the interaction region). 
Results are shown for the case of no heating and when some heating occurs as a 
result of 1-ns-long Thomson-scattering probe beam pulses centered at 27, 30 
and 37 ns after the start of the laser drive. These figures illustrate two key features of the experiment. 

First, the rapid collisional shock heating that occurs immediately after the interaction region forms causes the initial electron temperature to exceed 250 eV and the initial ion temperature to exceed 600 eV. This simultaneously decreases the resistivity and increases the viscosity, leading to $\mathrm{Pm} = \mathrm{Rm}/\mathrm{Re} > 1$ at early times. 
We note that the exceptionally high ion temperatures (> 1 keV) obtained in the simulations just after the interaction region forms are not likely to be physical; at these early times, the plasma is not fully collisional, and thus the collisional electron-ion heating model assumed in a one-fluid MHD code such as FLASH (which presumes that ions are viscously heated predominantly over electrons) is not strictly applicable. As a comparative reference, panels a) and b) of Figure \ref{fig:FLASH3b3evolve} show an alternative evolutionary history of the electron and ion temperatures derived under two assumptions: that evolution of the total thermal energy in the plasma is captured accurately by the FLASH simulations, but with fixed $T_e = T_i$ at all times. In this alternative model, we find that the electron and ion temperatures just after the interaction region forms (at $t$ = 25.5 ns) are $T_e = T_i \approx 450 \, \mathrm{eV}$. This is in close agreement with the reported experimental values, suggesting that as the interaction region is forming, collisional electron heating could be significant. Such an effect has indeed been seen previously in laser-plasma experiments involving counter-streaming CH plasma jets~\citep{Ross2012}.

Secondly, the Thomson-scattering probe beam heating has a non-trivial effect on the interaction-region plasma's dynamics, particularly at late times. For the earliest time at which the Thomson-scattering beam heating was introduced in the FLASH simulations ($t$ = 27 ns), the electron temperature and turbulent velocity are increased by $\sim 30 \%$ over their unperturbed values [see panels a) and c)] in the full AFV, resulting in an increased $\mathrm{Rm}$  [see panel e)]. At later times, the probe-beam heating has an increasingly significant effect on the electron and ion temperatures [see panel b)], decreasing the fluid Reynolds number $\mathrm{Re}$ [see panel d)] and increasing the magnetic Reynolds number $\mathrm{Rm}$ of the turbulent plasma.
Overall, the heating has only a modest effect on the turbulent kinetic energy density [see panel g)] and virtually 
no effect on the magnetic-energy density [see panel h)]. 
Panel f) shows that $\mathrm{Pm} > 1$ during 
the 1-ns duration of the Thomson-scattering probe beams at later times.
We emphasize that the initial amplification of the magnetic field always occurs in the $\mathrm{Pm} > 1$ regime, irrespective of the Thomson-scattering probe beam heating. 

\color{black}

\subsection*{The growth rate of the magnetic field in the FLASH simulations}

\begin{figure}
  \centering
\includegraphics[width=0.7\linewidth]{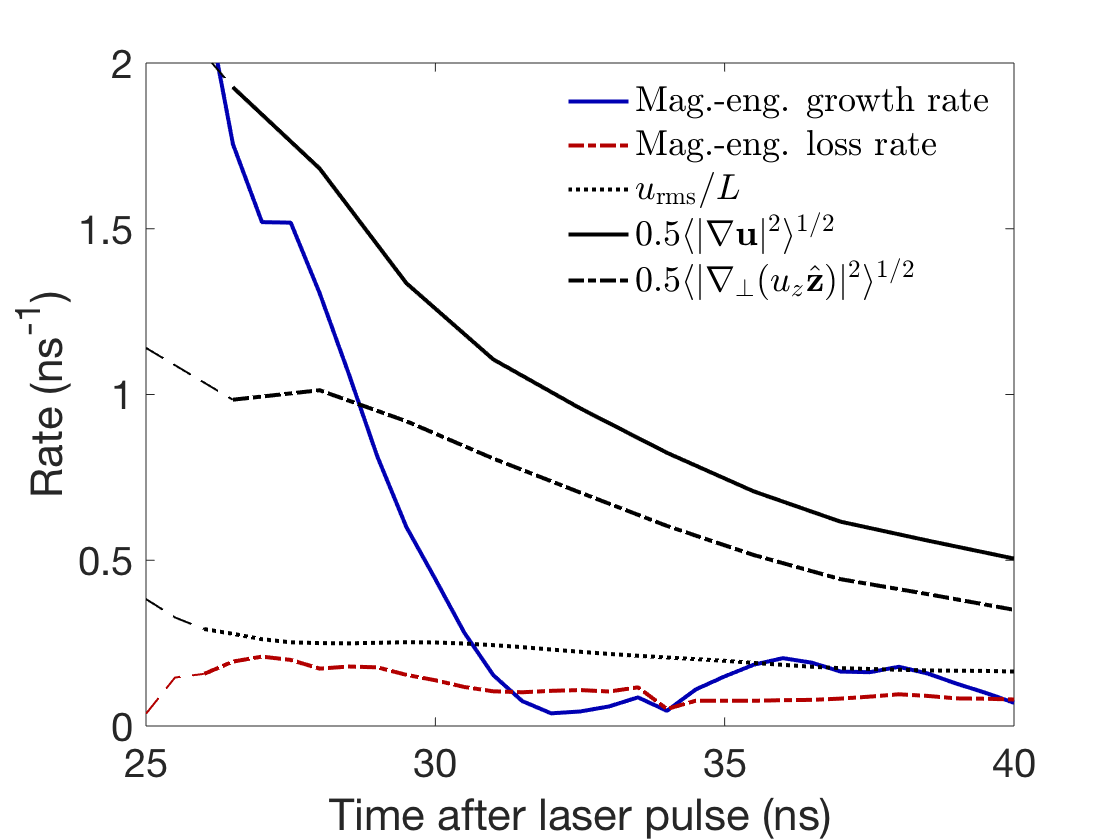}
\caption[FLASH simulation results for the rates of change for three physical 
quantities in the interaction region.]
{\textbf{FLASH simulation results for the rates of change of five physical 
quantities in the interaction region.}
Shown are the FLASH simulation results for the rate of magnetic-energy 
growth $\mathrm{d} \log{E_B}/\mathrm{d}t$ (which we define as the increase in the magnetic energy in a given time period, subtracting the magnetic energy advected in by the two plasma flows and adding the magnetic energy advected out by the bulk flow perpendicular to the line of centers in the same time period), 
the magnetic-energy loss rate due to advection out by the bulk flow perpendicular to the line of centers (which is determined by calculating the flux of magnetic energy through the four perpendicular faces of the  volume being sampled), the turbulent eddy turnover rate at the outer scale $L = 400 \, \mu \mathrm{m}$, the RMS rate of strain [defined here as $\langle |\nabla \boldsymbol{u}|^{2}\rangle^{1/2}$; the 0.5 factor is introduced for convenience of plotting], and the RMS rate of perpendicular strain of the component of velocity parallel to the line of centers [defined as $\langle |\nabla u_z \hat{\boldsymbol{z}}|^{2}\rangle^{1/2}$] -- all for the full AFV.}
\label{fig:rates}
\end{figure}

Figure~\ref{fig:rates} compares (1) the rate of growth of the magnetic energy 
in the interaction region (subtracting the magnetic energy advected into it by the two plasma flows and adding the magnetic energy advected out of it by the bulk flow perpendicular to the line of centers in that time period), 
(2) the rate at which magnetic energy is advected out of the interaction region by the bulk flow perpendicular to the line of centers, 
(3) the turbulent eddy turnover rate at the outer scale $L = 400 \, \mu \mathrm{m}$ in the interaction region given by the validated FLASH simulations, and (4) the RMS rate of strain of the FLASH-simulated velocity field (a fifth quantity, the RMS rate of perpendicular strain of the component of FLASH-simulated velocity parallel to the line of centers, is also shown, and is discussed in the next section). 
The results for all four quantities are for the full AFV. 
Figure~\ref{fig:rates} shows that, at early times, the growth rate of the 
magnetic energy is much greater than the rate at which the magnetic field 
energy is advected out of the interaction region, while at late times, the
former is comparable to, or somewhat larger than, the latter. 

Of particular interest is that, at early times, the magnetic energy grows at a rate $\gamma \approx 2 \times 10^9$ s$^{-1}$ (which is equivalent to a growth time of 0.5 ns). This rate, which is similar to that obtained experimentally, is comparable to the RMS rate of strain, but is ${\sim}5$--$10$ times larger than the outer-scale eddy turnover rate $\gamma_{\rm L} = {u_{\rm rms}}/{L} \approx 2.5 \times 10^8$ s$^{-1}$, where $L$ is the outer scale of the turbulence and $u_{\rm rms}$ is the RMS velocity at that scale. This rate is equivalent to a growth time $\tau_L \approx 4$ ns. As discussed in the main text, based on periodic-box MHD simulations of the fluctuation dynamo at comparable magnetic Reynolds and magnetic Prandtl numbers~\citep{SCTM04,S07}, the growth rate of the magnetic energy is expected to be smaller than what we see here. 
%, if turbulent mixing at the outer scale alone was the mechanism powering the fluctuation dynamo~\citep{SCTM04,S07}. 
This suggests that there is another growth mechanism also at play (see next section).

\subsection*{The role of directed shear flows in our experiment}

In the main text, we suggest that the interaction between stochastic motions and directed shear flows in the interaction-region plasma could explain the fast growth of the magnetic field that is observed in the experiment and in FLASH simulations; here, we use the latter to help provide evidence for this claim.

\begin{figure}
  \centering
\includegraphics[width=\linewidth]{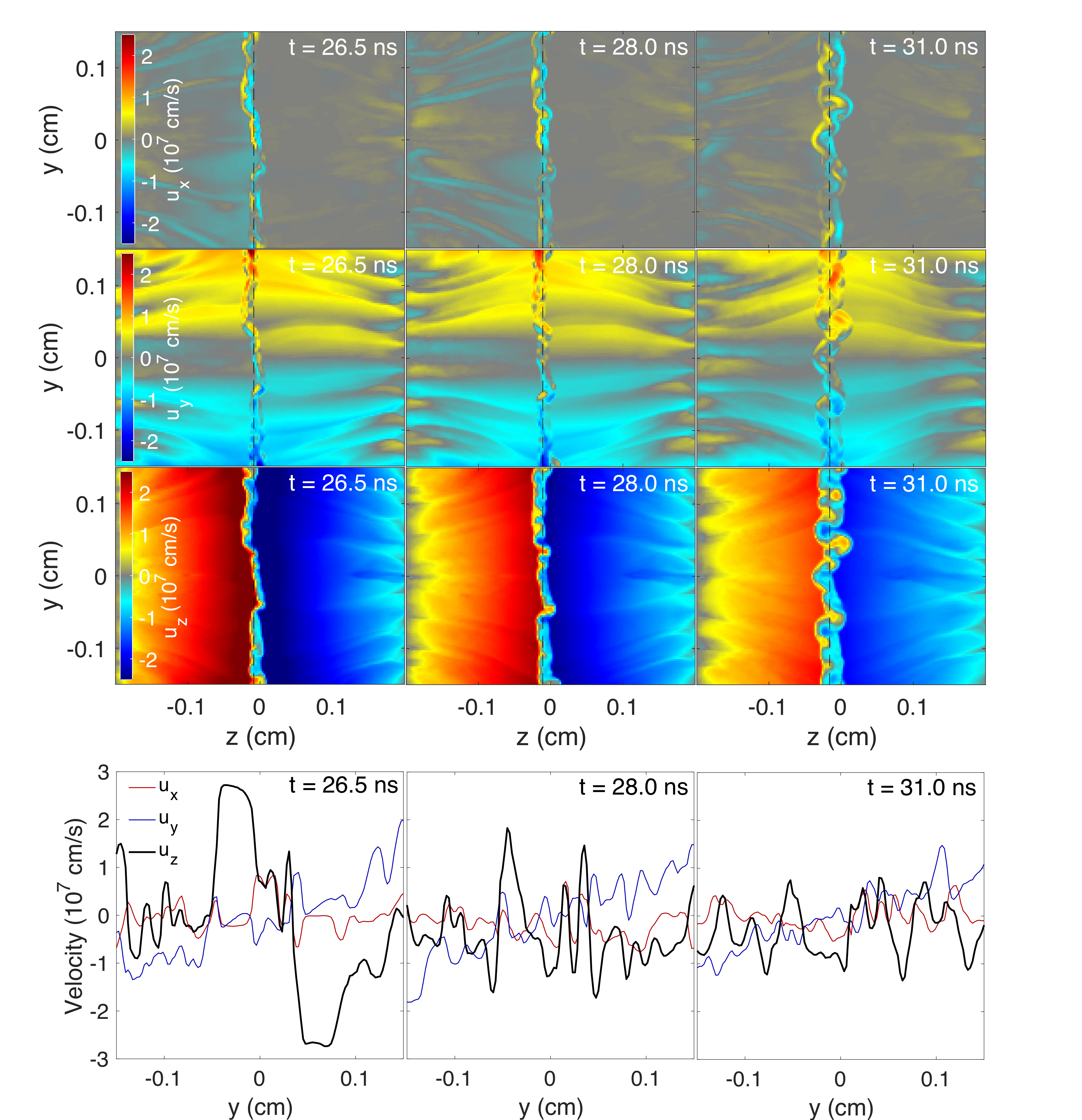}
\caption[The velocity field in the FLASH simulations.]
{\textbf{The velocity field in the FLASH simulations.}
Top grid: pseudo-color two-dimensional slice plots [in the ($y$,$z$) plane) of the three components of velocity in the FLASH simulations at $t = 26.5$ ns (left column), $t = 28.0$ ns (middle column) and $t = 31.0$ ns (right column). Bottom row: lineouts (in the $y$ direction) of the three components of velocity, at $t = 26.5$ ns (left column), $t = 28.0$ ns (middle column) and $t = 31.0$ ns (right column). The lineouts are taken at the positions indicated on the top grid of slice plots by a dashed line.
}
\label{fig:flashvelocity}
\end{figure}

\begin{figure}
  \centering
\includegraphics[width=\linewidth]{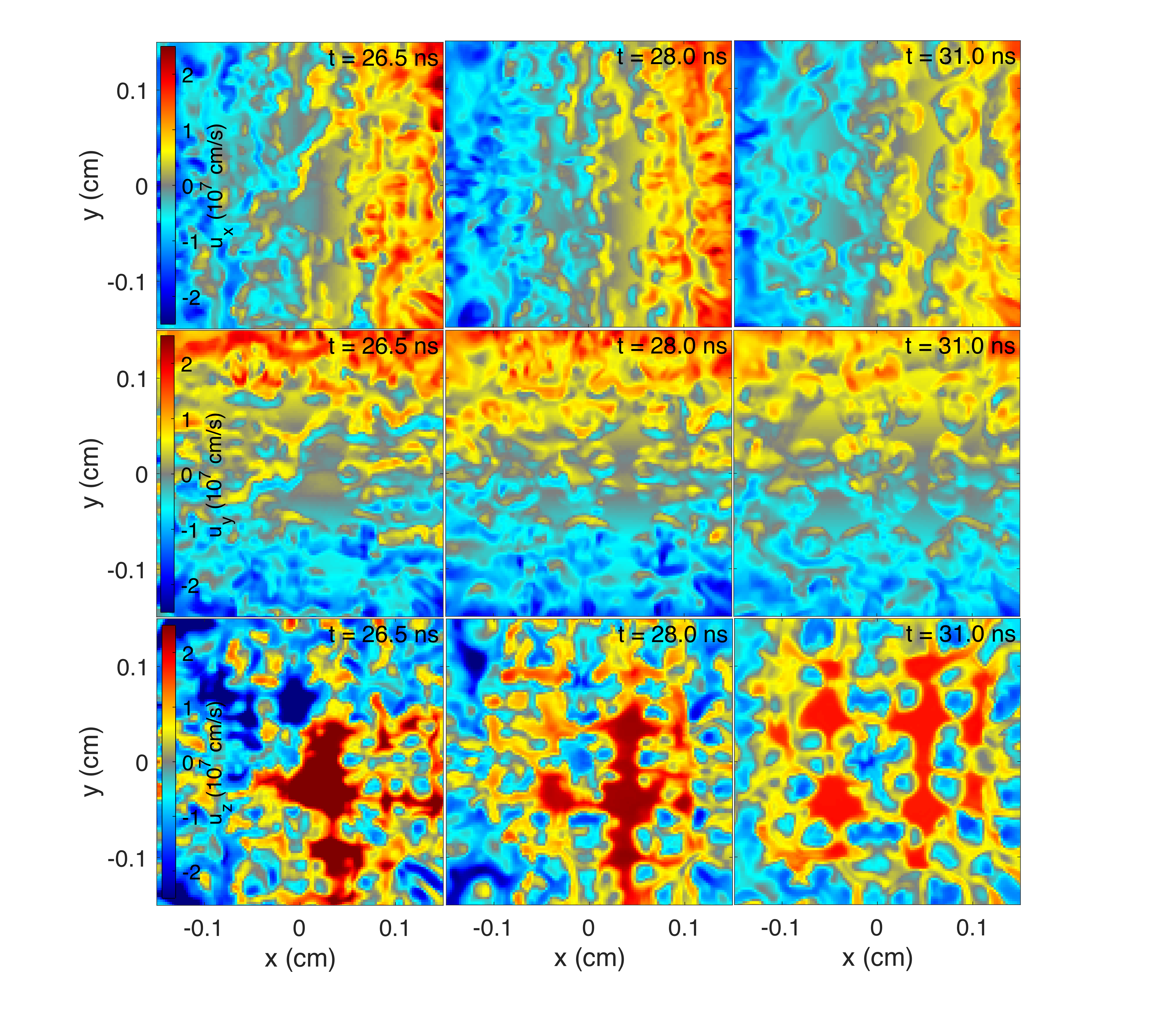}
\caption[Perpendicular slices of the velocity field in the FLASH simulations.]
{\textbf{Perpendicular slices of the velocity field in the FLASH simulations.}
Pseudo-color two-dimensional slices plots [in the ($x$,$y$) plane) of the three components of velocity in the FLASH simulations at $t = 26.5$ ns (left column), $t = 28.0$ ns (middle column) and $t = 31.0$ ns (right column). The slices are taken at the position on the $z$ axis coincident with the lineouts shown in Figure \ref{fig:flashvelocity}. 
}
\label{fig:flashvelocityperp}
\end{figure}

Figures \ref{fig:flashvelocity} and \ref{fig:flashvelocityperp} show the velocity field in the FLASH simulations at three times during the 6 ns time interval after the interaction region forms.
The presence of directed shear flows in addition to stochastic motions is most clearly illustrated by taking lineouts of the velocity field in the plane of the interaction-region plasma (see Figure \ref{fig:flashvelocity}, bottom row): changes in velocity $\Delta u_z \approx 200$--$300$ km$\, \mathrm{s}^{-1}$ are seen in the component of velocity ($u_z$) parallel to the line of centers over a characteristic reversal scale $\Delta x_{\perp} \approx 75$--$125 \; \mu$m in the directions that are perpendicular to the line of centers. By comparison, changes in the other two velocity components over the same length scale are much smaller ($\Delta u_x, \Delta u_y < 100$ km$\, \mathrm{s}^{-1}$). These shear flows are seen across the whole perpendicular plane of the interaction-region plasma (see Figure \ref{fig:flashvelocityperp}). 

The physical origin of these directed shear flows is the asymmetry of the counter-propagating plasma ``fingers'', or jets, originating from the offset of the two grids. In perpendicular spatial locations coincident with the holes in grid A, the density of the jets originating from grid A is much higher than grid B; an analogous statement holds for spatial locations coincident with the holes in grid B. On the other hand, the $u_z$ component of velocity is close to uniform across each jet (see Figure \ref{fig:flashvelocity}, bottom row of top grid). Conservation of momentum therefore dictates that when the jets collide, the bulk velocity of plasma in these perpendicular spatial locations will be directed towards grid B or grid A, respectively. The counter-propagating, interleaving jets will shear, forming shear layers along the line of centers (i.e., the z axis in the FLASH simulations). Even though the Kelvin-Helmholtz (KH) instability will likely act to destabilize these shears (see next section), the continual re-supply of plasma by the plasma jets helps to sustain these flows for at least one driving-scale eddy turnover time (see Figure \ref{fig:flashvelocity}, final column).

Given the parameter regime of our experiment -- specifically, the relatively large viscosity in the interaction-region plasma at times $t \leq 27 \, \mathrm{ns}$ ($\nu \approx 3 \times 10^{3} \; \mathrm{cm}^{2} \, \mathrm{s}^{-1}$) -- the FLASH simulations resolve the transverse scale of these shear layers, apart from at very early times in the experiment. Consider the evolution of a simple unsteady viscous shear
layer with $\boldsymbol{u} = u_z(y,t) \hat{\boldsymbol{z}}$.  The flow profile $u_z(y,t)$ in such a layer is given by~\citep{B67}
\begin{equation}
u_z(y,t) = \frac{\Delta u_z}{2} \, \mathrm{erf}{\left(\frac{y}{\sqrt{4 \nu t}}\right)} \approx \frac{\Delta u_z}{2} \tanh{\left(\frac{y}{\sqrt{\pi \nu t}}\right)}, 
\end{equation}
where the second expression will be used in the analysis of the KH instability in the next section. A reasonable estimate for the width of the shear layer at time $\Delta t$ after it forms is therefore twice the viscous scale length, viz., $\Delta x_{\perp} \approx 2 \sqrt{\pi \nu \Delta t}$. Conservatively choosing $\Delta t \gtrsim 1$ ns (corresponding to the relevant time delay after the interaction-region formation at $t = 25$--$26$ ns), we find $\Delta x_{\perp} \gtrsim 60 \; \mu$m. The resolution of the FLASH simulations is $\Delta x_{\rm res} = 25 \; \mu$m, implying that such reversals are indeed captured in the simulations. As a result, the strain rate of the shear flow in the experiment can be reasonably estimated as being similar to that found in the simulations. Note that, because $\nu$ decreases considerably as the plasma's ions cool, the shear layers are unlikely to broaden further due to viscous effects after $t \approx 27$ ns. Therefore, in the initial phases of the magnetic-field evolution within the turbulent interaction-region plasma, we have sustained, directed shear flows, which contribute to the total rate of strain that drives the dynamo action responsible for the magnetic-field amplification. 

More quantitatively, at $t\sim 27$ ns the strain rate of the directed shear flows $S \equiv \langle |\nabla_{\perp} (u_z \hat{\boldsymbol{z}}|^{2} \rangle^{1/2}$ is $S \approx \Delta u_z/\Delta x_{\perp} \approx 2 \times 10^{9} \; \mathrm{s}^{-1}$ (see also Figure \ref{fig:rates}, where the exact value of this quantity for the central AFV is calculated). This is ${\gtrsim}\,50\%$ of the total measured RMS rate of strain shown in Figure \ref{fig:rates}, supporting the claim that these directed flows contribute to the fast amplification rate of the magnetic field. It becomes apparent that the fluctuation dynamo realized in our experiment is enhanced by shear dynamo [see~\citep{R19} and references therein] at early times. 
This is further evidenced by the time evolution of the rate of strain $\gamma_{\ell_{\nu}} \sim \mathrm{Re}^{1/2} u_{\rm rms}/L$ of viscous-scale turbulent eddies,  which is anticipated to be proportional to the growth rate of the magnetic field if fluctuation dynamo were the only mechanism at play. At early times, this is not the case. 
In many astrophysical cases, such as the merger of galaxies and the growth of galaxies clusters through
cluster mergers, the capturing of groups and the infall of filaments, strong shear flows are present~\citep{S19}. In this sense, the experiments we have conducted could be more relevant to the astrophysical case than previous numerical simulations
of turbulent dynamo, which typically generate the turbulent spectrum by an isotropic forcing mechanism~\citep{S07} (although whether this is actually the case ultimately depends on the true magnitude of $\mathrm{Re}$ in astrophysical systems -- a quantity of some uncertainty~\citep{Z18} -- as well as the properties of astrophysical analogues of colliding jets). 

In spite of the important role of the directed shear flows, the presence of the stochastic fluid motions -- which are responsible for fluctuation dynamo in both experiments and simulations -- is essential. This is because a quasi-2D unidirectional shear flow cannot drive sustained amplification of magnetic fields by itself. This can be demonstrated by considering the MHD induction equation for a velocity field of the form $\boldsymbol{u}(\boldsymbol{r}) = U(x,y) \hat{\boldsymbol{z}}$; writing out separately the induction equation's components perpendicular and parallel to $\hat{\boldsymbol{z}}$, we have
\begin{eqnarray}
    \frac{\partial \boldsymbol{B}_{\perp}}{\partial t} + U \frac{\partial \boldsymbol{B}_{\perp}}{\partial z} & = & \eta \nabla^{2} \boldsymbol{B}_{\perp} \, , \\
    \frac{\partial B_{z}}{\partial t} + U \frac{\partial {B}_{z}}{\partial z} & = & \boldsymbol{B}_{\perp} \cdot \nabla_{\perp} U + \eta \nabla^{2} {B}_{z} \, ,
\end{eqnarray}
where $\nabla \equiv \partial/\partial \boldsymbol{r}$. 
It follows that the shear flow cannot stretch $\boldsymbol{B}_{\perp}$, which must, therefore, decay on resistive timescales. The remaining component of the magnetic field, $B_z$, can be amplified transiently, but must eventually also decay, because $\boldsymbol{B}_{\perp}$ does. If, on the other hand, $\boldsymbol{u}(\boldsymbol{r}) = U(x,y) \hat{\boldsymbol{z}} + \delta \boldsymbol{u}(\boldsymbol{r})$, where $\delta \boldsymbol{u}(\boldsymbol{x})$ is some stochastic component of the velocity, the MHD induction equation instead becomes 
\begin{eqnarray}
    \frac{\partial \boldsymbol{B}_{\perp}}{\partial t} + U \frac{\partial \boldsymbol{B}_{\perp}}{\partial z} + \delta \boldsymbol{u} \cdot \nabla \boldsymbol{B}_{\perp}  & = & \boldsymbol{B} \cdot \nabla \delta \boldsymbol{u}_{\perp} + \eta \nabla^{2} \boldsymbol{B}_{\perp} \, , \\
    \frac{\partial B_{z}}{\partial t} + U \frac{\partial {B}_{z}}{\partial z} + \delta \boldsymbol{u} \cdot \nabla {B}_{z} & = & \boldsymbol{B}_{\perp} \cdot \nabla_{\perp} U + \boldsymbol{B} \cdot \nabla \delta {u}_{z} + \eta \nabla^{2} {B}_{z} \, .
\end{eqnarray}
It is then possible for stochastic motions in the perpendicular direction to stretch $\boldsymbol{B}$, and thus couple the growth of $\boldsymbol{B}_{\perp}$ and $B_z$. 
%(at least, in principle). 
In our experiment, the perpendicular stochastic motions originate from the complex perpendicular flow profiles arising due to the interaction between jets extending from adjacent grid holes, and the aforementioned KH instabilities (which grow at a rate greater than the eddy turnover rate at the outer scale, as shown in the next section). 

In short, the dynamo action realized in our experiments amplifies the magnetic field at a rate consistent with the total rate of strain predicted by the validated FLASH simulations. A significant component of that strain rate is attributed to the presence of multiple directed shear flows, which are not scale separated from the turbulence. The amplification mechanism is therefore consistent with a fluctuation dynamo that is enhanced by a shear dynamo at early times. This hybrid configuration reconciles the fast amplification we obtain with respect to periodic-box MHD simulations, and may be more relevant for realistic astrophysical scenarios in the ICM where shear is present.

%In short, it seems likely in this set-up that we are dealing with a hybrid between a shear dynamo [see~\citep{R19} and references therein]
%and a fluctuation dynamo, except with multiple shear flows that are not scale separated from the turbulence. %We defer more detailed discussion of the dynamo mechanism in a flow such as that present in our experiment to a future publication. 

\subsection*{The Kelvin-Helmholtz instability in our experiment}

In this section, we estimate the growth rate of the KH instability associated with the directed shear flows in the interaction-region plasma, and show that motions perpendicular to the shear flows should grow at a greater rate than the outer-scale eddy-turnover rate.  

%Indeed, because it can be shown in the linear regime of the KH instability (that is, when the perturbed velocity $\delta u$ associated with the unstable KH mode is much smaller than the magnitude of the velocity shear $\Delta U$) that the perturbed magnetic field is proportional to the perturbed velocity, the growth rate of the hydrodynamic KH growth rate will be the same as the magnetic field's initial growth rate. 
The magnetic fields in the interaction region are not dynamically important initially, so we focus on the hydrodynamic KH instability. 
The theory of the hydrodynamic KH instability is well established. In an inviscid fluid containing a simple shear layer with a discontinuous velocity profile (change in velocity $\Delta u_z$), the wavevector of the fastest-growing modes is parallel to the fluid velocity, and the linear growth rate $\gamma_{\mathrm{KH},0}$ of such modes as a function of their wavenumber $k$ is given 
by~\citep{C61}
\begin{equation}
\gamma_{\mathrm{KH},0} = \frac{1}{2} k \Delta u_z.
\end{equation}
In a viscous fluid, a reduction of the growth rate is caused by the finite width of the shear layer~\citep{FJ01}. As an
example, \cite{MP82} show that, for a hyperbolic-tangent profile with a
characteristic reversal scale $\Delta x_{\perp}$, the peak growth (for
$M_s \equiv \Delta u_z/c_{\rm s} \approx 1$, as in our experiment) occurs at the wavenumber $k_{\rm max}$ satisfying
$k_{\rm max} \Delta x_{\perp}\approx 0.8$, with
\begin{equation}
\gamma_{\rm max} \approx 0.14 {\frac{\Delta u_{z}}{\Delta x_{\perp}}} \approx 0.2k_{\rm max} \Delta u_z \approx 0.4 \gamma_{\mathrm{KH},0}(k_{\rm max}). \label{KHgrowthrate}
\end{equation}
The finite width of the shear layer suppresses growth for wavenumbers greater
than $k_{c} \approx 1.8 \Delta x_{\perp}^{-1}$. From the FLASH simulations, we estimate $\Delta u_z \approx 250 \; \mathrm{km} \, \mathrm{s}^{-1}$ and $\Delta x_{\perp} \approx 60 \; \mu$m
at $t \approx 27$ ns. We therefore determine from Eq. [\ref{KHgrowthrate}] that $\gamma_{\rm max} \approx 6 \times 10^{8} \; \mathrm{s}^{-1} > L/u_{\rm rms}$ at $k_{\rm max} \approx 100 \; \mathrm{cm}^{-1}$. This confirms our original claim that the KH growth time is comparable to the outer-scale eddy-turnover rate. 

For completeness, we discuss a few caveats relating to this calculation, as well as an important implication. First, we have assumed that the KH growth rates of a finite initial perturbation of a shear layer superimposed on a turbulent environment can be estimated using linear theory derived for a simple shear profile. This assumption is not justified \textit{a priori} -- but it seems reasonable to assume that the true growth rate is not vastly different from our estimate. Secondly, our estimate also assumes that the shear layer driving the instability has a constant thickness: if instead the initial shear layers that develop in the interaction-region plasma have a smaller reversal scale, then both the KH-instability growth rate and the wavenumber of the fastest-growing KH mode could be considerably larger. Thirdly, for the parameters assumed here, we obtain the (perhaps surprising) result that the fastest-growing mode is at a wavenumber corresponding to the grid periodicity ($k_{\rm max} \approx 2 \pi/L$), rather than at smaller scales. This finding, which follows directly from the finite width of the shear layers, suggests that the effect of the KH instability on the velocity field is to convert the energy in the shear flows into outer-scale motions; in other words, our characterization in the main paper of the grid periodicity as the `driving scale' is an appropriate one. Note, however, that the most unstable KH modes have a half-wavelength ${\sim}L/2$ which is a (large) order-unity factor (${\sim}4$) greater than the characteristic reversal scale of the shears $\Delta x_{\perp}$, and so $\Delta u_{z}/\Delta x_{\perp} \approx 8 u_{\rm rms}/L$; this explains why the typical shear in the directed flows is measurably greater than the outer-scale eddy-turnover rate

\color{black}

%The viscous time $t$ satisfies $\gamma t_{\rm visc} \approx 1$, where $\gamma$ is the inviscid KH growth rate at wavenumber $k_c$.  
%Combining these conditions, it can be shown that
%
%\begin{equation}
%k_c \approx 0.25 \frac{U_{\rm S}}{\nu}, \quad t_{\rm visc} \approx 4{\frac{\nu}{U_{\rm S}}} .
%\end{equation}

%(equivalently, $k_{\rm max} \approx 0.12 U_{\rm S}/\nu$),
%we find
%\begin{equation}
%\gamma_{\rm max} \approx 0.05 \frac{U_{\rm S}^2}{\nu} 
%\end{equation}

%The growth rate for the KH instability is even larger (and the growth time scale even shorter) during the time periods when heating by the Thomson-scattering beam occurs, since $Re \approx 200 - 300$ during these times. The corresponding growth time at these values of $Re$ is $\sim 0.1 - 0.2$ ns.

%ADD COMMENT ABOUT MANY E-FOLDING TIMES

\subsection*{Dynamical significance of the magnetic fields in the FLASH simulations}

\begin{figure}
  \centering
\includegraphics[width=\linewidth]{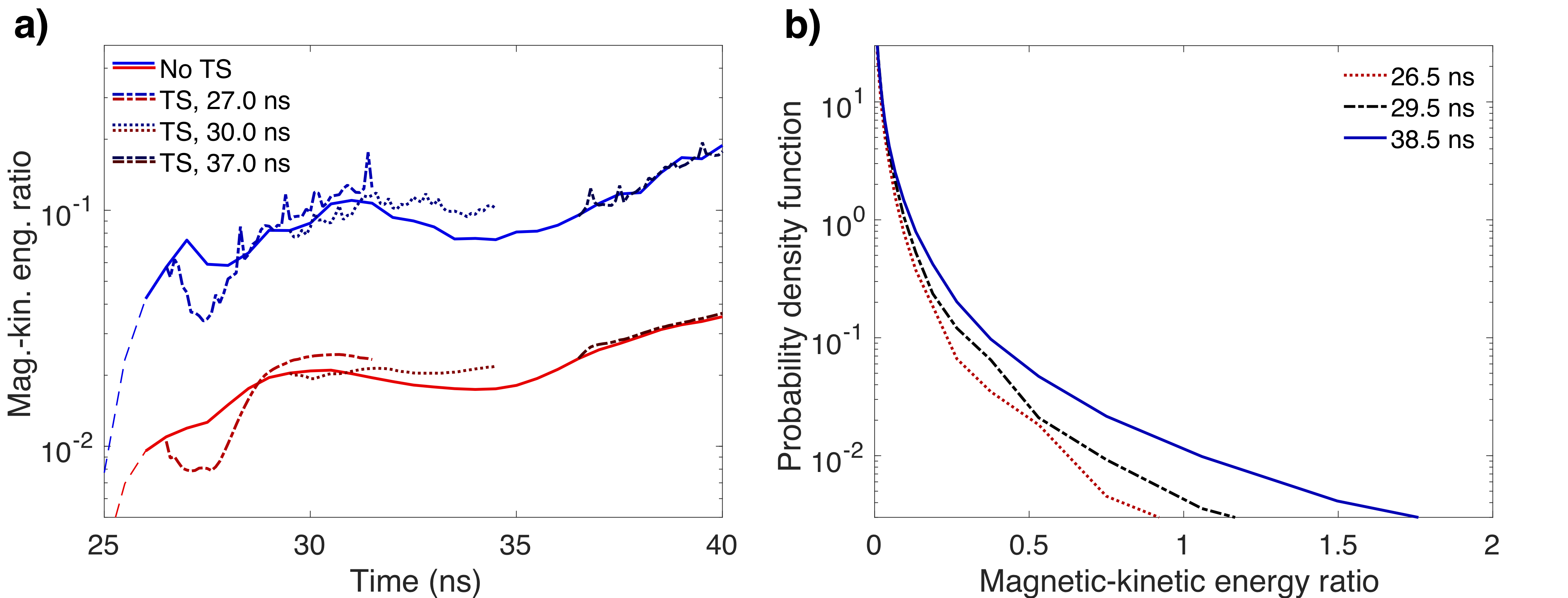}
\caption[Some properties of the magnetic field energy relative to the turbulent
kinetic energy as a
function of time in the adaptive volume in the interaction region from the FLASH
simulations.]
{\textbf{Some properties of the magnetic energy relative to the turbulent
kinetic energy as a
function of time in the adaptive volume in the interaction region from the FLASH
simulations.}
\textbf{a)} Global ratio of the total magnetic energy and the total turbulent
kinetic energy (red lines) and average of
the local ratios of the magnetic energy and the turbulent kinetic energy.
\textbf{b)} Probability density functions for the ratio of the magnetic energy
and the turbulent kinetic energy
in the spatial cells in the full adaptive volume at
different times during the experiment.}
\label{fig:magneticfieldproperties}
\end{figure}

The red curves in Figure~\ref{fig:magneticfieldproperties}a show the overall ratio of the sum of the magnetic energy in all of the cells and the sum of the turbulent kinetic energy in all of them as a function of time in the full AFV for the validated FLASH simulations. 
The ratio rises rapidly at first and then increases at a slower rate, reaching a value ${\sim}3$--$5\%$ at 40 ns. 

As discussed in the main text, the small value of global magnetic-kinetic energy ratio is quite small can, in fact, still be consistent with the magnetic field being dynamically significant in the FLASH simulations. We illustrate this in two ways. Firstly, the magnetic-kinetic energy ratio attains larger values locally than globally: the blue curves in Figure~\ref{fig:magneticfieldproperties}a show the sum of 
the ratio of the magnetic energy and the turbulent kinetic energy in each 
spatial cell in the full adaptive fiducial volume. The former is approximately five times larger than the latter. More directly, 
Figure~\ref{fig:magneticfieldproperties}b shows the probability density
function of the ratio of the magnetic energy and the turbulent kinetic energy in each spatial cell. It reveals that at later times, the ratio of the magnetic energy and the turbulent kinetic energy is comparable to, and even greater than, unity in some spatial cells, and, therefore, that the magnetic field is dynamically important at these locations in the interaction region. 
%The validated FLASH simulations also provide insights into the nature of the intermittency in the experiments.

\subsection*{Confirming the difference in magnetic-field integral scale between FLASH simulations and experiment by direct analysis of proton images}

\begin{figure}
  \centering
\includegraphics[width=\linewidth]{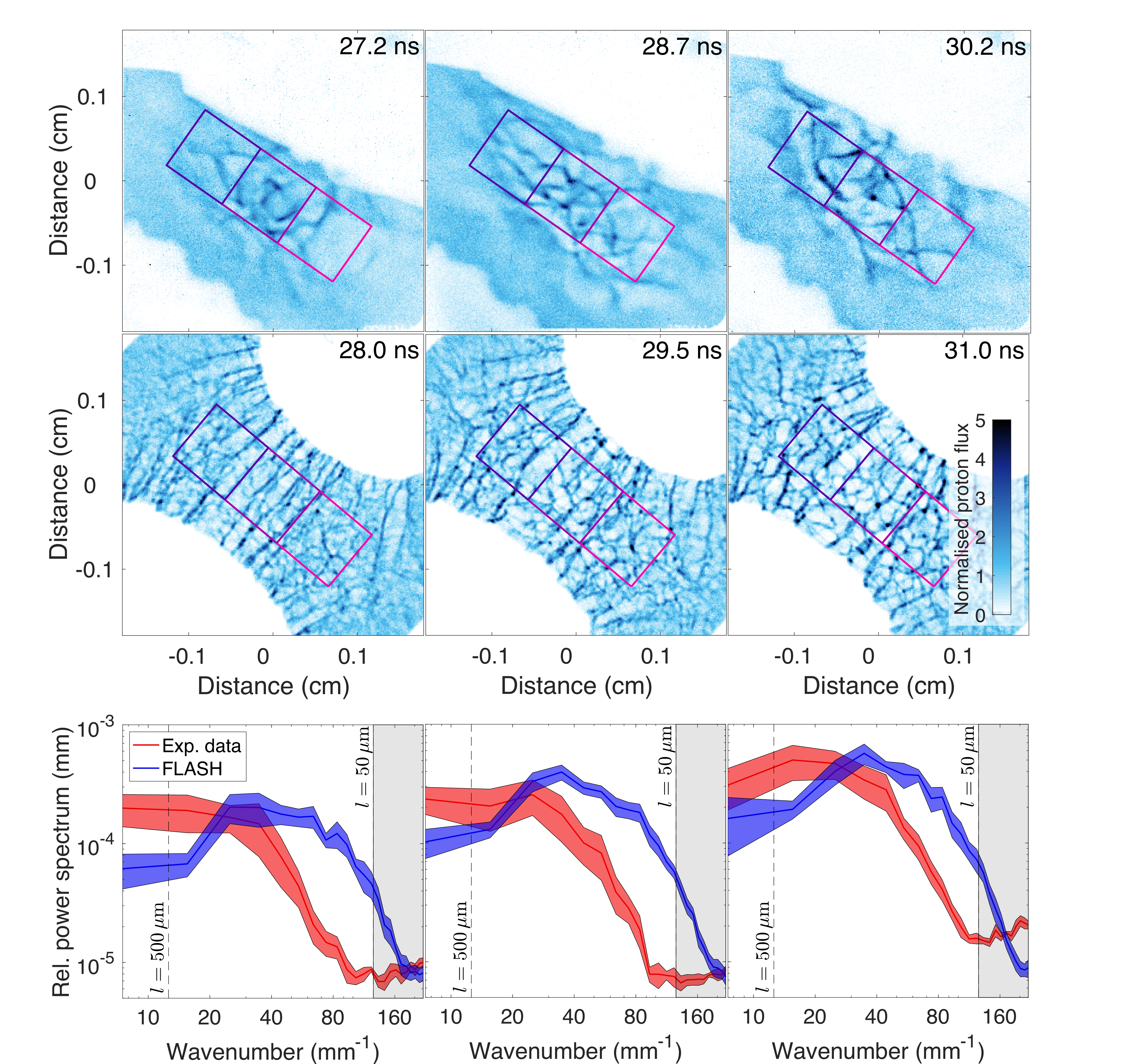}
\caption[Investigating the characteristic scale of magnetic fluctuations by direct analysis of proton images.]
{\textbf{Investigating the characteristic scale of magnetic fluctuations by direct analysis of proton images.} Top row: experimental proton images at specified times, along with an indication of regions to which spectral analysis is applied. Middle row: simulated 15.0 MeV FLASH proton images at specified times. The proton source and imaging set-up is modelled to be the same as the experiment's; in particular, the source has a finite size of $50 \, \mu$m. Bottom row: spectra of `relative' (i.e., normalized to the mean) proton flux variations. The spectra in a given plot are calculated for the experimental and simulation data from the three regions depicted in the images directly above that plot. The uncertainty in the spectra is derived from the uncertainty in the spectra across the three regions.}
\label{fig:correlationscale}
\end{figure}

In the main text, we claim that the integral scale $L_{\rm int,B}$ of the magnetic fields obtained from the experimental proton-imaging data was a factor of a few larger than the equivalent values obtained from the FLASH (and other MHD) simulations, and also that the amount of power in the measured magnetic-energy energy spectrum at wavenumbers close to the driving wavenumber was significantly larger than anticipated from simulations. However, as mentioned in the main text, it is known that the accurate recovery of the magnetic-energy spectrum from proton images via a (path-integrated) field-reconstruction algorithm is only possible in a certain imaging-parameter regime: \color{black}{specifically, when the magnetic field strength $B$ of magnetic structures with correlation scale $\ell_B$ satisfies 
\begin{equation}
B \lesssim 125 \left[\frac{\mathcal{M}}{\mathcal{M}-1}\right] \left[\frac{\mu_c}{1/2}\right] \left[\frac{W_p(\mathrm{MeV})}{15.0 \, \mathrm{MeV}}\right]^{1/2}
\left[\frac{r_i(\mathrm{cm})}{1 \, \mathrm{cm}}\right]^{-1}
\left[\frac{\ell(\mathrm{cm})}{0.01 \, \mathrm{cm}}\right]^{1/2} \left[\frac{l_p(\mathrm{cm})}{0.05 \, \mathrm{cm}}\right]^{-1/2} \mathrm{kG} \, ,
\end{equation}
where $\mathcal{M}$ is the magnification of the imaging set-up, $W_p$ the proton energy, $\mu_c$ the contrast parameter, and $r_i$ the distance between the magnetic fields being imaged and the proton source~\citep{AGGS16}.  
In this experiment, we find this condition is only marginally satisfied for the largest magnetic structures; it is therefore the case that the spectrum recovered at wavenumbers $k \gtrsim 3 \pi/L_{\rm int,B}$ will be inaccurate (viz., suppressed compared to the true spectrum) if the true spectrum follows a power law shallower than $k^{-2}$.}
\color{black}  
To confirm that the difference in integral scales is nevertheless a physical result, we calculate the one-dimensional spectrum of flux variations in the experimental proton images, and then re-apply the analysis to simulated proton images of the FLASH-simulated magnetic fields. Working with the proton-imaging data directly removes any possibility that it is the algorithm recovering the path-integrated fields that causes the discrepancy in characteristic field scale. Figure \ref{fig:correlationscale} shows the results of such an analysis at three different times after stochastic fields have been amplified in the experiment. At all times, an absence of power in the FLASH-simulated relative-flux spectra at wavenumbers $k \lesssim 25 \, \mathrm{mm}^{-1}$ is indeed observed. Thus, the discrepancy between the spatial structure of the experimentally measured and simulated fields is real.

%%% Add this line AFTER all your figures and tables
\FloatBarrier

%\section*{Heading}
%\subsection*{Subhead}
%Type or paste text here. You may break this section up into subheads as needed (e.g., one section on ``Materials'' and one on ``Methods'').

%\subsection*{Materials}
%Add a Materials subsection if you need to.

%\subsection*{Methods}
%Add a Methods subsection if you need to.

%%% Each figure should be on its own page
%\begin{figure}
%\centering
%\includegraphics[width=\textwidth]{example-image}
%\caption{First figure}
%\end{figure}

%\begin{figure}
%\centering
%\includegraphics[width=\textwidth]{frog}
%\caption{Second figure}
%\end{figure}

%\begin{table}\centering
%\caption{This is a table}

%\begin{tabular}{lrrr}
%Species & CBS & CV & G3 \\
%\midrule
%1. Acetaldehyde & 0.0 & 0.0 & 0.0 \\
%2. Vinyl alcohol & 9.1 & 9.6 & 13.5 \\
%3. Hydroxyethylidene & 50.8 & 51.2 & 54.0\\
%\bottomrule
%\end{tabular}
%\end{table}

%%% Add this line AFTER all your figures and tables
%\FloatBarrier

%\movie{Type caption for the movie here.}

%\movie{Type caption for the other movie here. Adding longer text to show what happens, to decide on alignment and/or indentations.}

%\movie{A third movie, just for kicks.}

%\dataset{dataset_one.txt}{Type or paste caption here.}

%\dataset{dataset_two.txt}{Type or paste caption here. Adding longer text to show what happens, to decide on alignment and/or indentations for multi-line or paragraph captions.}

%\bibliography{pnas-sample.bib}